\documentclass[pra,aps,amsmath,amssymb,amsfonts,twocolumn,nofootinbib,longbibliography,superscriptaddress,floatfix]{revtex4-1}
\usepackage{amssymb}
\usepackage{bm,mathrsfs}
\usepackage{graphicx}
\usepackage{epsfig}
\usepackage{amsmath,bbm}
\usepackage{amsfonts,amssymb}
\usepackage{times}
\usepackage[mathscr]{eucal}
\usepackage{verbatim}
\usepackage[sort&compress]{natbib}
\usepackage{amsmath}
\usepackage{bm}
\usepackage[colorlinks,breaklinks,linkcolor=blue,anchorcolor=blue,citecolor=blue,urlcolor=blue]{hyperref}

\begin{document}

\title{A Programmable Rydberg Quantum Bus for Nonlocal Connectivity}

\author{X. Jin}
\affiliation{Center for Quantum Science and School of Physics, Northeast Normal University, Changchun 130024,  China}

\author{F. Yang}
\email{Contact author: fan.yang@zju.edu.cn}
\affiliation{School of Physics and Zhejiang Key Laboratory of Micro-nano Quantum Chips \\ and Quantum Control, Zhejiang University, Hangzhou 310027, China}

\author{Weibin Li}
\email{Contact author: weibin.li@nottingham.ac.uk}
\affiliation{School of Physics and Astronomy, and Centre for the Mathematics and Theoretical Physics of Quantum Non-equilibrium Systems, The University of Nottingham, Nottingham NG7 2RD, United Kingdom}

\author{X. Q. Shao}
\email{Contact author: xqshao@nenu.edu.cn}
\affiliation{Center for Quantum Science and School of Physics, Northeast Normal University, Changchun 130024, China}
\affiliation{Institute of Quantum Science and Technology, Yanbian University, Yanji 133002, China}

\begin{abstract}

Scalable quantum networks require processing nodes with flexible internal connectivity, yet neutral-atom architectures remain constrained by the strong spatial dependence of native Rydberg interactions. Here we show that a Rydberg atom chain can act as a coherent quantum bus, converting a locally connected one-dimensional architecture into an effectively nonlocal interaction network. Virtual excitations in the dispersive regime mediate controllable interactions between spatially separated data units, which we derive analytically using a Green's-function continued-fraction method. The resulting mechanism is not restricted to single-excitation dynamics and supports several distinct functionalities, including Floquet-engineered chiral transport, remote entanglement of mechanical oscillators, and destructive interference for selectively suppressing unwanted dipole exchange. Simulations incorporating full long-range Rydberg interactions, atomic position fluctuations, and finite Rydberg-state lifetimes show that the mediated dynamics remain robust under experimentally relevant conditions. These results establish Rydberg chains as programmable coherent mediators for extending the internal connectivity of neutral-atom quantum nodes toward quasi-all-to-all coupling, providing a hardware-level route toward scalable and reconfigurable quantum-network architectures.

\end{abstract}

\maketitle

\section{introduction}
Nonlocal interactions play a central role in quantum information processing and quantum simulation.
By enabling coupling between spatially separated qubits, they facilitate the efficient generation of long-range entanglement~\cite{PhysRevLett.87.137901,PhysRevLett.78.3221,majer2007coupling}, significantly reduce the circuit depth required for multi-qubit operations~\cite{bravyi2018quantum,nishi2021implementation}, and support the development of scalable, modular, and distributed quantum architectures~\cite{bravyi2024high,PhysRevX.14.041030,aghaee2025scaling,zhong2021deterministic,daiss2021quantum,singh2025modular}. 
In the context of quantum many-body physics, such interactions also give rise to qualitatively distinct dynamical behavior, including accelerated information spreading and the formation of complex correlation patterns~\cite{PhysRevLett.119.170503,richerme2014non,jurcevic2014quasiparticle,RevModPhys.95.035002}. 
These properties are particularly beneficial for variational quantum algorithms and quantum optimization tasks, where the efficient exploration of high-dimensional Hilbert spaces is essential~\cite{pagano2020quantum,PhysRevA.99.052332}. 

To date, nonlocal connectivity has been successfully engineered across several prominent quantum platforms.
For instance, trapped-ion systems naturally exploit collective motion modes to achieve all-to-all connectivity~\cite{PhysRevLett.74.4091,PhysRevLett.82.1971,debnath2016demonstration,RevModPhys.93.025001}.
Superconducting circuits exploit microwave resonators or coplanar waveguide bus to mediate coherent long-range interactions between distant qubits~\cite{majer2007coupling,scarlino2019coherent}. In optical cavity-QED architectures, shared resonant modes or fiber-based photonic links provide a natural mechanism for establishing nonlocal connectivity over macroscopic distances~\cite{PhysRevLett.78.3221,ritter2012elementary,kato2019observation,grinkemeyer2025error}. 
In these schemes, the mediating modes provide a flexible and coherent channel for information transfer, allowing nonlocal interactions to be implemented in a relatively natural and controllable manner. 
In contrast, Rydberg atom arrays lack an equally natural and quantum mediating channel for coherently coupling spatially separated atoms. Although optical cavities and waveguides can provide such a channel, overcoming the inherently weak atom-photon coupling to achieve efficient and uniform interaction remains experimentally challenging.

Rydberg atom arrays have emerged as a premier platform for quantum simulation and computation due to their exceptional scalability, long coherence times, and highly controllable local interactions~\cite{shao2024rydberg,PhysRevLett.128.113602,PhysRevLett.131.170601,PhysRevA.103.022410,PhysRevLett.115.093002,PhysRevLett.128.013603,PRXQuantum.3.020303,PhysRevResearch.4.L032046,PRXQuantum.3.030325,PhysRevA.110.042612,adams2020rydberg,browaeys2020many}.
Recent breakthroughs have enabled the realization of high-fidelity gates~\cite{PhysRevLett.123.170503,evered2023high,PhysRevA.111.022420}, the preparation of large-scale entanglement states~\cite{graham2022multi,omran2019generation,PhysRevLett.123.230501,PhysRevA.95.052330,bluvstein2022quantum,bornet2023scalable,wdqt-tpwz,senoo2026high}, the application to quantum optimization tasks~\cite{ebadi2022quantum,PRXQuantum.4.010316,516n-6wtx,yeo2025approximating,3sjz-gfmr,PRXQuantum.6.020306}, and the simulation of exotic phases and dynamical phenomena~\cite{o2023entanglement,semeghini2021probing,PhysRevLett.129.090401,zhang2025probing,de2019observation,PhysRevB.101.220304,jr7l-2cfb,5qhh-322q,PhysRevA.101.043421,labuhn2016tunable,PhysRevLett.131.080403,bluvstein2021controlling,bernien2017probing,PhysRevLett.132.206503,746s-fv7x,scholl2021quantum,2gwz-65w1,zf2q-gxr1,r54t-myhc,6722-tf9c,ctwz-rhdv}. 
Such achievements demonstrate the capability of Rydberg platforms to handle complex large-scale quantum tasks and explore the frontiers of many-body physics. 
However, the native interactions between Rydberg atoms, governed by vdW interactions and dipole-dipole interactions, decay rapidly with distance~\cite{RevModPhys.82.2313,urban2009observation}.
This sharp spatial decay limits the direct interaction range, yielding negligibly weak coupling between distant modules and representing a major bottleneck for scaling up modular Rydberg architectures and to the simulation of quantum models with all-to-all connectivity.

Therefore, it is desirable to employ whether a Rydberg array itself can provide a stationary mediator for nonlocal interactions between distant data units.
In this work, we propose a theoretical model to achieve programmable nonlocal interactions mediated by a Rydberg atom chain. By exploiting dipole-dipole interactions and leveraging virtual processes in the large-detuning regime, we show that the Rydberg atom chain can generate an equivalent nonlocal coupling between data units.
On the theoretical front, we introduce the Green's function continued fraction method to derive the effective Hamiltonian in Sec.~\ref{sec2}, which avoids direct diagonalization of large matrices.
Building on this framework, we present three applications.
In Sec.~\ref{sec3}, we realize chiral transport in a 1D atomic chain with Floquet engineering. 
In Sec.~\ref{sec4}, we extend the scheme to hybrid quantum systems, where it enables entanglement generation between distant mechanical oscillators. 
In Sec.~\ref{sec5}, we consider the case where the data units are Rydberg atoms and discover a destructive interference mechanism that suppresses next-nearest-neighbor couplings without using the magic-angle configuration.
We also analyze the robustness of the proposed scheme under realistic experimental conditions in Sec.~\ref{sec6} and Sec.~\ref{sec7}, where
the numerical simulations verify the experimental feasibility of our scheme.
Finally, we present the conclusion in Sec.~\ref{sec8}.

\section{The ideal chain model}
\label{sec2}
\begin{figure}
    \centering
    \includegraphics[width=1.0\linewidth,keepaspectratio]{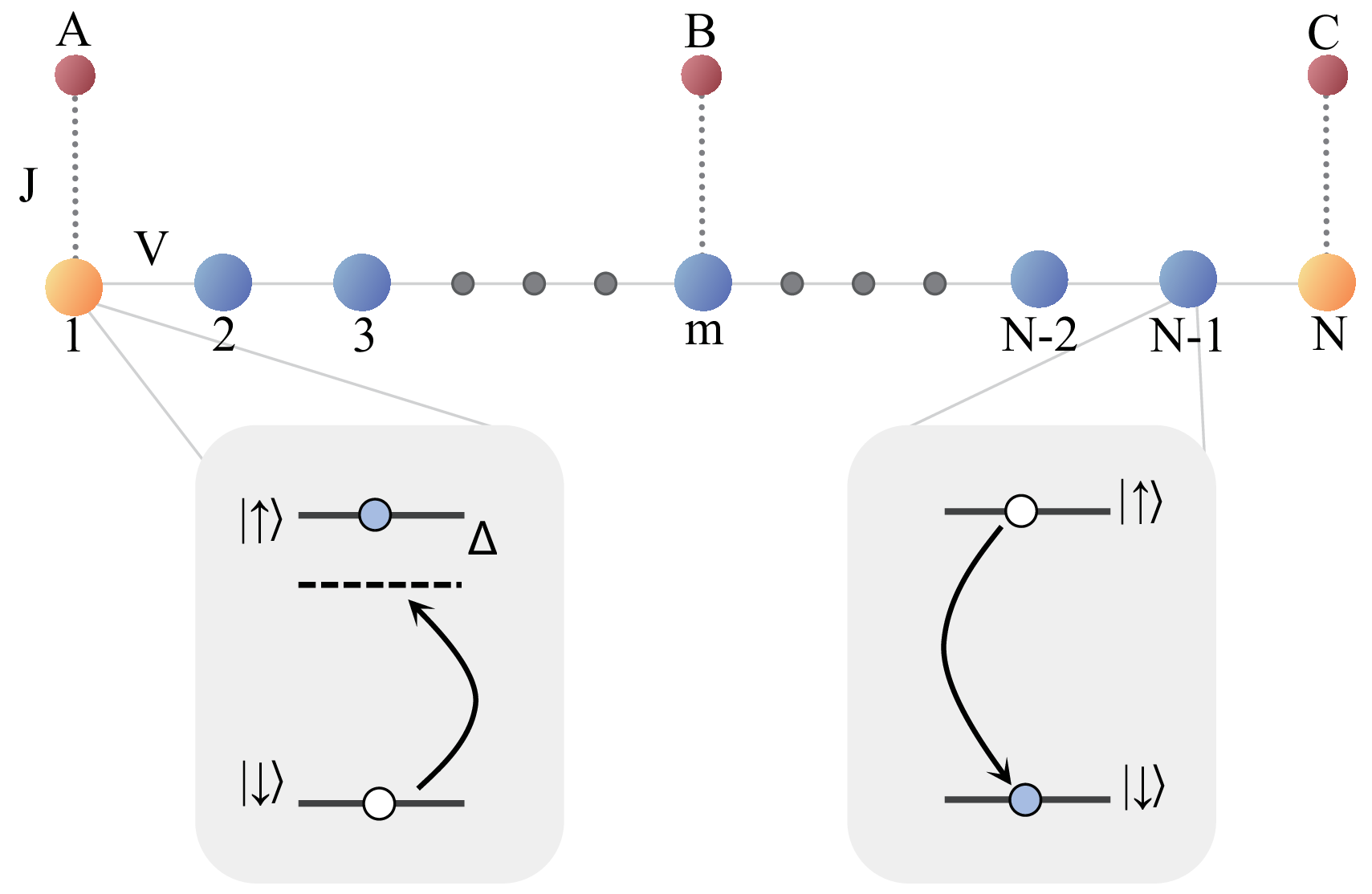}
    \caption{Schematic of the ideal model. The system consists of an atomic chain of $N$ sites, where $N$ is an odd integer. Each atom is labeled sequentially. Three data units, labeled $A$, $B$, and $C$ and depicted as red spheres, are coupled respectively to the three odd-numbered sites $1$, $m$, and $N$ of the chain. Each atom is modeled as a two-level system, with two Rydberg states denoted by $|{\uparrow\rangle}$ and $|{\downarrow\rangle}$. Within the chain, the interaction of atoms is nearest-neighbor dipole-exchange interaction with strength $V$. Sites $1$ and $N$, highlighted in orange, are subject to a detuning $\Delta$. Each of the data units couples to its nearest chain site with strength $J$.} 
    \label{fig1}
\end{figure}
We consider a quantum system consisting of a 1D chain of $N$ Rydberg atoms, as shown in Fig.~\ref{fig1}, where $N$ is an odd integer. The atomic sites in the chain are indexed sequentially by ${i \in \{1,\dots,N\}}$. Three data units, labeled by ${\mu \in \{A,B,C\}}$, are coupled to the sites $1$, $m$, and $N$ of the chain, respectively, where $m$ is an odd integer. 
Each atom is modeled as a two-level system with two Rydberg states, denoted by $|{\uparrow}\rangle$ and $|{\downarrow}\rangle$. 
In the idealized model, we retain only nearest-neighbor dipole-dipole interactions to capture the essential physics of the system. 
We note that interactions in a realistic Rydberg atom array are inherently long-range and not limited to nearest neighbors. To assess the performance of the proposed scheme under more realistic conditions, we include all pairwise interactions in the numerical simulations presented below. In addition, a local energy detuning is applied to the boundary atoms of the chain.
\begin{figure*}[t] 
    \centering  
    \includegraphics[width=1.0\linewidth,keepaspectratio]{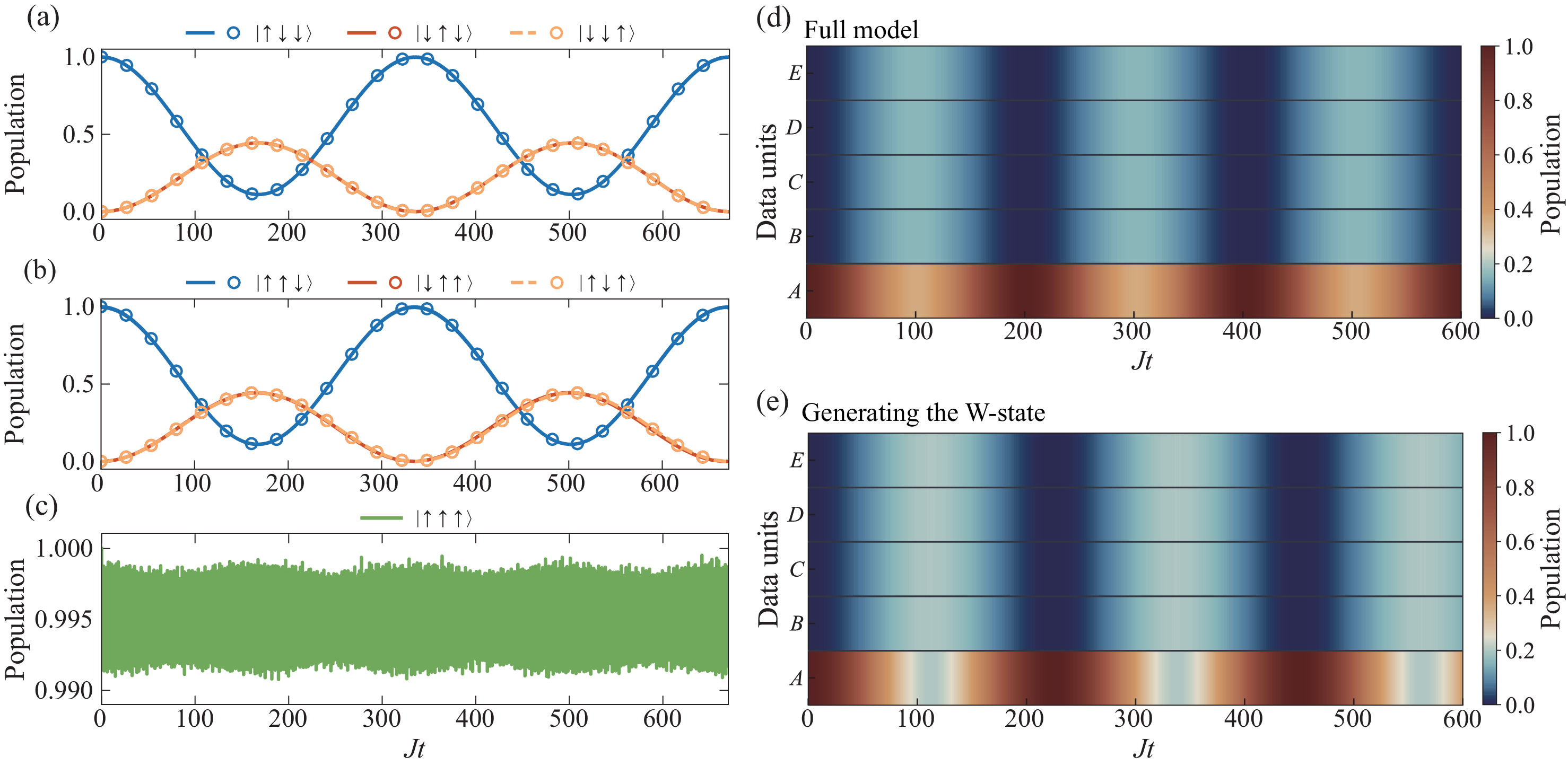}
    \caption{(a)-(c) Time evolution of the populations of the data units in a nine-atom chain coupled to three data units. (a) Dynamics in the single-excitation manifold. Blue solid, red solid, and orange dashed lines represent the populations of the excited states of data units $A$, $B$, and $C$, respectively. Initially, only site $A$ is excited to $|{\uparrow}\rangle$, while all other sites are in $|{\downarrow}\rangle$. (b) Dynamics in the double-excitation manifold, where blue solid, red solid, and orange dashed lines denote the populations of the double-excitation states. Initially, sites $A$ and $B$ are excited to $|{\uparrow}\rangle$, while all other sites are in $|{\downarrow}\rangle$. (c) Dynamics in the triple-excitation manifold, with all data units initially excited. Open circles in (a) and (b) represent the results obtained from the effective Hamiltonian.
    (d)-(e) Time evolution of the data-unit populations in a nine-atom chain coupled to five data units. 
    (d) Dynamics under the full Hamiltonian, demonstrating the effective qusi-all-to-all connectivity among the data units.
    (e) Dynamics with additional detuning compensation applied to the data units to prepare a W-state, where the detailed compensation scheme is described in Appendix~\ref{appendixB}. 
    The vertical axis labels the data units.
    The horizontal axis denotes the dimensionless time. The parameters are set to $V/J=40$ and $\Delta/J=80$.}
    \label{fig2}
\end{figure*}

In the interaction picture, the system Hamiltonian reads
\begin{equation}\label{E1}
\begin{split}
    H = \,&  \sum_{i \in \{1,N\}} \Delta_{i} n_{i} 
     + V \sum_{i=1}^{N-1} \left( \sigma^{+}_{i}\sigma^{-}_{i+1} + \text{H.c.} \right) \\
    & + \left( J_{A}\sigma^{+}_{A}\sigma^{-}_{1} +J_{B}\sigma^{+}_{B}\sigma^{-}_{m}+J_{C}\sigma^{+}_{C}\sigma^{-}_{N}+\text{H.c.}\right),
\end{split}
\end{equation}
where $n_{i}=|{\uparrow}\rangle\langle{\uparrow}|$ is the projection operator, and $\sigma^{+}_{i}=|{\uparrow}\rangle\langle{\downarrow}|$ and $\sigma^{-}_{i}=|{\downarrow}\rangle\langle{\uparrow}|$ are the raising and lowering operators of the $i$-th site. Here, $V$ represents the interaction strength between the chain atoms, and $J_A$, $J_B$, and $J_C$ denote the coupling strengths between the data units $A$, $B$, $C$ and the chain sites $1$, $m$, and $N$, respectively.

The total excitation number operator is defined as
${N}_{\text{tot}} = \sum_{i} n_i+\sum_{\mu} n_\mu$.
The dynamics can therefore be restricted to $N_\mathrm{tot}$-conserved subspaces. 
We focus on the single-excitation subspace, spanned by the basis
$\{|i\rangle,|\mu\rangle\}$,
where $|i\rangle$ ($|\mu\rangle$) denotes that the chain atom $i$ (data unit $\mu$) is in the state $|{\uparrow}\rangle$, with all other sites in $|{\downarrow}\rangle$.
The Hamiltonian can be represented as a $(N+3)\times(N+3)$ matrix
\begin{equation}
H = \left(
\begin{array}{ccccccccc}
0 & 0 & 0 & J_{A} & 0 & \cdots & 0 & \cdots & 0 \\
0 & 0 & 0 & 0 & 0 & \cdots & J_{B} & \cdots & 0 \\
0 & 0 & 0 & 0 & 0 & \cdots & 0 & \cdots & J_{C} \\
J_{A} & 0 & 0 & \Delta_{1} & V & 0 & \cdots & \cdots & 0 \\
0 & 0 & 0 & V & 0 & V & \cdots & \cdots & 0 \\
\vdots & \vdots & \vdots & 0 & V & 0 & \ddots & \cdots & \vdots \\
0 & J_{B} & 0 & \vdots & \vdots & \ddots & 0 & \ddots & 0 \\
\vdots & \vdots & \vdots & \vdots & \vdots & \vdots & \ddots & 0 & V \\
0 & 0 & J_{C} & 0 & 0 & \cdots & 0 & V & \Delta_{N}
\end{array}
\right).
\end{equation}

To derive the effective interactions between the data units mediated by the chain, we employ the Feshbach partitioning technique~\cite{feshbach1958unified,PhysRevLett.111.050402,PhysRevLett.102.080405,PhysRevA.89.032110}. The single-excitation Hilbert space is divided into two parts: a target subspace $P$, spanned by $\{|\mu\rangle\}$, and an auxiliary subspace $Q$, consisting of $\{|i\rangle\}$. Accordingly, the Hamiltonian matrix is partitioned into a block from
\begin{equation}
H_{\text{single}} = \left(
\begin{array}{cccc}
H_{PP} & H_{PQ} \\
H_{QP} & H_{QQ}\\
\end{array}
\right),
\end{equation}
where $H_{PP}$ describes the bare energies of the data units, $H_{QQ}$ represents the dynamics of the chain, and $H_{PQ}(H_{QP})$ denotes the coupling between the two subspaces.
By eliminating the $Q$ subspace, we obtain the effective Hamiltonian in the $P$ subspace
\begin{equation}
H_{\text{eff}}=H_{PP}+\Sigma(z),
\end{equation}
where $\Sigma(z)=H_{PQ}G_{QQ}(z)H_{QP}$ is the self-energy, which characterizes the effective modification of the system's properties induced by the coupling between the data units and the atom chain. $G_{QQ}(z) = (z{I}-H_{QQ})^{-1}$ is the bare Green’s function of the isolated chain, where $z = E+i\eta$ is the complex energy scan parameter and $\eta$ is an infinitesimal positive constant ensuring the retarded nature of the Green's function.

The mathematical behavior of the Green's function $G_{QQ}(z)$ is governed by the discrete spectrum $\{E_{k}\}$ of the mediated chain. While a uniform chain follows the dispersion relation ${E_{k}^{(0)}=2V\cos[k\pi/(N+1)]}$, where $k$ is the mode index, boundary detunings $\Delta$ renormalize these energy levels to ${E_{k}=\bar{\epsilon}+E_{k}^{(0)}}$, where $\bar{\epsilon}$ represents the energy shift. 
Crucially, this ensures that the resonance at ${z=0}$ is situated within a spectral gap. It is worth mentioning that even a single-ended detuning is sufficient to provide this spectral protection at the band center.
Building on this spectral separation, the elimination of the $Q$ subspace is justified in the dispersive regime, i.e., ${J_{\mu}\ll \text{min}_{k}|E_{k}|}$, which suppresses real population transfer into the chain. For the finite-size chains considered here, the spectral gap remains protected by the boundary detuning despite the increasing density of states with $N$.
Consequently, the weak energy dependence of $\Sigma(z)$ within the gap allows the self-energy to be evaluated at a zero-energy value.
The matrix elements of $\Sigma(0)$ can be expressed in terms of the bare Green’s function of the isolated chain. By explicit evaluation, we obtain the local Green’s functions at the relevant sites, 
\begin{align}
G_{11}(0) &= G_{mm}(0) = G_{NN}(0) = -\frac{1}{\Delta_{1}+\Delta_{N}}, \\
G_{1m}(0) &= -\frac{1}{\Delta_{1}+\Delta_{N}}\,(-1)^{{(m-1)}/{2}}, \\
G_{1N}(0) &= -\frac{1}{\Delta_{1}+\Delta_{N}}\,(-1)^{{(N-1)}/{2}},\\
G_{mN}(0) &= -\frac{1}{\Delta_{1}+\Delta_{N}}\,(-1)^{{(N-m)}/{2}}.
\end{align}
Using these results, the effective Hamiltonian in the single-excitation subspace can be constructed as
\begin{widetext}
\begin{equation}\label{E_eff_long}
H_{\text{eff}}^{\text{one}} = \frac{-1}{\Delta_{1}+\Delta_{N}} \begin{pmatrix}
J_A^2 & J_A J_B (-1)^{\frac{m-1}{2}} & J_A J_C (-1)^{\frac{N-1}{2}} \\
J_B J_A (-1)^{\frac{m-1}{2}} & J_B^2 & J_B J_C (-1)^{\frac{N-m}{2}} \\
J_C J_A (-1)^{\frac{N-1}{2}} & J_C J_B (-1)^{\frac{N-m}{2}} & J_C^2
\end{pmatrix}.
\end{equation}
\end{widetext}
The same calculation approach applies to higher excitation manifolds.
The detailed derivation is provided in Appendix~A.

To verify the accuracy of the derived effective Hamiltonian, we numerically simulate the time-dependent populations of the data units in different excitation manifolds for a chain of nine atoms, as illustrated in Fig. \ref{fig2}, where three data units are coupled to sites $1$, $5$, and $9$, respectively.
We set the coupling strength $J_{\mu}=J$, and scale all parameters accordingly, with $V/J=40$ and $\Delta/J=80$. 
In Fig.~\ref{fig2}, the solid lines represent the dynamics governed by the original Hamiltonian in Eq.~(\ref{E1}), and the open circles correspond to the evolution under the effective Hamiltonian.
The numerical results demonstrate an excellent agreement between the original Hamiltonian and the effective Hamiltonian. 
The population dynamics of data units $B$ and $C$ in the single-excitation manifold overlap almost perfectly, which demonstrates that the chain mediated couplings between data units are identical. A similar behavior is observed in the double-excitation manifold. 
For the triple-excitation case, the population remains nearly stationary due to the full occupancy of the data units. 

Our model can be extended to a larger number of data units. To illustrate this scalability, we consider a nine-atom chain with five data units coupled to the odd sites $1$, $3$, $5$, $7$, and $9$, respectively. Starting with a single excitation in data unit $A$, the population dynamics of all five data units exhibit qusi-all-to-all connectivity, as shown in Fig.~\ref{fig2}(d). By introducing appropriate detuning compensation, the system can further be used to prepare a W-state, as shown in Fig~\ref{fig2}(e). The detailed derivation of the compensation is provided in Appendix~\ref{appendixB}.
This example illustrates the scalability of our model and its potential for collective-state preparation. 

\section{chiral transport via Floquet engineering}
\label{sec3}

\begin{figure}
    \centering
    \includegraphics[width=1.0\linewidth]{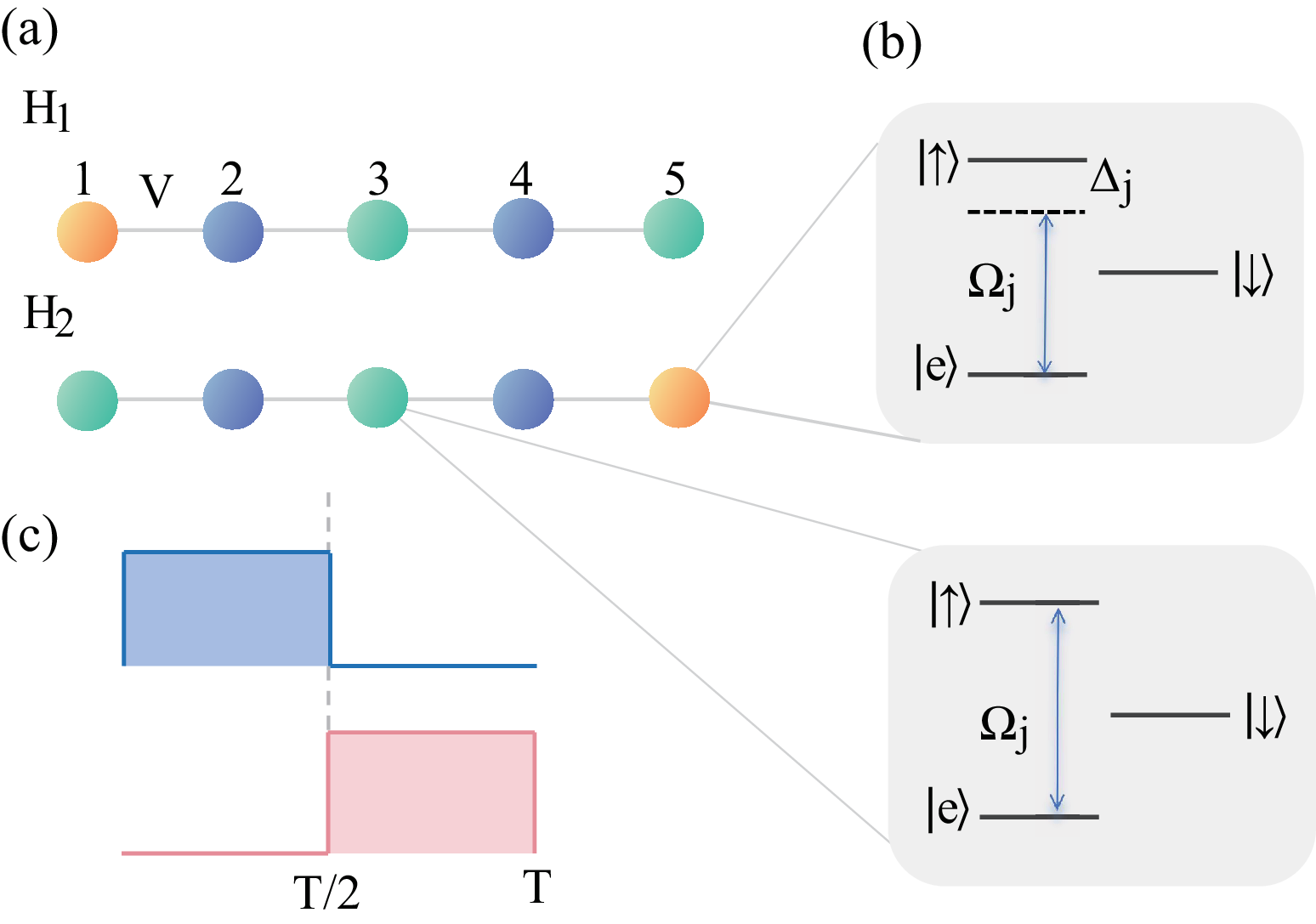}
    \caption{Schematic illustration of chiral transport via Floquet engineering. (a) A 1D chain consists of five atoms labeled from 1 to 5. Site 1, 3, and 5 are modeled as three-level systems comprising a low-lying state $|{e}\rangle$ and two Rydberg state $|{\uparrow}\rangle$ and $|{\downarrow}\rangle$, while sites 2 and 4 are two-level Rydberg atoms. Green spheres indicate sites coupled by laser fields with Rabi frequency $\Omega_{j}$, and orange spheres denote sites with a local detuning $\Delta_{j}$. The staggered configurations for $H_{1}$ and $H_{2}$ stages are shown. (b) Energy level diagrams of the driven atoms, illustrating the Rabi frequency $\Omega_{j}$ and the local detuning $\Delta_{j}$ applied to the Rydberg state. (c) The Floquet driving protocol with period $T$. The system alternates between $H_{1}$ and $H_{2}$ every half-cycle $\tau=T/2$.}
    \label{fig3}
\end{figure}

\begin{figure*}[t]  
    \centering
    \includegraphics[width=1.0\textwidth]{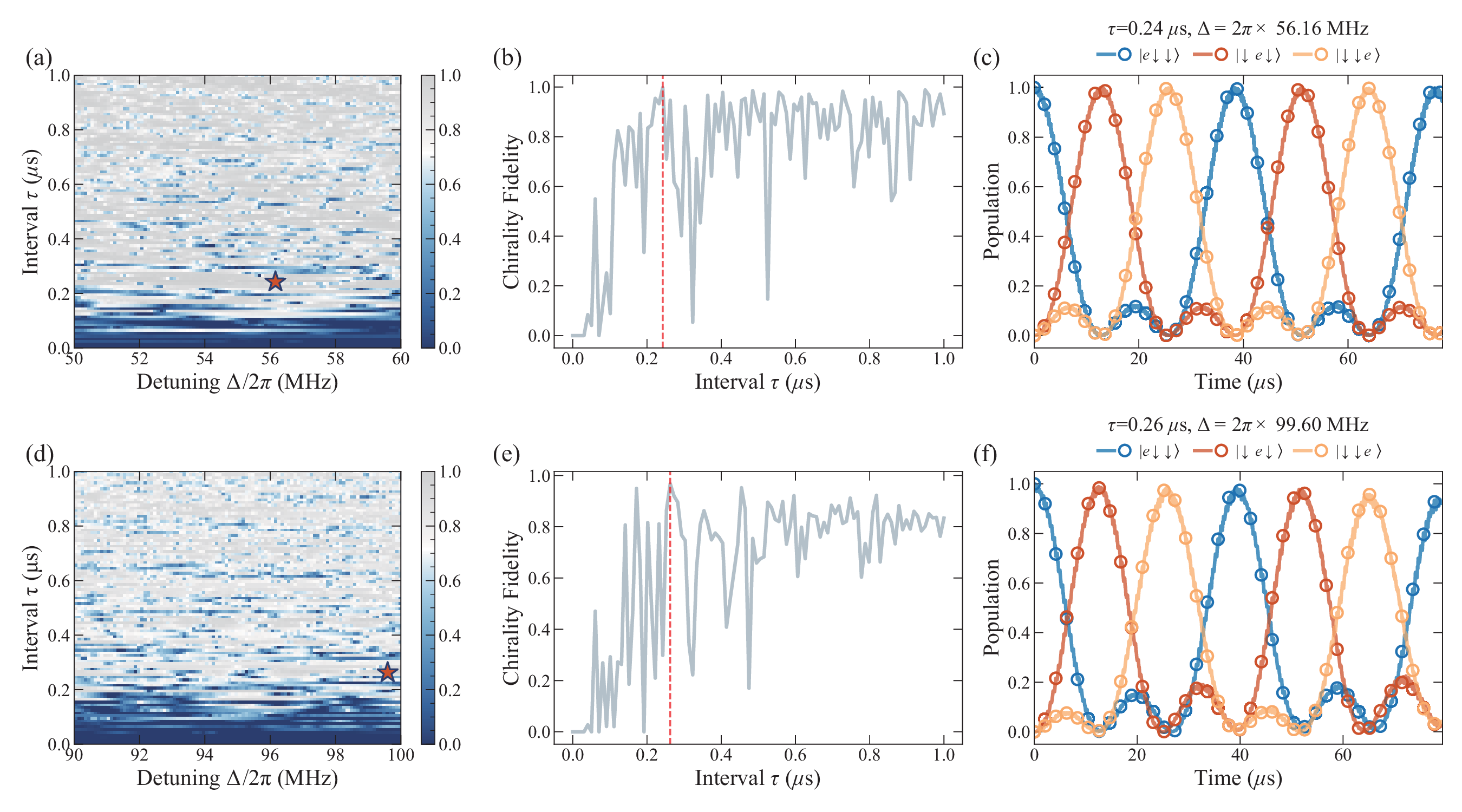} 
    \caption{The dynamics of chiral transport via Floquet engineering. Panels (a)-(c) are obtained using the ideal Hamiltonian with only nearest-neighbor interactions, while (d)-(f) use the full Hamiltonian that incorporates all long-range interactions.
    (a) and (d) Two-dimensional heat map of the fidelity as a function of the interval $\tau$ and the detuning $\Delta$. The red star marks the global maximum of the fidelity within the scanned parameter range. 
    (b) and (e) Chirality fidelity as a function of $\tau$ at the optimal detuning extracted from the heat map. The optimal interval is indicated by the red dashed line. 
    (c) and (f) Time evolution of the population at the best $\tau$. Solid lines represent the evolution under the full Hamiltonian, while open circles denote the results governed by the effective Hamiltonian $H_{\textbf{F}}$. Blue, red and orange colors correspond to the populations of atom 1, 3, and 5 being in the state $|{e}\rangle$, respectively. Parameters are set to $\Omega/2\pi=1~\text{MHz}$, $V/2\pi=30~\text{MHz}$.}
    \label{fig4}
\end{figure*}

Building on the chain-mediated interactions established above, we show that the same mechanism can be further engineered to realize chiral transport~\cite{PhysRevResearch.4.L032046,PhysRevX.10.021031,PhysRevA.105.032417}. 
In this section, we consider a 1D chain of five $^{87}\mathrm{Rb}$ atoms, as illustrated in Fig.~\ref{fig3}, where the atoms are indexed sequentially by ${i \in \{1,\dots,N\}}$. A magnetic field is applied along the chain, defining the quantization axis.

Atoms at sites 1, 3, and 5 are driven by lasers. The relevant level structure consists of a ground state $|e\rangle = |5S_{1/2}, F = 1, m_F = 0\rangle$, an intermediate state $|p\rangle = |5P_{1/2}, F = 2, m_F = 0\rangle$, and two Rydberg states $|{\uparrow}\rangle = |60 S_{1/2}, m_{j} = 1/2\rangle$ and $|{\downarrow}\rangle = |60P_{1/2}, m_j = -1/2\rangle$. 
The transition between the ground state $|e\rangle$ and the Rydberg state $|{\uparrow}\rangle$ is implemented via a two-photon process. The state $|e\rangle$ is first excited to the intermediate state $|p\rangle$ by a local $\pi$-polarized laser at a wavelength of $795~\text{nm}$, and this intermediate state is subsequently coupled to the Rydberg state $|{\uparrow}\rangle$ by another local $\pi$-polarized laser at a wavelength of $474~\text{nm}$. After adiabatically eliminating the intermediate state, the atoms at these sites can be effectively reduced to a three-level system comprising the states $|e\rangle$, $|{\uparrow}\rangle$, and $|{\downarrow}\rangle$. 
In this simplified framework, the effective coupling between $|{e}\rangle$ and $|{\uparrow}\rangle$ at sites $j\in\{1,3,5\}$ is described by a complex Rabi frequency $\Omega_{j}e^{i\phi_{j}}$, with site-dependent detuning $\Delta_{j}$.
The atoms at sites 2 and 4 are not directly driven by the lasers and thus remain within the Rydberg manifold, effectively forming a two-level system spanned by $|{\uparrow}\rangle$ and $|{\downarrow}\rangle$.

This physical implementation can be directly mapped onto the ideal model. The state $|{e}\rangle$ corresponds to the data units, and the laser-induced coupling between $|{e}\rangle$ and $|{\uparrow}\rangle$ maps onto the coupling between the data units and specific sites of the chain $J_{\mu}$. 
Based on this correspondence, we introduce two driving configurations by selectively applying detunings at different sites. Specifically, in configuration $H_1$, only site $1$ has detuning, whereas in configuration $H_2$, only site 5 has detuning.
These two configurations correspond to distinct static realizations of the same system and are described by Hamiltonians 
\begin{equation}
\begin{split}
    H_{1} =   & \Delta_{1} n_{1} 
     + V \sum_{i=1}^{4} \left( \sigma^{+}_{i}\sigma^{-}_{i+1} + \text{H.c.} \right) \\
    & + \sum_{j \in \{1,3,5\}}\left(\Omega_{j} e^{i\phi_{j}}  |{\uparrow}\rangle_{j}\langle{e}|_{j}+\text{H.c.}\right),
\end{split}
\end{equation}
and 
\begin{equation}
\begin{split}
    H_{2} =   & \Delta_{5} n_{5} 
     + V \sum_{i=1}^{4} \left( \sigma^{+}_{i}\sigma^{-}_{i+1} + \text{H.c.} \right) \\
    & + \sum_{j \in \{1,3,5\}}\left(\Omega_{j} |{\uparrow}\rangle_{j}\langle{e}|_{j}+\text{H.c.}\right).
\end{split}
\end{equation}

We now work in the dispersive regime, and restrict the dynamics to the single-excitation subspace. Following the same perturbative approach, we obtain the effective Hamiltonians.
By defining the relative laser phases $\phi_{mn}=\phi_{m}-\phi_{n}$, the effective Hamiltonians can be written as
\begin{equation}
H_{\text{eff}}^{1} = \frac{1}{\Delta_1} 
\begin{pmatrix}
-\Omega_1^2 & \Omega_1 \Omega_{3}e^{i\phi_{13}} & -\Omega_1 \Omega_5e^{i\phi_{15}} \\[6pt]
\Omega_3 \Omega_1e^{i\phi_{31}} & -\Omega_3^2 & \Omega_3 \Omega_5e^{i\phi_{35}} \\[6pt]
-\Omega_5 \Omega_1e^{i\phi_{51}} & \Omega_5 \Omega_3e^{\phi_{53}} & -\Omega_5^2
\end{pmatrix},
\end{equation}
and 
\begin{equation}
H_{\text{eff}}^{2} = \frac{1}{\Delta_5} 
\begin{pmatrix}
-\Omega_1^2 & \Omega_1 \Omega_3 & -\Omega_1 \Omega_5 \\[6pt]
\Omega_3 \Omega_1 & -\Omega_3^2 & \Omega_3 \Omega_5 \\[6pt]
-\Omega_5 \Omega_1 & \Omega_5 \Omega_3 & -\Omega_5^2
\end{pmatrix}.
\end{equation}
The effective Hamiltonians above are written in the basis $|{e}\rangle_{j}$, where the atom $j$ is in $|{e}\rangle$ and all other atoms are in $|{\downarrow}\rangle$.
Under the condition $\Omega_1 = \Omega_3= \Omega_5 = \Omega$, the effective Hamiltonians exhibits identical couplings mediated by the chain. The phases of the couplings are determined solely by the laser phases $\phi_j$. As a result, the 1D chain behaves like an enclosed equilateral triangle geometry, which provides the platform for chiral transport. 

We induce the required synthetic magnetic flux by dynamically switching between the two configurations, implemented via a piecewise time-dependent Hamiltonian. 
Specifically, within each Floquet cycle of duration $T$, the system evolves under
\begin{equation}
\label{original H}
H(t)=
\begin{cases}
H_1, & t \in [0, T/2) \\
H_2, & t \in [T/2, T)
\end{cases}
\end{equation}
where each Hamiltonian acts for a duration $\tau=T/2$. The evolution over one Floquet period is described by the unitary operator $U(T)={e}^{-iH_{\text{2}}\tau}{e}^{-iH_{\text{1}}\tau}$. In the short-period limit, the stroboscopic evolution can be approximated using the Trotter expansion as $U(T)\approx{e}^{-i(H_{\text{1}}+H_{\text{2}})\tau}$. This defines an effective Hamiltonian $H_{\text{eff}}$ via $U(T)={e}^{-iH_{\text{eff}}T}$, yielding $H_{\text{eff}}=(H_{\text{eff}}^{1}+H_{\text{eff}}^{2})/2$.

We set equal laser driving strength ${\Omega}_{1}={\Omega}_{3}={\Omega}_{5}={\Omega}$. The laser phases are as follows. 
For $H_{1}$, the phases are chosen as $\phi_{1}=0$, $\phi _{3}=2\pi/3$, and $\phi_{5}=4\pi/3$, while for $H_{2}$, all phases are set to zero. Under these conditions, the  Hamiltonians $H_{1}$ and $H_{2}$ effectively correspond to the two laser dressing fields (denoted as $A$ and $B$) introduced in Ref.~\cite{PhysRevResearch.4.L032046}.
By substituting $H_{\text{eff}}^{1}$ and $H_{\text{eff}}^{2}$ into $H_{\text{eff}}$, the effective coupling matrix element for the transition from $|{e}\rangle_{1}$ to $|{e}\rangle_{3}$ is expressed as
$\frac{\Omega^2}{2} \left( \frac{1}{\Delta_5} + \frac{e^{i\phi_{31}}}{\Delta_1} \right) = \frac{\Omega^2}{4\Delta_1\Delta_5} \left[ (2\Delta_1 - \Delta_5) + i\sqrt{3}\Delta_5 \right].$
The corresponding transition acquires a Peierls phase $\Phi$, given by 
\begin{equation}
\label{E21}
\tan\Phi = \frac{\sqrt{3}\Delta_{5}/\Delta_{1}}{2 - \Delta_{5}/\Delta_{1}}.
\end{equation}
For $\Delta_{5}/\Delta_{1}=1/2$, the Peierls phase reduces to $\Phi=\pi/6$, yielding a total synthetic magnetic flux $\Phi_{\text{tot}}=\pi/2$, resulting in clockwise 
chiral motion.

\begin{figure}
    \centering
    \includegraphics[width=1.0\linewidth]{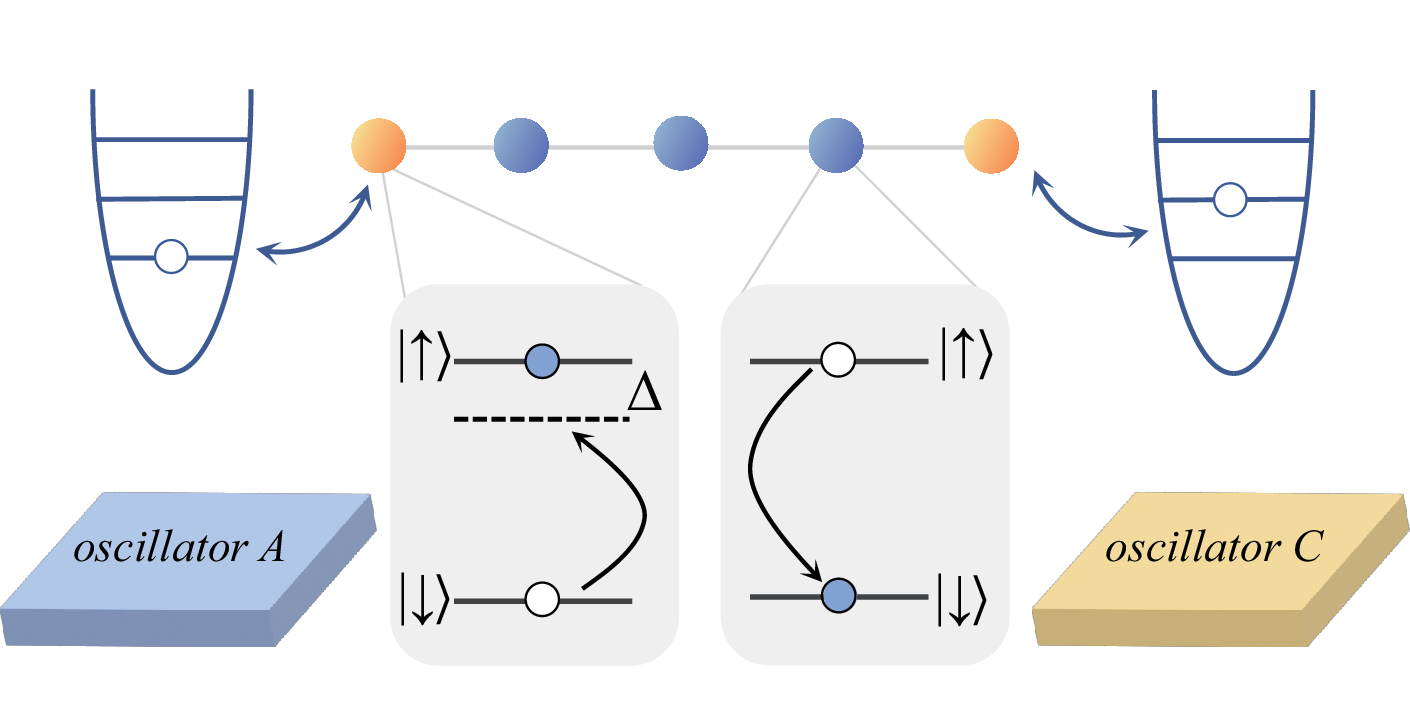}
    \caption{Schematic of the entanglement generation between mechanical oscillators mediated by a Rydberg atom chain. The 1D chain consists of five atoms, each modeled as a simple two-level system with two Rydberg states denoted as $|{\uparrow}\rangle$ and $|{\downarrow}\rangle$. The interaction is considered as a nearest-neighbor dipole-dipole interaction. The two edge atoms, highlighted in orange, exhibit a detuning $\Delta$. Two mechanical oscillators, labeled A and C, are coupled to the atoms at the boundaries of the chain, respectively.}
    \label{fig5}
\end{figure}

Building on the above analysis, we perform numerical simulations based on the original Hamiltonian Eq.~(\ref{original H}). To quantitatively characterize the chiral transport, we define a chirality fidelity function
\begin{equation}\label{E22}
    F=(p_{3,max}\times p_{5,max})-0.5\times|{p_{3,max}-p_{5,max}|},
\end{equation}
where $p_{j,max}$ denotes the maximum population that $|{e}\rangle_{j}$ state can reach during the evolution. The product term enhances simultaneous population of both sites, while the penalty term enforces symmetry.
A larger value of $F$ indicates a more efficient and symmetric population transfer.
We initialize the system in $|{e}\rangle_{1}$ and investigate the population of $|{e}\rangle_{3}$ and $|{e}\rangle_{5}$. In the simulations, the detuning ratio is fixed at $\Delta_{1}=2\Delta_{5}\equiv\Delta$. 
The interaction strength is set to $V/2\pi=30~\text{MHz}$, and the driven Rabi frequency is set to $\Omega/2\pi=1~\text{MHz}$.
We scan the detuning strength $\Delta$ and driving duration $\tau$. To reduce decoherence effects arising from long-time evolution, we restrict the duration $\tau$ to the range $[0~\mu\text{s},1~\mu\text{s}]$ and the detuning $\Delta/2\pi$ to the range $[50~\text{MHz},60~\text{MHz}]$.
The results are shown in Fig.~\ref{fig4}(a), where the red star marks the global optimal chirality point.
 Figure~\ref{fig4}(b) shows the fidelity as a function of $\tau$ at the optimal detuning, where the red dashed line indicates the optimal duration $\tau$.
Utilizing the optimal parameters obtained, the population dynamics for the clockwise sequence are presented in Fig.~\ref{fig4}(c).
The solid lines represent the evolution under the original Hamiltonian Eq.~(\ref{original H}), while the open circles denote the stroboscopic dynamics captured by the Floquet Hamiltonian$
H_{F} = \frac{i}{T} \ln\left(e^{-iH_{2}\tau} e^{-iH_{1}\tau}\right)$,
with $T=2\tau$~\cite{scully1998quantum}.
Meanwhile, to account for more realistic experimental conditions, we further include the inherent long-range interactions between Rydberg atoms. 
In this case, the above optimization procedure remains applicable, and the system retains the desired chiral transport behavior, as shown in Fig.~\ref{fig4}(d)-(f).

\section{realizing entanglement in mechanical oscillators}
\label{sec4}

\begin{figure}
    \centering
    \includegraphics[width=1.0\linewidth]{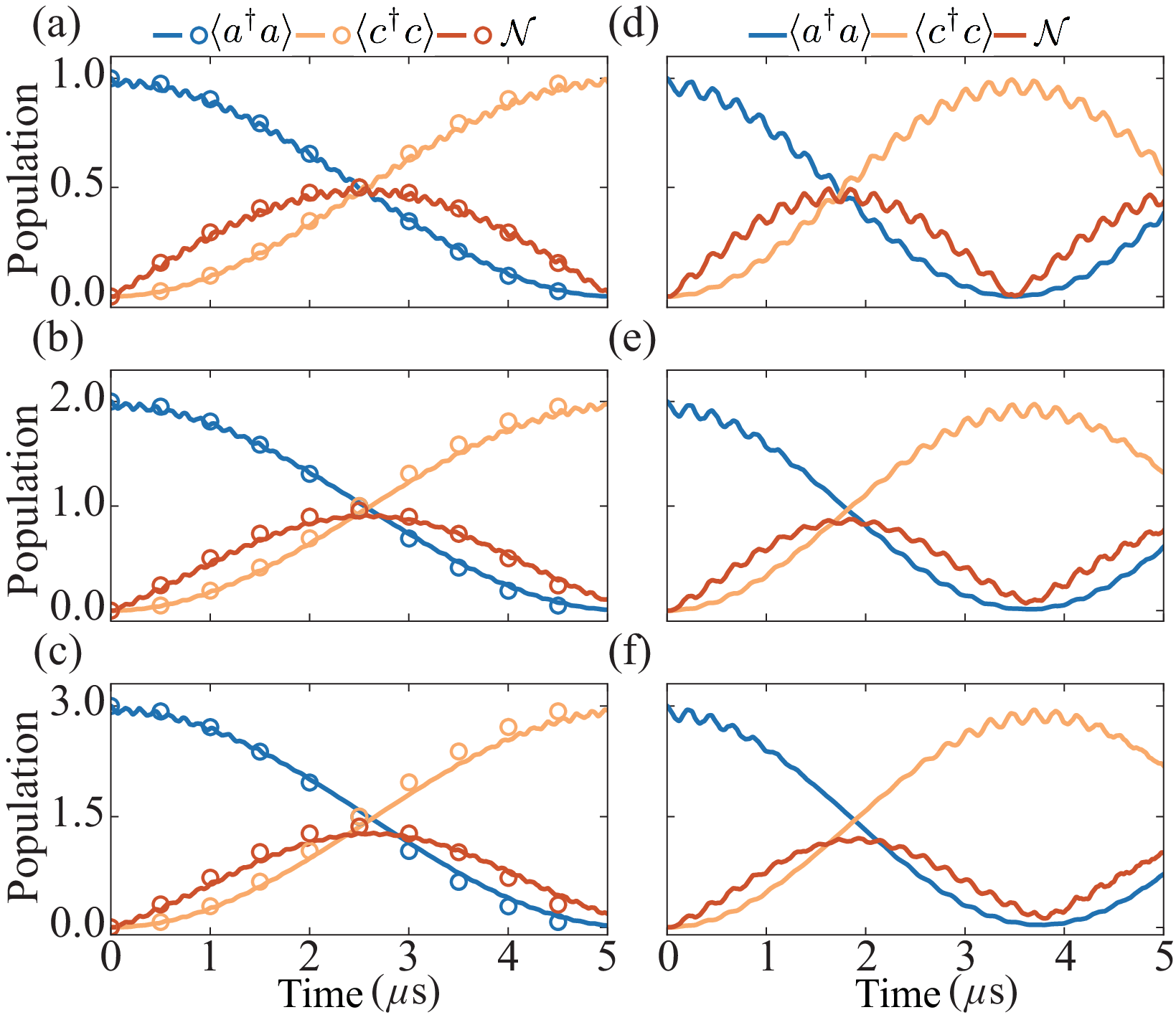}
    \caption{Time evolution of the mean excitation numbers and negativity. The plots show the evolution of mean excitation numbers $\langle{a^{\dagger}a\rangle}$ and $\langle{c^{\dagger}c\rangle}$ for mechanical oscillators A and C, along with the entanglement negativity. The blue and orange solid lines represent the mean excitation evolution of oscillators A and C, respectively, while the red solid lines denote the system negativity. Solid lines are the results obtained from the full Hamiltonian, whereas the open circles represent the evolution governed by the effective Hamiltonian. (a)-(c) show the evolution of the system in the $n$-excitation subspace($n$=1,2,3) under the ideal nearest-neighbor interaction model, where in all cases the initial state is prepared with $n$ excitations initially concentrated in oscillator A.
    (d)-(f) show the corresponding results with all long-range dipole-dipole interactions taken into account.
    Parameters are set to $J/2\pi=1~\text{MHz}$, $V/2\pi=15~\text{MHz}$, and $\Delta/2\pi=10~\text{MHz}$.}
    \label{fig6}
\end{figure}

The theoretical framework established in Sec.~\ref{sec2} shows that a Rydberg atomic chain can mediate nonlocal effective interactions. This mechanism serves as a resource for generating nonlocal quantum correlations. 
In this section, we show that this approach can be extended to a hybrid quantum platform, enabling the generation of quantum entanglement between two spatially separated mechanical oscillators~\cite{wind2025entanglement}.

We consider a 1D chain of five $^{87}\text{Rb}$ atoms, as illustrated in Fig.~\ref{fig5}. The atomic sites are indexed sequentially by $i \in \{1, \dots, 5\}$. 
Two mechanical oscillators, labeled by $\mu \in \{A, C\}$, are coupled to the edge atoms of the chain. 
Each atom is treated as a two-level system with Rydberg states $|{\uparrow}\rangle = |60 S_{1/2}, m_{j} = 1/2\rangle$ and $|{\downarrow}\rangle = |60P_{1/2}, m_j = -1/2\rangle$. 
The atoms at the boundaries of the chain are subject to a detuning $\Delta$. 
We consider nearest-neighbor dipole-dipole interactions between the atoms. In the interaction picture, the system Hamiltonian is given by 
\begin{equation}\label{E20}
\begin{split}
    H = \,& \sum_{i \in \{1,5\}} \Delta n_{i} 
     + V \sum_{i=1}^{4} \left( \sigma^{+}_{i}\sigma^{-}_{i+1} + \text{H.c.} \right) \\
    & + J \left( a^\dagger \sigma^{-}_{1} + c^\dagger \sigma^{-}_{5} + \text{H.c.} \right),
\end{split}
\end{equation}
where $a^\dagger(a)$ and $c^\dagger(c)$ are the creation (annihilation) operators for oscillators A and C, respectively. Here, $V$ represents the dipole-dipole interaction strength, and $J$ denotes the coupling strength between the oscillators and the edge atoms. The operator $n_{i}=|{\uparrow}\rangle\langle{\uparrow}|$ is the projection operator, while $\sigma^{+}_{i}=|{\uparrow}\rangle\langle{\downarrow}|$ and $\sigma^{-}_{i}=|{\downarrow}\rangle\langle{\uparrow}|$ are the raising and lowering operators of the atoms. 

This physical implementation can be directly mapped onto the ideal model introduced in Sec.~\ref{sec2}, where the mechanical oscillators serve as the data units.
In the dispersive regime, the same formalism developed in Sec.~\ref{sec2} can be applied to derive the effective Hamiltonian. 
However, in contrast to the two-level data units considered previously, the mechanical oscillators are bosonic modes, allowing multiple excitations to occupy each oscillator.
We define the total excitation number operator as
${N}_{\text{tot}} = \sum_{i} n_i+\sum_{\mu} n_\mu$.
In this section, we assume that the initial excitations are prepared in the mechanical oscillators, while the atomic chain remains in the ground state $|{\downarrow}\rangle$.
As a result, the atomic excitations are only virtually populated.
Within the subspace of a fixed total excitation number $N_{\mathrm{tot}}$, the effective Hamiltonian acts on a Hilbert space of dimension $N_{\mathrm{tot}}+1$, spanned by the bosonic states $\{|n_A, n_C\rangle\}$, where $n_A+n_C=N_{\text{tot}}$.
In this subspace, the effective Hamiltonian takes a universal form
\begin{widetext}
\begin{equation}
H_{\mathrm{eff}}^{(N_{\mathrm{tot}})} = \frac{-J^2}{2\Delta}
\begin{pmatrix}
N_{\mathrm{tot}} & \sqrt{N_{\mathrm{tot}}} & 0 & \cdots & 0 \\
\sqrt{N_{\mathrm{tot}}} & N_{\mathrm{tot}} & \sqrt{2(N_{\mathrm{tot}}-1)} & \cdots & 0 \\
0 & \sqrt{2(N_{\mathrm{tot}}-1)} & N_{\mathrm{tot}} & \cdots & 0 \\
\vdots & \vdots & \vdots & \ddots & \sqrt{N_{\mathrm{tot}}} \\
0 & 0 & 0 & \sqrt{N_{\mathrm{tot}}} & N_{\mathrm{tot}}
\end{pmatrix}.
\end{equation}
\end{widetext}

To evaluate the degree of entanglement between the two mechanical oscillators, we employ the negativity $\mathcal{N}(\rho_{ac})$. For oscillators A and C, the reduced density matrix $\rho_{ac}$ is obtained by tracing over the Rydberg chain. The negativity is defined by 
\begin{equation}
\mathcal{N}(\rho_{ac}) = \frac{1}{2} \left( ||{T_{a}} \text{tr}_{chain}(\rho) ||_1 - 1 \right),
\end{equation} 
where $\rho$ is the density matrix of the full system, ${T_{a}}$ indicates the partial transpose with respect to oscillator A, and $||\cdot||_{1}$ denotes the trace norm~\cite{PhysRevA.65.032314,RevModPhys.81.865,wind2025entanglement}. 
For a given excitation cutoff $N_{\text{tot}}$, the maximally entangled state yields $\mathcal{N} = N_{\text{tot}}/2$. 
This bound indicates that the initial excitation number fundamentally limits the maximum entanglement generated during the evolution.

We perform numerical simulations for the cases where oscillator A is initially prepared with single, double, and triple excitations, as shown in Fig.~\ref{fig6}. Figures~\ref{fig6} (a)-(c) correspond to the ideal nearest-neighbor interaction model. 
The maximum negativity reaches $0.4972$, $0.9118$, and $1.2799$ in the single, double, and triple-excitation manifolds, respectively, approaching the corresponding theoretical bounds.
When all long-range interactions are taken into account, our scheme still successfully generates entanglement between the mechanical oscillators, as shown in Figs.~\ref{fig6}(d)-(f). 
The corresponding maximal negativity reach $0.4946$, $0.8748$, and $1.2083$, respectively.

\section{Rydberg-array-mediated destructive interference of dipole exchange}
\label{sec5}

In this section, we investigate the case where the data units are Rydberg atoms. 
In such a system, the inherent long-range dipole-dipole interactions coexist with the nonlocal interactions mediated by the Rydberg chain. 
The interplay between these two interaction mechanisms gives rise to destructive interference, which can be exploited to suppress the next-nearest-neighbor couplings.
\begin{figure}
    \centering
    \includegraphics[width=1.0\linewidth]{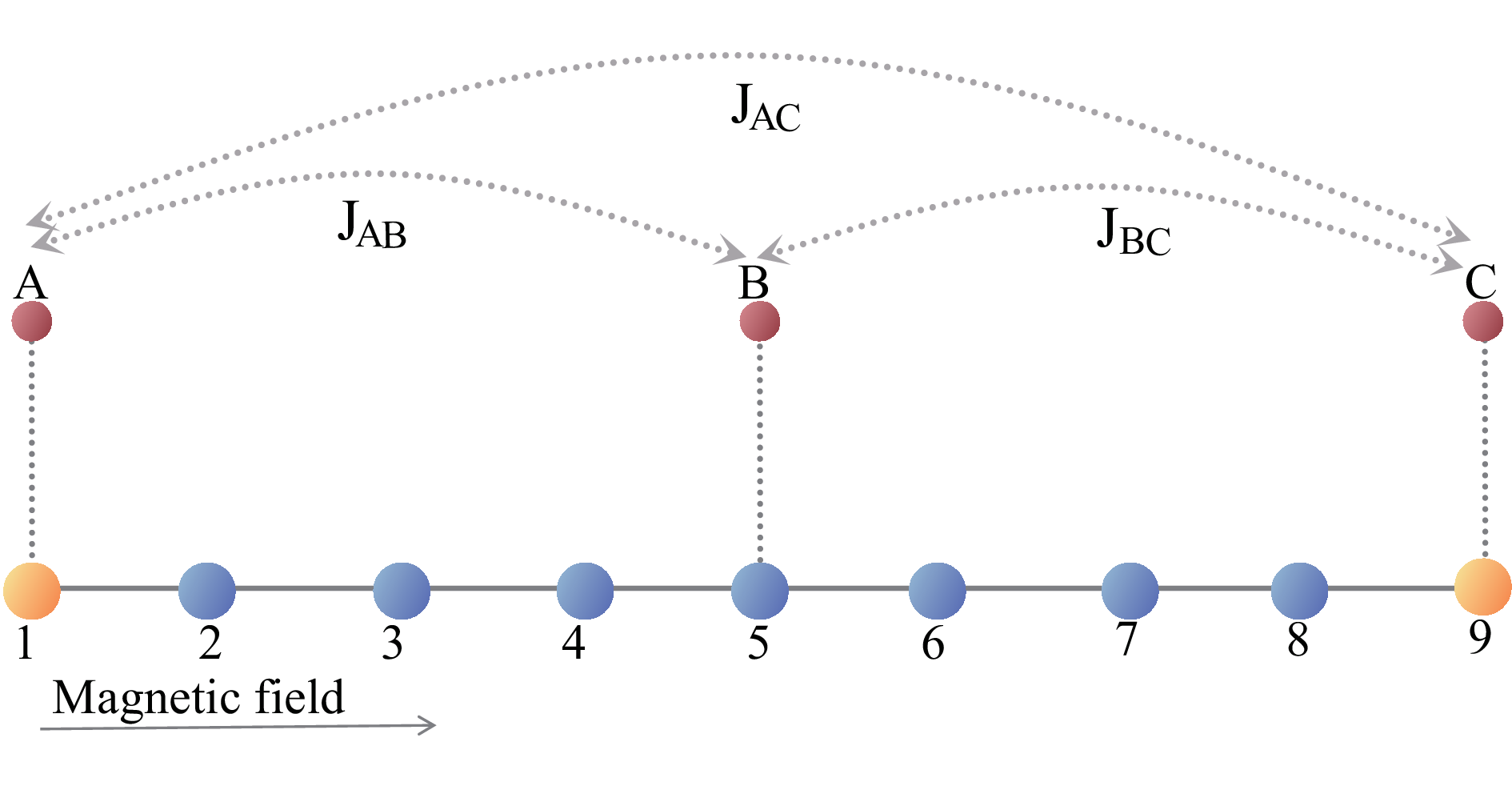}
    \caption{Schematic illustration of dipole-exchange transparency induced by the Rydberg atom chain. The $1\text{D}$ atomic chain consists of nine Rydberg atoms, with the two edge atoms(highlighted in orange) subject to a detuning $\Delta$. Each atom is labeled sequentially. Three data atoms, labeled as $A$, $B$, and $C$ and marked in red, are positioned directly above sites $1$, $5$, and $9$, respectively. Each atom is treated as a simple two-level system. Due to the long-range nature of the dipole-dipole interaction, each atomic pair is mutually coupled. The direct interactions among the three data atoms are denoted by $J_{AB}$, $J_{BC}$, and $J_{AC}$. The magnetic field is directed along the chain. }
    \label{fig7}
\end{figure}

In Rydberg atom arrays, the conventional approach to suppressing the dipole-dipole interaction between selected Rydberg atom pairs is to arrange the interatomic axis at the magic angle $\theta_m = \arccos(1/\sqrt{3})$ with respect to the quantization axis.
However, satisfying this magic-angle condition for non-nearest-neighbor pairs  becomes extremely challenging in complex two- or three dimensional arrays. 
To overcome this limitation, we propose a scheme that leverages the Rydberg chain as a tunable mediator to generate an additional interaction pathway between the data atoms. 
By properly choosing the detuning of the boundary atoms, the chain-mediated interaction can be engineered to cancel the direct dipole-exchange coupling via destructive interference, thereby selectively eliminating the next-nearest-neighbor coupling while preserving the desired nearest-neighbor exchange channels.

To elucidate the underlying mechanism, we consider a 1D Rydberg chain composed of nine $^{87}\mathrm{Rb}$ atoms labeled by $i,j \in \{1,\dots,9\}$. 
Three data atoms, denoted by $\mu,\nu \in \{A,B,C\}$, are positioned above the sites $1$, $5$, and $9$ of the chain, respectively, and lie along a straight line, as shown in Fig.~\ref{fig7}.
Each atom is modeled as an effective two-level system with Rydberg states $|{\uparrow}\rangle = |60S_{1/2}, m_J = 1/2\rangle$ and $|{\downarrow}\rangle = |60P_{1/2}, m_J = -1/2\rangle$.
We impose equal detunings on the boundary atoms, $\Delta_1 = \Delta_9 \equiv \Delta$. 
For simplicity, the chain is restricted to nearest-neighbor interactions, and each data atom couples only to its nearest chain site with equal strength $J$, i.e., $A$ ($B$, $C$) to sites $1$ ($5$, $9$), respectively. 
In addition to the chain-mediated coupling, the data atoms also subject to direct dipole-dipole exchange interactions.
The Hamiltonian of the system is given by
\begin{equation}
\begin{split}
H = \,&  \sum_{i \in \{1,9\}} \Delta \, n_{i} 
+ V \sum_{i=1}^{8} \left( \sigma^{+}_{i}\sigma^{-}_{i+1} + \text{H.c.} \right) \\
& + J\left( \sigma^{+}_{A}\sigma^{-}_{1} + \sigma^{+}_{B}\sigma^{-}_{5}+\sigma^{+}_{C}\sigma^{-}_{9}+\text{H.c.} \right) \\[6pt]
& + \sum_{\mu \neq \nu} J_{\mu\nu} \left( \sigma^{+}_{\mu}\sigma^{-}_{\nu} + \text{H.c.} \right).
\end{split}
\end{equation}
In the dispersive regime, following the approach outlined in Sec.~\ref{sec2}, we derive an effective Hamiltonian for the system. 
Notably, the data atoms exist directly coupled via dipole-dipole interactions. 
As a result, the Hamiltonian projected onto the data subspace contains nonvanishing off-diagonal terms. 
Within the single-excitation subspace, the effective Hamiltonian takes the form
%\begin{widetext}
\begin{equation}
H_{\text{eff}} =
\begin{pmatrix}
-\frac{J^2}{2\Delta} & J_{AB} - \frac{J^{2}}{2\Delta} & J_{AC} - \frac{J^{2}}{2\Delta} \\[6pt]
J_{BA} - \frac{J^{2}}{2\Delta} & -\frac{J^2}{2\Delta} & J_{BC} - \frac{J^{2}}{2\Delta} \\[6pt]
J_{CA} - \frac{J^{2}}{2\Delta} & J_{CB} - \frac{J^{2}}{2\Delta} & -\frac{J^2}{2\Delta}
\end{pmatrix}.
\end{equation}
%\end{widetext}
This result indicates that, by tuning the detuning to satisfy 
$J_{\text{eff}}^{AC}=J_{AC}-{J^{2}}/{2\Delta} = 0$, the direct dipole-dipole coupling between next-nearest-neighbor data atoms can be destructively interfered with by the chain-mediated interaction, resulting in the complete suppression of the next-nearest-neighbor coupling.

\begin{figure}
    \centering
    \includegraphics[width=1.0\linewidth]{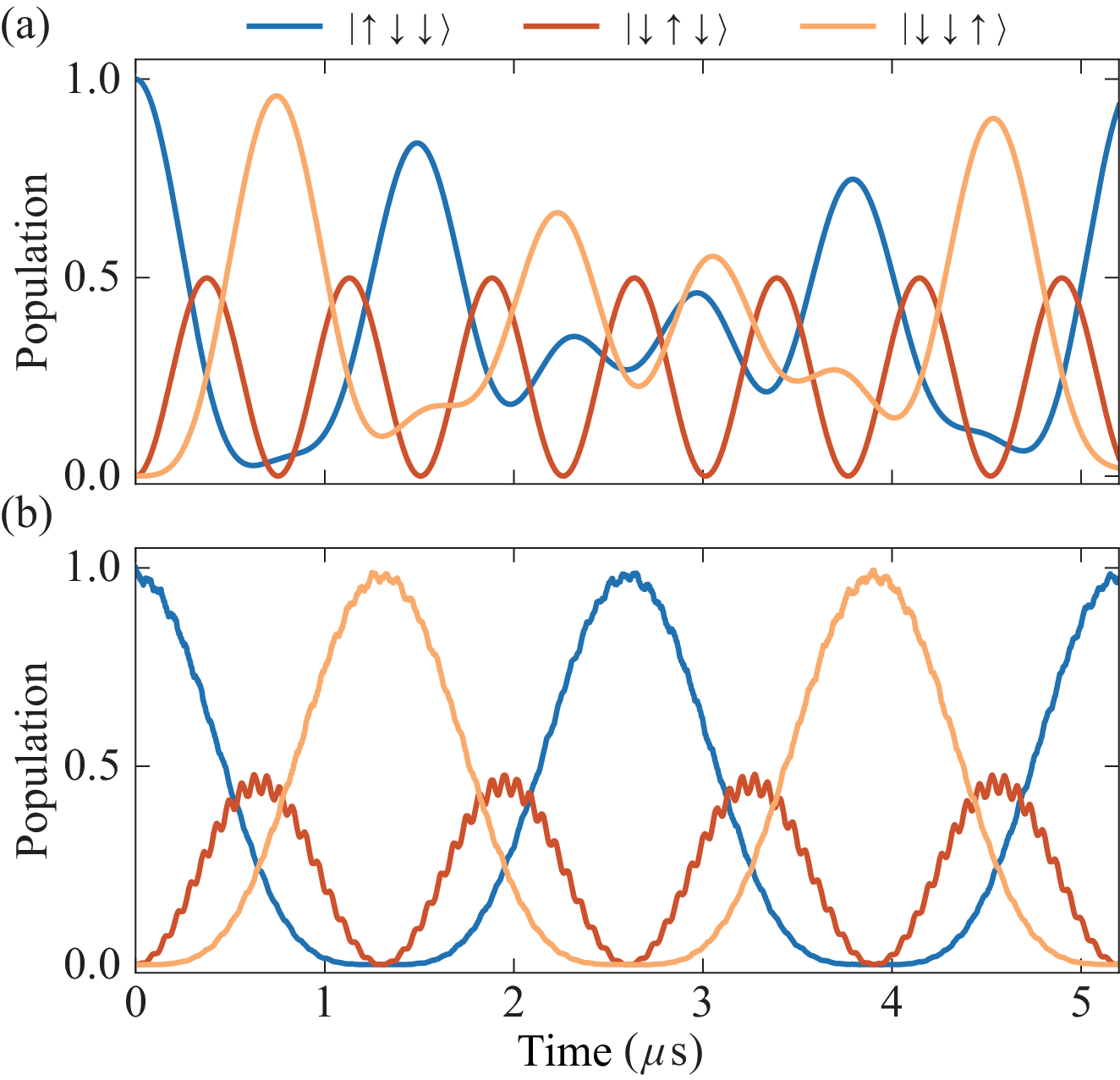}
    \caption{Population evolution of the data atoms. (a) Dynamics governed solely by the direct dipole-dipole coupling among the three data atoms, exhibiting a simple dipole-exchange process. (b) Population evolution of the three data atoms after the Rydberg chain is introduced. In both panels, the initial state is $|{\uparrow \downarrow \downarrow}\rangle$, with atom $A$ in the $|{\uparrow}\rangle$ state. The blue, red, and orange lines represent the populations of the three data atoms in the $|{\uparrow}\rangle$ state, respectively. The parameter are set to $V/2\pi=30~\text{MHz}$, $J/2\pi=5~\text{MHz}$, $\Delta/2\pi=101~\text{MHz}$, and $\Delta_{B}/2\pi=0.43~\text{MHz}$. }
    \label{fig8}
\end{figure}

\begin{figure*}
    \centering
    \includegraphics[width=1\linewidth]{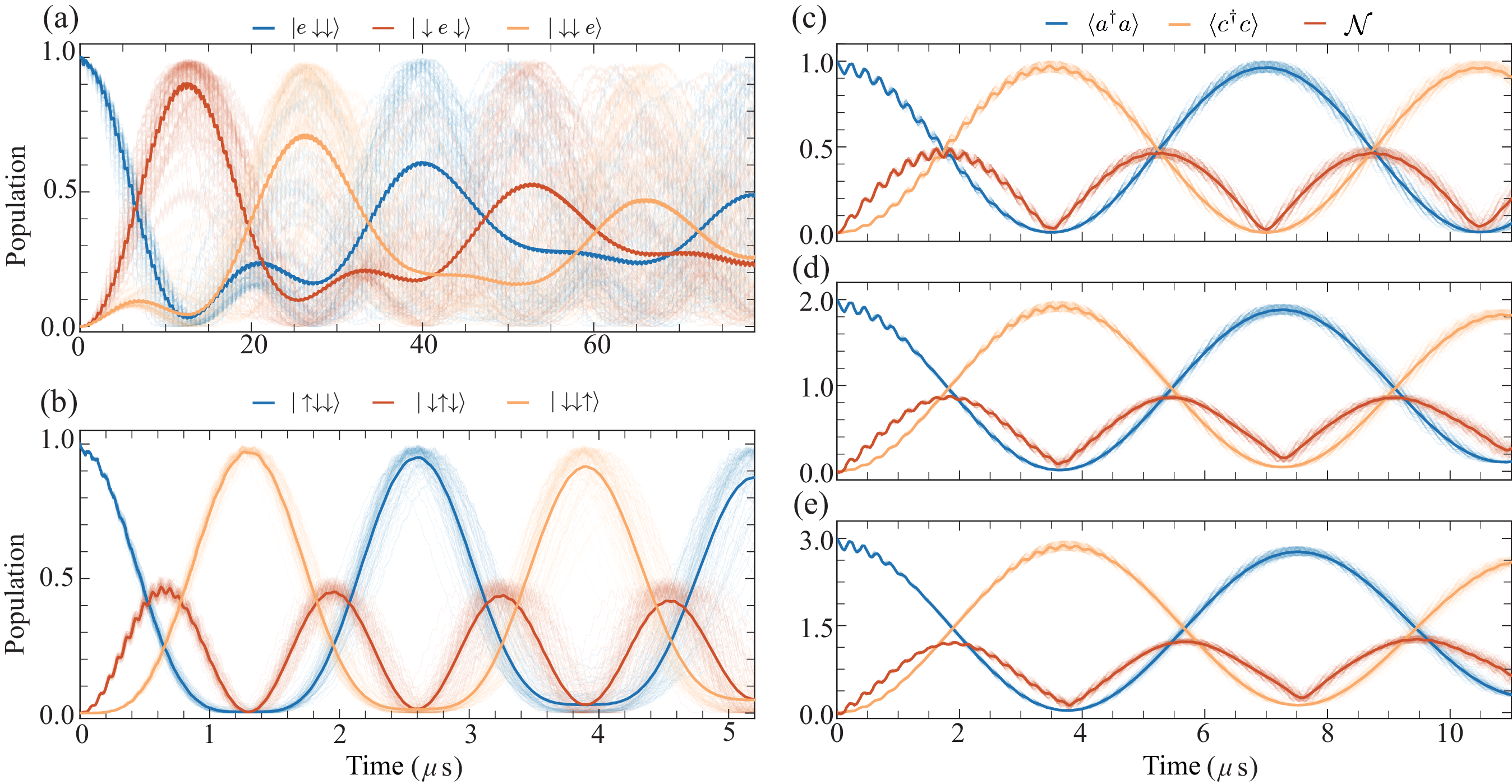}
    \caption{Monte Carlo evolution results under atomic position fluctuation. The figure displays the dynamical evolution of the system under 100 random realizations of atomic position fluctuation. The position disorder follows a Gaussian distribution with zero mean and a standard deviation of $0.02\mu\text{m}$, and all long-range interactions are fully taken into account in the simulations. The faint lines represent the individual evolution trajectories, while the bold solid lines denote the average over the 100 realizations. (a) Chiral transport via Floquet engineering, with the same parameter settings as in Fig.~\ref{fig4}(f). (b) Destructive interference of dipole exchange induced by the Rydberg atom array, with the same parameter settings as in Fig.~\ref{fig8}(b). (c)-(e) Entanglement evolution of mechanical oscillators mediated by the Rydberg atomic chain in the single-excitation, double-excitation, and triple-excitation subspaces, respectively, with the same parameter setting as in Fig.~\ref{fig6}.}
    \label{fig9}
\end{figure*}

To demonstrate the feasibility of the proposed mechanism, we perform numerical simulations, including all long-range interactions. 
The resonant exchange between atoms is mediated by the dipole-dipole interaction, with the interaction strength given by~\cite{PhysRevLett.114.113002,browaeys2016experimental}:
\begin{equation}
V_{ij}=\frac{d^{2}}{{4\pi\epsilon_0}R_{ij}^{3}}(1-3\cos^{2}\theta_{ij}),
\end{equation}
where $d$ is the transition dipole moment, $R_{ij}$ is the separation of atom $i$ and $j$, and $\theta_{ij}$ represents the angle between the quantization axis defined by the magnetic field and the interatomic separation vector.
In our setup, the magnetic field is aligned along the chain axis, such that $\theta_{ij}=0^\circ$ for atoms within the chain. 
Owing to the characteristic $1/R^3$ scaling and angular dependence of the interaction, the data atoms couple to all sites of the chain with varying strengths.
Considering all long-range interactions, the system Hamiltonian reads: 
\begin{equation}\label{E12}
\begin{split}
    H_{\text{full}} = \,&  \sum_{i \in \{1,9\}} \Delta n_{i} 
     + \sum_{i < j}^{9} V_{ij} \left( \sigma^{+}_{i}\sigma^{-}_{j} + \text{H.c.} \right) \\
    & + \sum_{\mu} \sum_{i=1}^{9} J_{\mu i} \left( \sigma^{+}_{\mu}\sigma^{-}_{i} + \text{H.c.} \right) \\
    & + \sum_{\mu \neq \nu} J_{\mu\nu} \left( \sigma^{+}_{\mu}\sigma^{-}_{\nu} + \text{H.c.} \right).
\end{split}
\end{equation}
The first term describes the detuning applied to the boundary atoms of the chain. The second term accounts for the full range of interactions within the chain. The third term represents the coupling between the data atoms and all chain atoms, while the final term  captures the direct interactions among the data atoms.
For the numerical simulations, the nearest-neighbor spacing of the chain is set to $4.65~\mu\text{m}$, corresponding to $V/2\pi = 30~\text{MHz}$. 
The data atoms are positioned $6.72~\mu\text{m}$ above the chain, yielding a dominant coupling strength $J/2\pi = 5~\text{MHz}$. 
By adjusting the detuning of the boundary atoms, the effective coupling between $A$ and $C$ vanishes at $\Delta/2\pi = 101~\text{MHz}$.
A small compensating detuning $\Delta_{B}/2\pi= 0.43~\text{MHz}$ is applied to atom $B$ to offset the induced level shift. 
We plot the population dynamics starting from an initial state where only atom $A$ is prepared in $|{\uparrow}\rangle$, and all other atoms remain in $|{\downarrow}\rangle$, as shown in Fig.~\ref{fig8}. 
In the absence of the chain, a clear direct exchange between $A$ and $C$ is observed. 
When the chain is included with a properly chosen detuning, the next-nearest-neighbor exchange is strongly suppressed, and the excitation propagates predominantly via nearest-neighbor hopping.
These results confirm destructive interference between the chain-mediated and direct coupling pathways, suppressing the coupling between $A$ and $C$ and leaving nearest-neighbor interactions dominant.

\section{The atomic position fluctuation}
\label{sec6}

\begin{figure*}
    \centering
    \includegraphics[width=1\linewidth]{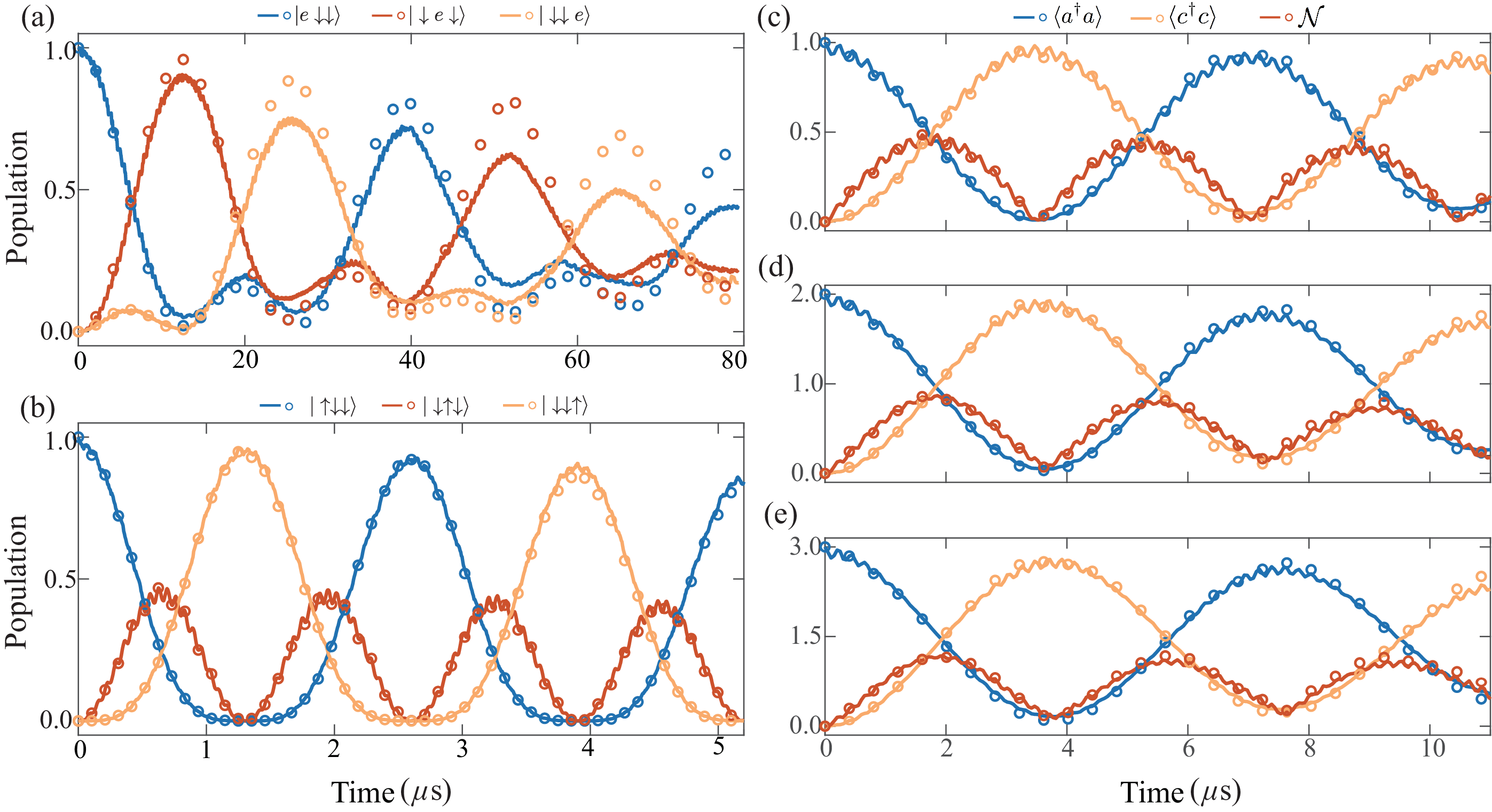}
    \caption{Monte Carlo evolution results with dissipation. The dynamical curves are obtained by averaging over $150$ independent random trajectories. The solid lines represent the evolution at $T=77~\text{K}$, while the open circles denote the baseline results under the idealized condition of $T=0~\text{K}$.
    (a) Chiral transport via Floquet engineering, with the same parameter settings as in Fig.~\ref{fig4}(f). (b) Destructive interference of dipole exchange induced by the Rydberg atom array, with the same parameter settings as in Fig.~\ref{fig8}(b). (c)-(e) Entanglement evolution of mechanical oscillators mediated by the Rydberg atomic chain in the single-excitation, double-excitation, and triple-excitation subspaces, respectively, with the same parameter setting as in Fig.~\ref{fig6}.}
    \label{fig10}
\end{figure*}

In realistic experimental setups, atoms exhibit finite spatial wave-packet distributions rather than being perfectly localized, leading to positional uncertainties and corresponding fluctuations in the interactions. 
Typical deviations range from $0.01~\mu\text{m}$ to $0.1\mu\text{m}$, depending on the trap frequency and cooling limits~\cite{PhysRevLett.118.063606,PhysRevLett.132.223201,PhysRevResearch.7.L022035,PhysRevX.15.011035}. 
To assess the robustness of our model under atomic position fluctuations, we introduce positional disorder via a two-dimensional Gaussian noise model in this section.

Specifically, each atomic coordinate is perturbed as ${x_{i}' = x_{i} + \delta x_{i}}$ and ${y_{i}' = y_{i} + \delta y_{i}}$, where $\delta x_{i}$ and $\delta y_{i}$ are independent Gaussian random variables with zero mean and a standard deviation of ${\sigma = 0.02~\mu\text{m}}$. The resulting interatomic distance and relative angle are modified to 
$R_{ij}' = \sqrt{(x_i' - x_j')^2 + (y_i' - y_j')^2}$ and 
$\theta_{ij}' = \arccos(|x_i' - x_j'|/R_{ij}')$, 
which are used to compute the interaction strengths in the presence of positional disorder.

Based on the discussions in the preceding sections, we investigate the impact of positional disorder in the presence of full long-range interactions. 
For each case, we perform 100 Monte Carlo simulations, with the results shown in Fig.~\ref{fig9}. The faint curves represent the individual evolution trajectories, and the bold curves denote the ensemble average. The results demonstrate that our scheme is robust against realistic random atomic position fluctuations. 
This robustness is rooted in the physical mechanism of our scheme. The effective interactions between the data units are mediated by a virtual excitation process. In this off-resonant coupling regime, the effective interaction strength is inherently insensitive to small variations in the local coupling strength~\cite{PhysRevLett.82.1971}. 
Therefore, our scheme exhibits robustness under realistic experimental conditions, confirming its high feasibility.

\section{effects of dissipation}
\label{sec7}
In a realistic experimental setting, Rydberg atoms suffer from spontaneous emission and blackbody-radiation-induced transitions, which limit the effective lifetime of the Rydberg state. In this section, we analyze the impact of such dissipation on the system. 

In principle, the dissipative dynamics can be described by the Lindblad master equation. However, the exponential growth of the Hilbert space with the number of atoms renders its direct numerical solution intractable for our system. To overcome this difficulty, we employ the Monte Carlo wavefunction method~\cite{PhysRevLett.68.580,molmer1993monte}, which unravels the master equation dynamics into an ensemble of independent random quantum trajectories.
When analyzing dissipation in our model, we do not explicitly track all possible atomic states populated following spontaneous emission or blackbody-radiation-induced transitions. Instead, these processes are treated as irreversible losses, with the Rydberg-state population leaving the computational subspace.
Within a sufficiently small time step $dt$, the evolution in the absence of a quantum jump is governed by the 
non-Hermitian effective Hamiltonian
\begin{equation}
    H_{\text{eff}} = H - \frac{i}{2} \sum_{k} L_k^\dagger L_k,
\end{equation}
where $L_k $ denotes the set of collapse operators describing the individual dissipation channels. 
The corresponding unnormalized trial state is 
\begin{equation}
|\psi(t+dt)\rangle=e^{-iH_{\mathrm{eff}}dt}|\psi(t)\rangle .
\end{equation}
Because $H_{\text{eff}}$ is non-Hermitian, the norm of this state decreases, with $P_{\text{no}}=\langle\psi(t+dt)|\psi(t+dt)\rangle$ giving the probability that no quantum jump occurs during this time step. The total jump probability is $P_{\text{jump}}=1-P_{\text{no}}$. For sufficiently small $dt$, $P_{\text{jump}}$ reduces to 
\begin{equation}
P_{\text{jump}}\approx{dt\sum_{k}\langle\psi(t)|L_k^\dagger L_k|\psi(t)\rangle}.
\end{equation}
In each time step, a uniformly distributed random number $r\in[0,1]$ is generated to determine whether a jump occurs. If $r\leq P _{\text{no}}$, the system evolves smoothly and must be renormalized. Conversely, a quantum jump is occurred. For system with explicitly dissipation channels, the jump channel is then selected according to the relative weights
\begin{equation}
w_k=|L_k|\psi(t)\rangle|^2=\langle\psi(t)|L_k^\dagger L_k|\psi(t)\rangle,
\end{equation} with
\begin{equation}
P_{\text{jump}}^{k}=\frac{w_k}{\sum_j w_j}.
\end{equation}
The state is subsequently updated according to the selected collapse operator and renormalized. 
For an atom in Rydberg state, the corresponding loss channel is constructed. We define $|d\rangle$ as a loss state. When a quantum jump occurs, the wavefunction is transferred to the loss state, after which the subsequent decay dynamics of the atom are not tracked. 

For the chosen Rydberg states ${|{\uparrow}\rangle = |60 S_{1/2}, m_{j} = 1/2\rangle}$ and ${|{\downarrow}\rangle = |60P_{1/2}, m_j = -1/2\rangle}$, we evaluate the impact of dissipation under both the idealized zero temperature limit ${T=0~\text{K}}$ and realistic experimental conditions ${T=77~\text{K}}$~\cite{PhysRevA.108.022805}. The corresponding lifetimes are ${\tau_{\text{S}}~(0)=229.95~\mu\text{s}}$ and ${\tau_{\text{P}}~(0)=499.15~\mu\text{s}}$ at ${0~\mathrm{K}}$, and they are reduced to ${\tau_{\text{S}}~(77)=173.59~\mu\text{s}}$ and ${\tau_{\text{P}}~(77)=302.35~\mu\text{s}}$ at $77~\text{K}$.
The corresponding dissipative decay rates in the two temperature regimes are defined as $\gamma_S = 1/\tau_S$ and $\gamma_P = 1/\tau_P$.
We independently simulate $150$ random dynamical trajectories, as shown in Fig.~\ref{fig10}. The solid lines represent the evolution results at $77~\text{K}$, and the open circles denote those under the idealized condition of $0~\text{K}$. The results show that the system is able to complete the coherent transfer of quantum states before dissipative losses set in. 
These results indicate that the protocol can be implemented under experimentally accessible conditions, without requiring extreme cryogenic temperatures.

\section{conclusion}
\label{sec8}

In conclusion, we have developed a theoretical framework for engineering nonlocal interactions mediated by a Rydberg atom chain in the dispersive regime. In this regime, the mediator chain remains only virtually populated and acts as a coherent quantum bus that induces effective couplings between spatially separated data units. To characterize this mechanism, we employ a Green’s-function continued-fraction approach to derive effective Hamiltonians in different excitation subspaces without requiring direct diagonalization of the full mediator Hamiltonian.

Building on this framework, we demonstrate several distinct functionalities enabled by the mediated interactions, including Floquet-engineered chiral transport, entanglement generation between distant mechanical oscillators, and interference-based suppression of unwanted dipole-exchange pathways without relying on a magic-angle configuration. We further assess the robustness of the scheme by incorporating long-range dipole-dipole interactions, random atomic position fluctuations, and Rydberg-state dissipation, and find that the characteristic dynamics remain well preserved under experimentally relevant conditions.

These results establish the Rydberg atom chain as a programmable coherent mediator for extending the native connectivity of neutral-atom and hybrid quantum architectures beyond geometrically local interactions, providing a route toward quasi-all-to-all quantum connectivity.

\section*{acknowledgment}
This work was supported by the National Natural Science
Foundation (Grant No. 12174048). W.L. acknowledges support from the EPSRC through Grant No. EP/W015641/1.

\appendix
\section{Derivation of the Effective Hamiltonian}

In this appendix, we provide the detailed derivation of the effective Hamiltonian introduced in Sec.~\ref{sec2}.
We employ a recursive Green’s function approach to systematically eliminate the degrees of freedom of the atom chain and obtain the effective interactions within the $P$ subspace.

To derive the matrix elements of the effective Hamiltonian, we define 
$g_{i\to N}(z)$ as the local Green’s function at site $i$ for a truncated 
sub-chain extending to site $N$. 
The response at site $i$ is determined by its onsite energy and the self-energy induced by the rest of the chain. 
For nearest-neighbor coupling, the recursion relation reads
\begin{equation}
    g_{i\to N}(z)=\frac{1}{z-\epsilon_{i}-V^{2}g_{i+1\to N}(z)},
\end{equation}
where $\epsilon_{i}$ is the bare energy of site $i$. 

We first evaluate the diagonal matrix elements of the effective Hamiltonian. 
For data units $A$ coupled to site $1$, the derivation follows a recursive approach starting from the boundary site $N$. 
Owing to the local detuning $\Delta_{N}$ at site $N$, the local Green’s function is expressed as $g_{N\to N}(z)=1/(z-\Delta_{N})$. 
For intermediate sites $i\in \{2,3,...,{N-1}\}$, without additional detuning, the recursion relation takes the uniform form
$g_{i\to N}={1}/({z-V^2g_{i+1\to N}})$.
At site $1$, with detuning $\Delta_{1}$, the Green's function $G_{11}$ reads $G_{11}(z)=1/(z-\Delta_{1}-V^{2}g_{2\to N}(z))$. Expanding this recursion relation, we obtain $G_{11}$ in the form of an $N$-level continued fraction
\begin{equation}
    \begin{split}
        G_{11}(z) &= \cfrac{1}{z - \Delta_{1} - \cfrac{V^2}{z - \cfrac{V^2}{z - \cfrac{V^2}{\ddots - \cfrac{V^2}{z - \Delta_{N}}}}}}.
    \end{split}
\end{equation}
This expression captures the cumulative self-energy feedback from the entire chain onto site $1$.

In the dispersive regime discussed in the main text, the self-energy is well captured by its zero-energy value. Accordingly, we set $z=0$ in the following.
The continued fraction exhibits a clear parity dependence on the length of the truncated sub-chain. 
Denoting by $l = N - i + 1$ the number of sites 
from $i$ to $N$, one obtains
\begin{equation}
\label{A3}
g_{i\to N}(0)=
\begin{cases}
-1/\Delta_N, & \text{for odd } l,\\
\Delta_N/V^2, & \text{for even } l.
\end{cases}
\end{equation}
For site $2$, $g_{2\to N}(0)$ corresponds to a segment of $N-1$. Given that $N$ is an odd integer, the sub-chain length $N-1$ is even, which yields $g_{2 \to N}(0) = \Delta_{N}/{V^2}$. Consequently, the full-chain Green’s function at site $1$ is determined as 
\begin{equation}
G_{11}(0)=-\frac{1}{(\Delta_{1}+\Delta_{N})}.
\end{equation}
Projecting onto the $P$ subspace, the first diagonal element of the effective Hamiltonian is renormalized to $\Sigma_{AA}(0) = -{J_{A}^2}/{(\Delta_{1}+\Delta_{N})}.$
By symmetry, for unit $C$, one obtains $\Sigma_{CC}(0) = -J_{C}^{2}/(\Delta_{1} + \Delta_{N})$.

For the data units $B$, coupled to site $m$, the response is determined by contributions from both sides of the chain.
A bidirectional folding approach is required. 
To this end, we partition the chain into three regions: the left sub-chain $L$ (sites $1$ to $m-1$), the central site $m$, and the right sub-chain $R$ (sites $m+1$ to $N$). 
The Green's function at site $m$ can then be written as
\begin{equation}
    G_{mm}(0)=-\frac{1}{\Sigma_{L}(0)+\Sigma_{R}(0)}.
\end{equation}
We evaluate the feedback from both sub-chains. Since the left sub-chain has an even length $m-1$, the parity result gives $g_{(m-1)\to 1}(0) = \Delta_1/V^2$, leading to $\Sigma_L(0) = \Delta_1$. Similarly, the right sub-chain has an even length $N-m$, yielding $g_{(m+1)\to N}(0) = \Delta_N/V^2$ and hence $\Sigma_R(0) = \Delta_N$. Therefore,
\begin{equation}
G_{mm}(0) = -\frac{1}{(\Delta_1 + \Delta_N)}.
\end{equation}
Accordingly, the effective self-energy of unit $B$ is $\Sigma_{BB}(0) = -{J_{B}^2}/({\Delta_{1} + \Delta_{N}})$.

Having determined the diagonal terms, we now evaluate the off-diagonal elements, which describe the effective interactions mediated by the chain. 
To this end, we employ a recursive relation derived from the Dyson equation for block matrices, expressed as
\begin{equation}
    G_{ij}(z) = G_{ii}(z) \prod_{k=i+1}^{j} [V g_{k \to N}(z)],
\end{equation}
where $G_{ij}(z)$ is the Green's function from site $i$ to site $j$.
In the case $z=0$, the excitation transport exhibits a parity-dependent structure.
We define $\mathcal{T}_{k}=V g_{k \to N}(0)$ as the single-step excitation transfer factor.
According to Eq.~(\ref{A3}), its value depends on the parity of the truncated sub-chain length,
\begin{equation}
\mathcal{T}_{k} =
\begin{cases}
- V/\Delta_N, & \text{for odd } l, \\
\Delta_N/V, & \text{for even } l.
\end{cases}
\end{equation}
As a consequence, a two-step transfer factor between sites of the same parity satisfies
$\mathcal{T}_{k}\mathcal{T}_{k+1} = -1$.
This means that a two-step hopping process across the chain maintains a constant amplitude absolute value, accompanied solely by a $\pi$ phase shift.
We consider the effective coupling between units $A$ and $B$. 
The propagation from site $1$ to site $m$ involves $(m-1)/2$ such two-step processes, leading to
\begin{equation}
G_{1m}(0) = -\frac{1}{\Delta_{1}+\Delta_{N}}\, (-1)^{({m-1})/{2}}.
\end{equation}
Applying the same procedure, we obtain
\begin{align}
G_{1N}(0) &= -\frac{1}{\Delta_{1}+\Delta_{N}}\,(-1)^{{(N-1)}/{2}},\\
G_{mN}(0) &= -\frac{1}{\Delta_{1}+\Delta_{N}}\,(-1)^{{(N-m)}/{2}}.
\end{align}
Projecting onto the $P$ subspace, the effective Hamiltonian in the single-excitation sector reads
\begin{widetext}
\begin{equation}\label{E_eff_long}
H_{\text{eff}}^{\text{one}} = \frac{-1}{\Delta_{1}+\Delta_{N}} 
\begin{pmatrix}
J_A^2 & J_A J_B (-1)^{\frac{m-1}{2}} & J_A J_C (-1)^{\frac{N-1}{2}} \\
J_B J_A (-1)^{\frac{m-1}{2}} & J_B^2 & J_B J_C (-1)^{\frac{N-m}{2}} \\
J_C J_A (-1)^{\frac{N-1}{2}} & J_C J_B (-1)^{\frac{N-m}{2}} & J_C^2
\end{pmatrix}.
\end{equation}
\end{widetext}

The above analysis can be directly extended to the multi-excitation sectors. 
Owing to the linear dipole-exchange interaction, multiple spin excitations behave as independent quasiparticles, allowing the effective Hamiltonian to be constructed from the single-excitation self-energies.
We first consider the double-excitation subspace. The basis is given by $\{|ij\rangle\}$, where $|ij\rangle$ denotes the state with excitations at sites $i$ and $j$. The corresponding $P$ subspace is spanned by the data-spin pair states $\{|AB\rangle, |BC\rangle, |AC\rangle\}$.
The diagonal matrix elements are additive. 
For the off-diagonal elements, one excitation acts as a spectator, while the other undergoes the same mediated process as in the single-excitation case. For instance, the transition between $|AB\rangle$ and $|BC\rangle$ reduces to an effective transfer from $A$ to $C$, yielding a coupling $\Sigma_{AC}(0)$.
Following this procedure, the effective Hamiltonian in the double-excitation subspace reads
\begin{widetext}
\begin{equation}\label{E_eff_long}
H_{\text{eff}}^{\text{double}} = \frac{-1}{\Delta_{1}+\Delta_{N}} 
\begin{pmatrix}
2J_A^2 & J_A J_B (-1)^{\frac{m-1}{2}} & J_A J_C (-1)^{\frac{N-1}{2}} \\
J_B J_A (-1)^{\frac{m-1}{2}} & 2J_B^2 & J_B J_C (-1)^{\frac{N-m}{2}} \\
J_C J_A (-1)^{\frac{N-1}{2}} & J_C J_B (-1)^{\frac{N-m}{2}} & 2J_C^2
\end{pmatrix}.
\end{equation}
\end{widetext}

Extending the analysis to the triple-excitation subspace, the $P$ subspace becomes one-dimensional, spanned by the state $|{ABC}\rangle$. 
The effective Hamiltonian thus reduces to a scalar, 
\begin{equation}
H_{\text{eff}}^{\text{triple}} = \Sigma_{AA}(0) + \Sigma_{BB}(0) + \Sigma_{CC}(0).
\end{equation}
In this case, the mediated exchange processes vanish, as three data units are occupied.
%%%%%%%%%%
\section{Generating the W-state}
\label{appendixB}
In Sec.~\ref{sec2}, we show that a Rydberg atom chain can serve as a quantum bus, generating effective all-to-all coupling between spatially separated data units. Here, we further show that this qusi-all-to-all connectivity can be combined with local detuning compensation to enable single-step preparation of a W-state in the single-excitation subspace.

Consider a 1D chain consisting of $N$ atoms, where $N$ is odd and $M=(N+1)/2$ atoms are located at odd sites. Each odd site contains three states, including the ground state $|e\rangle$ and two Rydberg states $|{\uparrow}\rangle$ and $|{\downarrow}\rangle$, while each even site contains only the two Rydberg states. A detuning $\Delta$ is applied to the $|{\uparrow}\rangle$ states of the two edge atoms. The odd sites are driven to couple $|e\rangle$ and $|{\uparrow}\rangle$, with Rabi frequency magnitude $J$ and phases alternating between $0$ and $\pi$. The $|e\rangle$ states at the odd sites therefore serve as the data units. 

To prepare a W state, we further apply local detuning compensation to the
$|e\rangle$ states. The detuning is chosen as
\begin{equation}
\delta_k =
\begin{cases}
-{J^2}/{2\Delta}, & k=1,\\[6pt]
\dfrac{(M-1)J^2}{2\Delta}, & k\in\{3,5,\ldots,N\}.
\end{cases}
\end{equation}
Under ideal nearest-neighbor dipole-dipole interactions, the system Hamiltonian is given by 
\begin{equation}
\begin{split}
    H = &\sum_{i \in \{1,N\}} \Delta_{i} n_{i} 
     + V \sum_{i=1}^{N-1} \left( \sigma^{+}_{i}\sigma^{-}_{i+1} + \text{H.c.} \right) \\
    & + \sum_{k=1}J\left(e^{i\phi_{k}} |{\uparrow}\rangle_{k}\langle{e}|_{k}+\text{H.c.}\right)+\delta_{k}|e\rangle_k\langle e|_k,
\end{split}
\end{equation}
Within the dispersive regime discussed above, the Green's function continued-fraction method in the single-excitation subspace yields the following $M\times M$ effective Hamiltonian in the basis where one of the odd sites is in the $|e\rangle$ state and all other atoms are in the $|{\downarrow}\rangle$ state
\begin{equation}
H_{\mathrm{eff}}
=
-\frac{J^2}{2\Delta}
\begin{pmatrix}
2 & 1 & 1 & \cdots & 1 \\[6pt]
1 & -(M-2) & 1 & \cdots & 1 \\[6pt]
1 & 1 & -(M-2) & \cdots & 1 \\[6pt]
\vdots & \vdots & \vdots & \ddots & \vdots \\[6pt]
1 & 1 & 1 & \cdots & -(M-2)
\end{pmatrix}.
\end{equation}
Except for site 1, the remaining odd sites have identical diagonal energies and equal coupling strengths to site 1, giving rise to a full permutation symmetry. We define the normalized fully symmetric state of the remaining $M-1$ odd sites as 
\begin{equation}
|S\rangle=\frac{1}{\sqrt{M-1}}\sum_{k\neq 1}|e\rangle_{k}.
\end{equation}
In the basis $\{|e\rangle_{1},|S\rangle\}$, the Hamiltonian reduces to the following $2\times2$ matrix:
\begin{equation}
H_{\mathrm{eff}}^{(2\times 2)}
=
-\frac{J^2}{2\Delta}
\begin{pmatrix}
2 & \sqrt{M-1} \\[6pt]
\sqrt{M-1} & 0
\end{pmatrix}.
\end{equation}
The effective Rabi frequency of this two-level system is given by $\Omega_{R}=\frac{J^2}{2\Delta}\sqrt{M}$.
Starting from the initial state $|\psi(0)\rangle=|e\rangle_1$, the evolution time required
to prepare the W state is
\begin{equation}
t_W =\frac{\pi\Delta}{J^2\sqrt{M}}=\frac{\pi\Delta}{J^2\sqrt{(N+1)/2}}.
\end{equation}
At $t_W$ the system evolves to 
\begin{equation}
|\psi(t_W)\rangle=-i\left(\frac{1}{\sqrt{M}}|1\rangle+
\sqrt{\frac{M-1}{M}}|S\rangle\right)=
-i|W_M\rangle,
\end{equation}
where the global phase $-i$ can be neglected, while $|W_M\rangle$ is exactly the target $M$-atom W-state
\begin{equation}
|W_M\rangle=
\frac{1}{\sqrt{M}}
\sum_{k=1}|e\rangle_k.
\end{equation}
This provides a scalable approach for single-step W-state generation.

\bibliography{main.bbl}

\begin{thebibliography}{96}%
\makeatletter
\providecommand \@ifxundefined [1]{%
 \@ifx{#1\undefined}
}%
\providecommand \@ifnum [1]{%
 \ifnum #1\expandafter \@firstoftwo
 \else \expandafter \@secondoftwo
 \fi
}%
\providecommand \@ifx [1]{%
 \ifx #1\expandafter \@firstoftwo
 \else \expandafter \@secondoftwo
 \fi
}%
\providecommand \natexlab [1]{#1}%
\providecommand \enquote  [1]{``#1''}%
\providecommand \bibnamefont  [1]{#1}%
\providecommand \bibfnamefont [1]{#1}%
\providecommand \citenamefont [1]{#1}%
\providecommand \href@noop [0]{\@secondoftwo}%
\providecommand \href [0]{\begingroup \@sanitize@url \@href}%
\providecommand \@href[1]{\@@startlink{#1}\@@href}%
\providecommand \@@href[1]{\endgroup#1\@@endlink}%
\providecommand \@sanitize@url [0]{\catcode `\\12\catcode `\$12\catcode `\&12\catcode `\#12\catcode `\^12\catcode `\_12\catcode `\%12\relax}%
\providecommand \@@startlink[1]{}%
\providecommand \@@endlink[0]{}%
\providecommand \url  [0]{\begingroup\@sanitize@url \@url }%
\providecommand \@url [1]{\endgroup\@href {#1}{\urlprefix }}%
\providecommand \urlprefix  [0]{URL }%
\providecommand \Eprint [0]{\href }%
\providecommand \doibase [0]{http://dx.doi.org/}%
\providecommand \selectlanguage [0]{\@gobble}%
\providecommand \bibinfo  [0]{\@secondoftwo}%
\providecommand \bibfield  [0]{\@secondoftwo}%
\providecommand \translation [1]{[#1]}%
\providecommand \BibitemOpen [0]{}%
\providecommand \bibitemStop [0]{}%
\providecommand \bibitemNoStop [0]{.\EOS\space}%
\providecommand \EOS [0]{\spacefactor3000\relax}%
\providecommand \BibitemShut  [1]{\csname bibitem#1\endcsname}%
\let\auto@bib@innerbib\@empty
%</preamble>
\bibitem [{\citenamefont {D\"ur}\ \emph {et~al.}(2001)\citenamefont {D\"ur}, \citenamefont {Vidal}, \citenamefont {Cirac}, \citenamefont {Linden},\ and\ \citenamefont {Popescu}}]{PhysRevLett.87.137901}%
  \BibitemOpen
  \bibfield  {author} {\bibinfo {author} {\bibfnamefont {W.}~\bibnamefont {D\"ur}}, \bibinfo {author} {\bibfnamefont {G.}~\bibnamefont {Vidal}}, \bibinfo {author} {\bibfnamefont {J.~I.}\ \bibnamefont {Cirac}}, \bibinfo {author} {\bibfnamefont {N.}~\bibnamefont {Linden}}, \ and\ \bibinfo {author} {\bibfnamefont {S.}~\bibnamefont {Popescu}},\ }\bibfield  {title} {\enquote {\bibinfo {title} {Entanglement capabilities of nonlocal hamiltonians},}\ }\href {\doibase 10.1103/PhysRevLett.87.137901} {\bibfield  {journal} {\bibinfo  {journal} {Phys. Rev. Lett.}\ }\textbf {\bibinfo {volume} {87}},\ \bibinfo {pages} {137901} (\bibinfo {year} {2001})}\BibitemShut {NoStop}%
\bibitem [{\citenamefont {Cirac}\ \emph {et~al.}(1997)\citenamefont {Cirac}, \citenamefont {Zoller}, \citenamefont {Kimble},\ and\ \citenamefont {Mabuchi}}]{PhysRevLett.78.3221}%
  \BibitemOpen
  \bibfield  {author} {\bibinfo {author} {\bibfnamefont {J.~I.}\ \bibnamefont {Cirac}}, \bibinfo {author} {\bibfnamefont {P.}~\bibnamefont {Zoller}}, \bibinfo {author} {\bibfnamefont {H.~J.}\ \bibnamefont {Kimble}}, \ and\ \bibinfo {author} {\bibfnamefont {H.}~\bibnamefont {Mabuchi}},\ }\bibfield  {title} {\enquote {\bibinfo {title} {Quantum state transfer and entanglement distribution among distant nodes in a quantum network},}\ }\href {\doibase 10.1103/PhysRevLett.78.3221} {\bibfield  {journal} {\bibinfo  {journal} {Phys. Rev. Lett.}\ }\textbf {\bibinfo {volume} {78}},\ \bibinfo {pages} {3221--3224} (\bibinfo {year} {1997})}\BibitemShut {NoStop}%
\bibitem [{\citenamefont {Majer}\ \emph {et~al.}(2007)\citenamefont {Majer}, \citenamefont {Chow}, \citenamefont {Gambetta}, \citenamefont {Koch}, \citenamefont {Johnson}, \citenamefont {Schreier}, \citenamefont {Frunzio}, \citenamefont {Schuster}, \citenamefont {Houck}, \citenamefont {Wallraff} \emph {et~al.}}]{majer2007coupling}%
  \BibitemOpen
  \bibfield  {author} {\bibinfo {author} {\bibfnamefont {Johannes}\ \bibnamefont {Majer}}, \bibinfo {author} {\bibfnamefont {JM}~\bibnamefont {Chow}}, \bibinfo {author} {\bibfnamefont {JM}~\bibnamefont {Gambetta}}, \bibinfo {author} {\bibfnamefont {Jens}\ \bibnamefont {Koch}}, \bibinfo {author} {\bibfnamefont {BR}~\bibnamefont {Johnson}}, \bibinfo {author} {\bibfnamefont {JA}~\bibnamefont {Schreier}}, \bibinfo {author} {\bibfnamefont {L}~\bibnamefont {Frunzio}}, \bibinfo {author} {\bibfnamefont {DI}~\bibnamefont {Schuster}}, \bibinfo {author} {\bibfnamefont {Andrew~Addison}\ \bibnamefont {Houck}}, \bibinfo {author} {\bibfnamefont {Andreas}\ \bibnamefont {Wallraff}},  \emph {et~al.},\ }\bibfield  {title} {\enquote {\bibinfo {title} {Coupling superconducting qubits via a cavity bus},}\ }\href {\doibase https://doi.org/10.1038/nature06184} {\bibfield  {journal} {\bibinfo  {journal} {Nature}\ }\textbf {\bibinfo {volume} {449}},\ \bibinfo {pages} {443--447} (\bibinfo {year} {2007})}\BibitemShut {NoStop}%
\bibitem [{\citenamefont {Bravyi}\ \emph {et~al.}(2018)\citenamefont {Bravyi}, \citenamefont {Gosset},\ and\ \citenamefont {K{\"o}nig}}]{bravyi2018quantum}%
  \BibitemOpen
  \bibfield  {author} {\bibinfo {author} {\bibfnamefont {Sergey}\ \bibnamefont {Bravyi}}, \bibinfo {author} {\bibfnamefont {David}\ \bibnamefont {Gosset}}, \ and\ \bibinfo {author} {\bibfnamefont {Robert}\ \bibnamefont {K{\"o}nig}},\ }\bibfield  {title} {\enquote {\bibinfo {title} {Quantum advantage with shallow circuits},}\ }\href {\doibase 10.1126/science.aar3106} {\bibfield  {journal} {\bibinfo  {journal} {Science}\ }\textbf {\bibinfo {volume} {362}},\ \bibinfo {pages} {308--311} (\bibinfo {year} {2018})}\BibitemShut {NoStop}%
\bibitem [{\citenamefont {Nishi}\ \emph {et~al.}(2021)\citenamefont {Nishi}, \citenamefont {Kosugi},\ and\ \citenamefont {Matsushita}}]{nishi2021implementation}%
  \BibitemOpen
  \bibfield  {author} {\bibinfo {author} {\bibfnamefont {Hirofumi}\ \bibnamefont {Nishi}}, \bibinfo {author} {\bibfnamefont {Taichi}\ \bibnamefont {Kosugi}}, \ and\ \bibinfo {author} {\bibfnamefont {Yu-ichiro}\ \bibnamefont {Matsushita}},\ }\bibfield  {title} {\enquote {\bibinfo {title} {Implementation of quantum imaginary-time evolution method on nisq devices by introducing nonlocal approximation},}\ }\href {\doibase https://doi.org/10.1038/s41534-021-00409-y} {\bibfield  {journal} {\bibinfo  {journal} {npj Quantum Information}\ }\textbf {\bibinfo {volume} {7}},\ \bibinfo {pages} {85} (\bibinfo {year} {2021})}\BibitemShut {NoStop}%
\bibitem [{\citenamefont {Bravyi}\ \emph {et~al.}(2024)\citenamefont {Bravyi}, \citenamefont {Cross}, \citenamefont {Gambetta}, \citenamefont {Maslov}, \citenamefont {Rall},\ and\ \citenamefont {Yoder}}]{bravyi2024high}%
  \BibitemOpen
  \bibfield  {author} {\bibinfo {author} {\bibfnamefont {Sergey}\ \bibnamefont {Bravyi}}, \bibinfo {author} {\bibfnamefont {Andrew~W}\ \bibnamefont {Cross}}, \bibinfo {author} {\bibfnamefont {Jay~M}\ \bibnamefont {Gambetta}}, \bibinfo {author} {\bibfnamefont {Dmitri}\ \bibnamefont {Maslov}}, \bibinfo {author} {\bibfnamefont {Patrick}\ \bibnamefont {Rall}}, \ and\ \bibinfo {author} {\bibfnamefont {Theodore~J}\ \bibnamefont {Yoder}},\ }\bibfield  {title} {\enquote {\bibinfo {title} {High-threshold and low-overhead fault-tolerant quantum memory},}\ }\href {\doibase https://doi.org/10.1038/s41586-024-07107-7} {\bibfield  {journal} {\bibinfo  {journal} {Nature}\ }\textbf {\bibinfo {volume} {627}},\ \bibinfo {pages} {778--782} (\bibinfo {year} {2024})}\BibitemShut {NoStop}%
\bibitem [{\citenamefont {Wu}\ \emph {et~al.}(2024)\citenamefont {Wu}, \citenamefont {Yan}, \citenamefont {Andersson}, \citenamefont {Anferov}, \citenamefont {Chou}, \citenamefont {Conner}, \citenamefont {Grebel}, \citenamefont {Joshi}, \citenamefont {Li}, \citenamefont {Miller}, \citenamefont {Povey}, \citenamefont {Qiao},\ and\ \citenamefont {Cleland}}]{PhysRevX.14.041030}%
  \BibitemOpen
  \bibfield  {author} {\bibinfo {author} {\bibfnamefont {Xuntao}\ \bibnamefont {Wu}}, \bibinfo {author} {\bibfnamefont {Haoxiong}\ \bibnamefont {Yan}}, \bibinfo {author} {\bibfnamefont {Gustav}\ \bibnamefont {Andersson}}, \bibinfo {author} {\bibfnamefont {Alexander}\ \bibnamefont {Anferov}}, \bibinfo {author} {\bibfnamefont {Ming-Han}\ \bibnamefont {Chou}}, \bibinfo {author} {\bibfnamefont {Christopher~R.}\ \bibnamefont {Conner}}, \bibinfo {author} {\bibfnamefont {Joel}\ \bibnamefont {Grebel}}, \bibinfo {author} {\bibfnamefont {Yash~J.}\ \bibnamefont {Joshi}}, \bibinfo {author} {\bibfnamefont {Shiheng}\ \bibnamefont {Li}}, \bibinfo {author} {\bibfnamefont {Jacob~M.}\ \bibnamefont {Miller}}, \bibinfo {author} {\bibfnamefont {Rhys~G.}\ \bibnamefont {Povey}}, \bibinfo {author} {\bibfnamefont {Hong}\ \bibnamefont {Qiao}}, \ and\ \bibinfo {author} {\bibfnamefont {Andrew~N.}\ \bibnamefont {Cleland}},\ }\bibfield  {title} {\enquote {\bibinfo {title} {Modular quantum processor with an all-to-all reconfigurable
  router},}\ }\href {\doibase 10.1103/PhysRevX.14.041030} {\bibfield  {journal} {\bibinfo  {journal} {Phys. Rev. X}\ }\textbf {\bibinfo {volume} {14}},\ \bibinfo {pages} {041030} (\bibinfo {year} {2024})}\BibitemShut {NoStop}%
\bibitem [{\citenamefont {Aghaee~Rad}\ \emph {et~al.}(2025)\citenamefont {Aghaee~Rad}, \citenamefont {Ainsworth}, \citenamefont {Alexander}, \citenamefont {Altieri}, \citenamefont {Askarani}, \citenamefont {Baby}, \citenamefont {Banchi}, \citenamefont {Baragiola}, \citenamefont {Bourassa}, \citenamefont {Chadwick} \emph {et~al.}}]{aghaee2025scaling}%
  \BibitemOpen
  \bibfield  {author} {\bibinfo {author} {\bibfnamefont {H}~\bibnamefont {Aghaee~Rad}}, \bibinfo {author} {\bibfnamefont {Thomas}\ \bibnamefont {Ainsworth}}, \bibinfo {author} {\bibfnamefont {Rafael~N}\ \bibnamefont {Alexander}}, \bibinfo {author} {\bibfnamefont {Brandon}\ \bibnamefont {Altieri}}, \bibinfo {author} {\bibfnamefont {Mohsen~F}\ \bibnamefont {Askarani}}, \bibinfo {author} {\bibfnamefont {R}~\bibnamefont {Baby}}, \bibinfo {author} {\bibfnamefont {Leonardo}\ \bibnamefont {Banchi}}, \bibinfo {author} {\bibfnamefont {Ben~Q}\ \bibnamefont {Baragiola}}, \bibinfo {author} {\bibfnamefont {J~Eli}\ \bibnamefont {Bourassa}}, \bibinfo {author} {\bibfnamefont {RS}~\bibnamefont {Chadwick}},  \emph {et~al.},\ }\bibfield  {title} {\enquote {\bibinfo {title} {Scaling and networking a modular photonic quantum computer},}\ }\href {\doibase https://doi.org/10.1038/s41586-024-08406-9} {\bibfield  {journal} {\bibinfo  {journal} {Nature}\ }\textbf {\bibinfo {volume} {638}},\ \bibinfo {pages} {912--919} (\bibinfo {year}
  {2025})}\BibitemShut {NoStop}%
\bibitem [{\citenamefont {Zhong}\ \emph {et~al.}(2021)\citenamefont {Zhong}, \citenamefont {Chang}, \citenamefont {Bienfait}, \citenamefont {Dumur}, \citenamefont {Chou}, \citenamefont {Conner}, \citenamefont {Grebel}, \citenamefont {Povey}, \citenamefont {Yan}, \citenamefont {Schuster} \emph {et~al.}}]{zhong2021deterministic}%
  \BibitemOpen
  \bibfield  {author} {\bibinfo {author} {\bibfnamefont {Youpeng}\ \bibnamefont {Zhong}}, \bibinfo {author} {\bibfnamefont {Hung-Shen}\ \bibnamefont {Chang}}, \bibinfo {author} {\bibfnamefont {Audrey}\ \bibnamefont {Bienfait}}, \bibinfo {author} {\bibfnamefont {{\'E}tienne}\ \bibnamefont {Dumur}}, \bibinfo {author} {\bibfnamefont {Ming-Han}\ \bibnamefont {Chou}}, \bibinfo {author} {\bibfnamefont {Christopher~R}\ \bibnamefont {Conner}}, \bibinfo {author} {\bibfnamefont {Joel}\ \bibnamefont {Grebel}}, \bibinfo {author} {\bibfnamefont {Rhys~G}\ \bibnamefont {Povey}}, \bibinfo {author} {\bibfnamefont {Haoxiong}\ \bibnamefont {Yan}}, \bibinfo {author} {\bibfnamefont {David~I}\ \bibnamefont {Schuster}},  \emph {et~al.},\ }\bibfield  {title} {\enquote {\bibinfo {title} {Deterministic multi-qubit entanglement in a quantum network},}\ }\href {\doibase https://doi.org/10.1038/s41586-021-03288-7} {\bibfield  {journal} {\bibinfo  {journal} {Nature}\ }\textbf {\bibinfo {volume} {590}},\ \bibinfo {pages} {571--575}
  (\bibinfo {year} {2021})}\BibitemShut {NoStop}%
\bibitem [{\citenamefont {Daiss}\ \emph {et~al.}(2021)\citenamefont {Daiss}, \citenamefont {Langenfeld}, \citenamefont {Welte}, \citenamefont {Distante}, \citenamefont {Thomas}, \citenamefont {Hartung}, \citenamefont {Morin},\ and\ \citenamefont {Rempe}}]{daiss2021quantum}%
  \BibitemOpen
  \bibfield  {author} {\bibinfo {author} {\bibfnamefont {Severin}\ \bibnamefont {Daiss}}, \bibinfo {author} {\bibfnamefont {Stefan}\ \bibnamefont {Langenfeld}}, \bibinfo {author} {\bibfnamefont {Stephan}\ \bibnamefont {Welte}}, \bibinfo {author} {\bibfnamefont {Emanuele}\ \bibnamefont {Distante}}, \bibinfo {author} {\bibfnamefont {Philip}\ \bibnamefont {Thomas}}, \bibinfo {author} {\bibfnamefont {Lukas}\ \bibnamefont {Hartung}}, \bibinfo {author} {\bibfnamefont {Olivier}\ \bibnamefont {Morin}}, \ and\ \bibinfo {author} {\bibfnamefont {Gerhard}\ \bibnamefont {Rempe}},\ }\bibfield  {title} {\enquote {\bibinfo {title} {A quantum-logic gate between distant quantum-network modules},}\ }\href {\doibase 10.1126/science.abe3150} {\bibfield  {journal} {\bibinfo  {journal} {Science}\ }\textbf {\bibinfo {volume} {371}},\ \bibinfo {pages} {614--617} (\bibinfo {year} {2021})}\BibitemShut {NoStop}%
\bibitem [{\citenamefont {Singh}\ \emph {et~al.}(2025)\citenamefont {Singh}, \citenamefont {Gu}, \citenamefont {De~Bone}, \citenamefont {Villase{\~n}or}, \citenamefont {Elkouss},\ and\ \citenamefont {Borregaard}}]{singh2025modular}%
  \BibitemOpen
  \bibfield  {author} {\bibinfo {author} {\bibfnamefont {Siddhant}\ \bibnamefont {Singh}}, \bibinfo {author} {\bibfnamefont {Fenglei}\ \bibnamefont {Gu}}, \bibinfo {author} {\bibfnamefont {Sebastian}\ \bibnamefont {De~Bone}}, \bibinfo {author} {\bibfnamefont {Eduardo}\ \bibnamefont {Villase{\~n}or}}, \bibinfo {author} {\bibfnamefont {David}\ \bibnamefont {Elkouss}}, \ and\ \bibinfo {author} {\bibfnamefont {Johannes}\ \bibnamefont {Borregaard}},\ }\bibfield  {title} {\enquote {\bibinfo {title} {Modular architectures and entanglement schemes for error-corrected distributed quantum computation},}\ }\href {\doibase https://doi.org/10.1038/s41534-025-01146-2} {\bibfield  {journal} {\bibinfo  {journal} {npj Quantum Information}\ } (\bibinfo {year} {2025}),\ https://doi.org/10.1038/s41534-025-01146-2}\BibitemShut {NoStop}%
\bibitem [{\citenamefont {Eldredge}\ \emph {et~al.}(2017)\citenamefont {Eldredge}, \citenamefont {Gong}, \citenamefont {Young}, \citenamefont {Moosavian}, \citenamefont {Foss-Feig},\ and\ \citenamefont {Gorshkov}}]{PhysRevLett.119.170503}%
  \BibitemOpen
  \bibfield  {author} {\bibinfo {author} {\bibfnamefont {Zachary}\ \bibnamefont {Eldredge}}, \bibinfo {author} {\bibfnamefont {Zhe-Xuan}\ \bibnamefont {Gong}}, \bibinfo {author} {\bibfnamefont {Jeremy~T.}\ \bibnamefont {Young}}, \bibinfo {author} {\bibfnamefont {Ali~Hamed}\ \bibnamefont {Moosavian}}, \bibinfo {author} {\bibfnamefont {Michael}\ \bibnamefont {Foss-Feig}}, \ and\ \bibinfo {author} {\bibfnamefont {Alexey~V.}\ \bibnamefont {Gorshkov}},\ }\bibfield  {title} {\enquote {\bibinfo {title} {Fast quantum state transfer and entanglement renormalization using long-range interactions},}\ }\href {\doibase 10.1103/PhysRevLett.119.170503} {\bibfield  {journal} {\bibinfo  {journal} {Phys. Rev. Lett.}\ }\textbf {\bibinfo {volume} {119}},\ \bibinfo {pages} {170503} (\bibinfo {year} {2017})}\BibitemShut {NoStop}%
\bibitem [{\citenamefont {Richerme}\ \emph {et~al.}(2014)\citenamefont {Richerme}, \citenamefont {Gong}, \citenamefont {Lee}, \citenamefont {Senko}, \citenamefont {Smith}, \citenamefont {Foss-Feig}, \citenamefont {Michalakis}, \citenamefont {Gorshkov},\ and\ \citenamefont {Monroe}}]{richerme2014non}%
  \BibitemOpen
  \bibfield  {author} {\bibinfo {author} {\bibfnamefont {Philip}\ \bibnamefont {Richerme}}, \bibinfo {author} {\bibfnamefont {Zhe-Xuan}\ \bibnamefont {Gong}}, \bibinfo {author} {\bibfnamefont {Aaron}\ \bibnamefont {Lee}}, \bibinfo {author} {\bibfnamefont {Crystal}\ \bibnamefont {Senko}}, \bibinfo {author} {\bibfnamefont {Jacob}\ \bibnamefont {Smith}}, \bibinfo {author} {\bibfnamefont {Michael}\ \bibnamefont {Foss-Feig}}, \bibinfo {author} {\bibfnamefont {Spyridon}\ \bibnamefont {Michalakis}}, \bibinfo {author} {\bibfnamefont {Alexey~V}\ \bibnamefont {Gorshkov}}, \ and\ \bibinfo {author} {\bibfnamefont {Christopher}\ \bibnamefont {Monroe}},\ }\bibfield  {title} {\enquote {\bibinfo {title} {Non-local propagation of correlations in quantum systems with long-range interactions},}\ }\href {\doibase https://doi.org/10.1038/nature13450} {\bibfield  {journal} {\bibinfo  {journal} {Nature}\ }\textbf {\bibinfo {volume} {511}},\ \bibinfo {pages} {198--201} (\bibinfo {year} {2014})}\BibitemShut {NoStop}%
\bibitem [{\citenamefont {Jurcevic}\ \emph {et~al.}(2014)\citenamefont {Jurcevic}, \citenamefont {Lanyon}, \citenamefont {Hauke}, \citenamefont {Hempel}, \citenamefont {Zoller}, \citenamefont {Blatt},\ and\ \citenamefont {Roos}}]{jurcevic2014quasiparticle}%
  \BibitemOpen
  \bibfield  {author} {\bibinfo {author} {\bibfnamefont {Petar}\ \bibnamefont {Jurcevic}}, \bibinfo {author} {\bibfnamefont {Ben~P}\ \bibnamefont {Lanyon}}, \bibinfo {author} {\bibfnamefont {Philipp}\ \bibnamefont {Hauke}}, \bibinfo {author} {\bibfnamefont {Cornelius}\ \bibnamefont {Hempel}}, \bibinfo {author} {\bibfnamefont {Peter}\ \bibnamefont {Zoller}}, \bibinfo {author} {\bibfnamefont {Rainer}\ \bibnamefont {Blatt}}, \ and\ \bibinfo {author} {\bibfnamefont {Christian~F}\ \bibnamefont {Roos}},\ }\bibfield  {title} {\enquote {\bibinfo {title} {Quasiparticle engineering and entanglement propagation in a quantum many-body system},}\ }\href {\doibase https://doi.org/10.1038/nature13461} {\bibfield  {journal} {\bibinfo  {journal} {Nature}\ }\textbf {\bibinfo {volume} {511}},\ \bibinfo {pages} {202--205} (\bibinfo {year} {2014})}\BibitemShut {NoStop}%
\bibitem [{\citenamefont {Defenu}\ \emph {et~al.}(2023)\citenamefont {Defenu}, \citenamefont {Donner}, \citenamefont {Macr\`{\i}}, \citenamefont {Pagano}, \citenamefont {Ruffo},\ and\ \citenamefont {Trombettoni}}]{RevModPhys.95.035002}%
  \BibitemOpen
  \bibfield  {author} {\bibinfo {author} {\bibfnamefont {Nicol\`o}\ \bibnamefont {Defenu}}, \bibinfo {author} {\bibfnamefont {Tobias}\ \bibnamefont {Donner}}, \bibinfo {author} {\bibfnamefont {Tommaso}\ \bibnamefont {Macr\`{\i}}}, \bibinfo {author} {\bibfnamefont {Guido}\ \bibnamefont {Pagano}}, \bibinfo {author} {\bibfnamefont {Stefano}\ \bibnamefont {Ruffo}}, \ and\ \bibinfo {author} {\bibfnamefont {Andrea}\ \bibnamefont {Trombettoni}},\ }\bibfield  {title} {\enquote {\bibinfo {title} {Long-range interacting quantum systems},}\ }\href {\doibase 10.1103/RevModPhys.95.035002} {\bibfield  {journal} {\bibinfo  {journal} {Rev. Mod. Phys.}\ }\textbf {\bibinfo {volume} {95}},\ \bibinfo {pages} {035002} (\bibinfo {year} {2023})}\BibitemShut {NoStop}%
\bibitem [{\citenamefont {Pagano}\ \emph {et~al.}(2020)\citenamefont {Pagano}, \citenamefont {Bapat}, \citenamefont {Becker}, \citenamefont {Collins}, \citenamefont {De}, \citenamefont {Hess}, \citenamefont {Kaplan}, \citenamefont {Kyprianidis}, \citenamefont {Tan}, \citenamefont {Baldwin} \emph {et~al.}}]{pagano2020quantum}%
  \BibitemOpen
  \bibfield  {author} {\bibinfo {author} {\bibfnamefont {Guido}\ \bibnamefont {Pagano}}, \bibinfo {author} {\bibfnamefont {Aniruddha}\ \bibnamefont {Bapat}}, \bibinfo {author} {\bibfnamefont {Patrick}\ \bibnamefont {Becker}}, \bibinfo {author} {\bibfnamefont {Katherine~S}\ \bibnamefont {Collins}}, \bibinfo {author} {\bibfnamefont {Arinjoy}\ \bibnamefont {De}}, \bibinfo {author} {\bibfnamefont {Paul~W}\ \bibnamefont {Hess}}, \bibinfo {author} {\bibfnamefont {Harvey~B}\ \bibnamefont {Kaplan}}, \bibinfo {author} {\bibfnamefont {Antonis}\ \bibnamefont {Kyprianidis}}, \bibinfo {author} {\bibfnamefont {Wen~Lin}\ \bibnamefont {Tan}}, \bibinfo {author} {\bibfnamefont {Christopher}\ \bibnamefont {Baldwin}},  \emph {et~al.},\ }\bibfield  {title} {\enquote {\bibinfo {title} {Quantum approximate optimization of the long-range ising model with a trapped-ion quantum simulator},}\ }\href {\doibase https://doi.org/10.1073/pnas.2006373117} {\bibfield  {journal} {\bibinfo  {journal} {Proceedings of the National Academy of
  Sciences}\ }\textbf {\bibinfo {volume} {117}},\ \bibinfo {pages} {25396--25401} (\bibinfo {year} {2020})}\BibitemShut {NoStop}%
\bibitem [{\citenamefont {Ho}\ \emph {et~al.}(2019)\citenamefont {Ho}, \citenamefont {Jonay},\ and\ \citenamefont {Hsieh}}]{PhysRevA.99.052332}%
  \BibitemOpen
  \bibfield  {author} {\bibinfo {author} {\bibfnamefont {Wen~Wei}\ \bibnamefont {Ho}}, \bibinfo {author} {\bibfnamefont {Cheryne}\ \bibnamefont {Jonay}}, \ and\ \bibinfo {author} {\bibfnamefont {Timothy~H.}\ \bibnamefont {Hsieh}},\ }\bibfield  {title} {\enquote {\bibinfo {title} {Ultrafast variational simulation of nontrivial quantum states with long-range interactions},}\ }\href {\doibase 10.1103/PhysRevA.99.052332} {\bibfield  {journal} {\bibinfo  {journal} {Phys. Rev. A}\ }\textbf {\bibinfo {volume} {99}},\ \bibinfo {pages} {052332} (\bibinfo {year} {2019})}\BibitemShut {NoStop}%
\bibitem [{\citenamefont {Cirac}\ and\ \citenamefont {Zoller}(1995)}]{PhysRevLett.74.4091}%
  \BibitemOpen
  \bibfield  {author} {\bibinfo {author} {\bibfnamefont {J.~I.}\ \bibnamefont {Cirac}}\ and\ \bibinfo {author} {\bibfnamefont {P.}~\bibnamefont {Zoller}},\ }\bibfield  {title} {\enquote {\bibinfo {title} {Quantum computations with cold trapped ions},}\ }\href {\doibase 10.1103/PhysRevLett.74.4091} {\bibfield  {journal} {\bibinfo  {journal} {Phys. Rev. Lett.}\ }\textbf {\bibinfo {volume} {74}},\ \bibinfo {pages} {4091--4094} (\bibinfo {year} {1995})}\BibitemShut {NoStop}%
\bibitem [{\citenamefont {S\o{}rensen}\ and\ \citenamefont {M\o{}lmer}(1999)}]{PhysRevLett.82.1971}%
  \BibitemOpen
  \bibfield  {author} {\bibinfo {author} {\bibfnamefont {Anders}\ \bibnamefont {S\o{}rensen}}\ and\ \bibinfo {author} {\bibfnamefont {Klaus}\ \bibnamefont {M\o{}lmer}},\ }\bibfield  {title} {\enquote {\bibinfo {title} {Quantum computation with ions in thermal motion},}\ }\href {\doibase 10.1103/PhysRevLett.82.1971} {\bibfield  {journal} {\bibinfo  {journal} {Phys. Rev. Lett.}\ }\textbf {\bibinfo {volume} {82}},\ \bibinfo {pages} {1971--1974} (\bibinfo {year} {1999})}\BibitemShut {NoStop}%
\bibitem [{\citenamefont {Debnath}\ \emph {et~al.}(2016)\citenamefont {Debnath}, \citenamefont {Linke}, \citenamefont {Figgatt}, \citenamefont {Landsman}, \citenamefont {Wright},\ and\ \citenamefont {Monroe}}]{debnath2016demonstration}%
  \BibitemOpen
  \bibfield  {author} {\bibinfo {author} {\bibfnamefont {Shantanu}\ \bibnamefont {Debnath}}, \bibinfo {author} {\bibfnamefont {Norbert~M}\ \bibnamefont {Linke}}, \bibinfo {author} {\bibfnamefont {Caroline}\ \bibnamefont {Figgatt}}, \bibinfo {author} {\bibfnamefont {Kevin~A}\ \bibnamefont {Landsman}}, \bibinfo {author} {\bibfnamefont {Kevin}\ \bibnamefont {Wright}}, \ and\ \bibinfo {author} {\bibfnamefont {Christopher}\ \bibnamefont {Monroe}},\ }\bibfield  {title} {\enquote {\bibinfo {title} {Demonstration of a small programmable quantum computer with atomic qubits},}\ }\href {\doibase https://doi.org/10.1038/nature18648} {\bibfield  {journal} {\bibinfo  {journal} {Nature}\ }\textbf {\bibinfo {volume} {536}},\ \bibinfo {pages} {63--66} (\bibinfo {year} {2016})}\BibitemShut {NoStop}%
\bibitem [{\citenamefont {Monroe}\ \emph {et~al.}(2021)\citenamefont {Monroe}, \citenamefont {Campbell}, \citenamefont {Duan}, \citenamefont {Gong}, \citenamefont {Gorshkov}, \citenamefont {Hess}, \citenamefont {Islam}, \citenamefont {Kim}, \citenamefont {Linke}, \citenamefont {Pagano}, \citenamefont {Richerme}, \citenamefont {Senko},\ and\ \citenamefont {Yao}}]{RevModPhys.93.025001}%
  \BibitemOpen
  \bibfield  {author} {\bibinfo {author} {\bibfnamefont {C.}~\bibnamefont {Monroe}}, \bibinfo {author} {\bibfnamefont {W.~C.}\ \bibnamefont {Campbell}}, \bibinfo {author} {\bibfnamefont {L.-M.}\ \bibnamefont {Duan}}, \bibinfo {author} {\bibfnamefont {Z.-X.}\ \bibnamefont {Gong}}, \bibinfo {author} {\bibfnamefont {A.~V.}\ \bibnamefont {Gorshkov}}, \bibinfo {author} {\bibfnamefont {P.~W.}\ \bibnamefont {Hess}}, \bibinfo {author} {\bibfnamefont {R.}~\bibnamefont {Islam}}, \bibinfo {author} {\bibfnamefont {K.}~\bibnamefont {Kim}}, \bibinfo {author} {\bibfnamefont {N.~M.}\ \bibnamefont {Linke}}, \bibinfo {author} {\bibfnamefont {G.}~\bibnamefont {Pagano}}, \bibinfo {author} {\bibfnamefont {P.}~\bibnamefont {Richerme}}, \bibinfo {author} {\bibfnamefont {C.}~\bibnamefont {Senko}}, \ and\ \bibinfo {author} {\bibfnamefont {N.~Y.}\ \bibnamefont {Yao}},\ }\bibfield  {title} {\enquote {\bibinfo {title} {Programmable quantum simulations of spin systems with trapped ions},}\ }\href {\doibase 10.1103/RevModPhys.93.025001}
  {\bibfield  {journal} {\bibinfo  {journal} {Rev. Mod. Phys.}\ }\textbf {\bibinfo {volume} {93}},\ \bibinfo {pages} {025001} (\bibinfo {year} {2021})}\BibitemShut {NoStop}%
\bibitem [{\citenamefont {Scarlino}\ \emph {et~al.}(2019)\citenamefont {Scarlino}, \citenamefont {Van~Woerkom}, \citenamefont {Mendes}, \citenamefont {Koski}, \citenamefont {Landig}, \citenamefont {Andersen}, \citenamefont {Gasparinetti}, \citenamefont {Reichl}, \citenamefont {Wegscheider}, \citenamefont {Ensslin} \emph {et~al.}}]{scarlino2019coherent}%
  \BibitemOpen
  \bibfield  {author} {\bibinfo {author} {\bibfnamefont {Pasquale}\ \bibnamefont {Scarlino}}, \bibinfo {author} {\bibfnamefont {David~J}\ \bibnamefont {Van~Woerkom}}, \bibinfo {author} {\bibfnamefont {Udson~C}\ \bibnamefont {Mendes}}, \bibinfo {author} {\bibfnamefont {Jonne~V}\ \bibnamefont {Koski}}, \bibinfo {author} {\bibfnamefont {Andreas~J}\ \bibnamefont {Landig}}, \bibinfo {author} {\bibfnamefont {Christian~Kraglund}\ \bibnamefont {Andersen}}, \bibinfo {author} {\bibfnamefont {Simone}\ \bibnamefont {Gasparinetti}}, \bibinfo {author} {\bibfnamefont {Christian}\ \bibnamefont {Reichl}}, \bibinfo {author} {\bibfnamefont {Werner}\ \bibnamefont {Wegscheider}}, \bibinfo {author} {\bibfnamefont {Klaus}\ \bibnamefont {Ensslin}},  \emph {et~al.},\ }\bibfield  {title} {\enquote {\bibinfo {title} {Coherent microwave-photon-mediated coupling between a semiconductor and a superconducting qubit},}\ }\href {\doibase https://doi.org/10.1038/s41467-019-10798-6} {\bibfield  {journal} {\bibinfo  {journal} {Nature
  communications}\ }\textbf {\bibinfo {volume} {10}},\ \bibinfo {pages} {3011} (\bibinfo {year} {2019})}\BibitemShut {NoStop}%
\bibitem [{\citenamefont {Ritter}\ \emph {et~al.}(2012)\citenamefont {Ritter}, \citenamefont {N{\"o}lleke}, \citenamefont {Hahn}, \citenamefont {Reiserer}, \citenamefont {Neuzner}, \citenamefont {Uphoff}, \citenamefont {M{\"u}cke}, \citenamefont {Figueroa}, \citenamefont {Bochmann},\ and\ \citenamefont {Rempe}}]{ritter2012elementary}%
  \BibitemOpen
  \bibfield  {author} {\bibinfo {author} {\bibfnamefont {Stephan}\ \bibnamefont {Ritter}}, \bibinfo {author} {\bibfnamefont {Christian}\ \bibnamefont {N{\"o}lleke}}, \bibinfo {author} {\bibfnamefont {Carolin}\ \bibnamefont {Hahn}}, \bibinfo {author} {\bibfnamefont {Andreas}\ \bibnamefont {Reiserer}}, \bibinfo {author} {\bibfnamefont {Andreas}\ \bibnamefont {Neuzner}}, \bibinfo {author} {\bibfnamefont {Manuel}\ \bibnamefont {Uphoff}}, \bibinfo {author} {\bibfnamefont {Martin}\ \bibnamefont {M{\"u}cke}}, \bibinfo {author} {\bibfnamefont {Eden}\ \bibnamefont {Figueroa}}, \bibinfo {author} {\bibfnamefont {Joerg}\ \bibnamefont {Bochmann}}, \ and\ \bibinfo {author} {\bibfnamefont {Gerhard}\ \bibnamefont {Rempe}},\ }\bibfield  {title} {\enquote {\bibinfo {title} {An elementary quantum network of single atoms in optical cavities},}\ }\href {\doibase https://doi.org/10.1038/nature11023} {\bibfield  {journal} {\bibinfo  {journal} {Nature}\ }\textbf {\bibinfo {volume} {484}},\ \bibinfo {pages} {195--200} (\bibinfo
  {year} {2012})}\BibitemShut {NoStop}%
\bibitem [{\citenamefont {Kato}\ \emph {et~al.}(2019)\citenamefont {Kato}, \citenamefont {N{\'e}met}, \citenamefont {Senga}, \citenamefont {Mizukami}, \citenamefont {Huang}, \citenamefont {Parkins},\ and\ \citenamefont {Aoki}}]{kato2019observation}%
  \BibitemOpen
  \bibfield  {author} {\bibinfo {author} {\bibfnamefont {Shinya}\ \bibnamefont {Kato}}, \bibinfo {author} {\bibfnamefont {Nikolett}\ \bibnamefont {N{\'e}met}}, \bibinfo {author} {\bibfnamefont {Kohei}\ \bibnamefont {Senga}}, \bibinfo {author} {\bibfnamefont {Shota}\ \bibnamefont {Mizukami}}, \bibinfo {author} {\bibfnamefont {Xinhe}\ \bibnamefont {Huang}}, \bibinfo {author} {\bibfnamefont {Scott}\ \bibnamefont {Parkins}}, \ and\ \bibinfo {author} {\bibfnamefont {Takao}\ \bibnamefont {Aoki}},\ }\bibfield  {title} {\enquote {\bibinfo {title} {Observation of dressed states of distant atoms with delocalized photons in coupled-cavities quantum electrodynamics},}\ }\href {\doibase https://doi.org/10.1038/s41467-019-08975-8} {\bibfield  {journal} {\bibinfo  {journal} {Nature communications}\ }\textbf {\bibinfo {volume} {10}},\ \bibinfo {pages} {1160} (\bibinfo {year} {2019})}\BibitemShut {NoStop}%
\bibitem [{\citenamefont {Grinkemeyer}\ \emph {et~al.}(2025)\citenamefont {Grinkemeyer}, \citenamefont {Guardado-Sanchez}, \citenamefont {Dimitrova}, \citenamefont {Shchepanovich}, \citenamefont {Mandopoulou}, \citenamefont {Borregaard}, \citenamefont {Vuleti{\'c}},\ and\ \citenamefont {Lukin}}]{grinkemeyer2025error}%
  \BibitemOpen
  \bibfield  {author} {\bibinfo {author} {\bibfnamefont {Brandon}\ \bibnamefont {Grinkemeyer}}, \bibinfo {author} {\bibfnamefont {Elmer}\ \bibnamefont {Guardado-Sanchez}}, \bibinfo {author} {\bibfnamefont {Ivana}\ \bibnamefont {Dimitrova}}, \bibinfo {author} {\bibfnamefont {Danilo}\ \bibnamefont {Shchepanovich}}, \bibinfo {author} {\bibfnamefont {G~Eirini}\ \bibnamefont {Mandopoulou}}, \bibinfo {author} {\bibfnamefont {Johannes}\ \bibnamefont {Borregaard}}, \bibinfo {author} {\bibfnamefont {Vladan}\ \bibnamefont {Vuleti{\'c}}}, \ and\ \bibinfo {author} {\bibfnamefont {Mikhail~D}\ \bibnamefont {Lukin}},\ }\bibfield  {title} {\enquote {\bibinfo {title} {Error-detected quantum operations with neutral atoms mediated by an optical cavity},}\ }\href {\doibase 10.1126/science.adr7075} {\bibfield  {journal} {\bibinfo  {journal} {Science}\ }\textbf {\bibinfo {volume} {387}},\ \bibinfo {pages} {1301--1305} (\bibinfo {year} {2025})}\BibitemShut {NoStop}%
\bibitem [{\citenamefont {Shao}\ \emph {et~al.}(2024)\citenamefont {Shao}, \citenamefont {Su}, \citenamefont {Li}, \citenamefont {Nath}, \citenamefont {Wu},\ and\ \citenamefont {Li}}]{shao2024rydberg}%
  \BibitemOpen
  \bibfield  {author} {\bibinfo {author} {\bibfnamefont {Xiao-Qiang}\ \bibnamefont {Shao}}, \bibinfo {author} {\bibfnamefont {Shi-Lei}\ \bibnamefont {Su}}, \bibinfo {author} {\bibfnamefont {Lin}\ \bibnamefont {Li}}, \bibinfo {author} {\bibfnamefont {Rejish}\ \bibnamefont {Nath}}, \bibinfo {author} {\bibfnamefont {Jin-Hui}\ \bibnamefont {Wu}}, \ and\ \bibinfo {author} {\bibfnamefont {Weibin}\ \bibnamefont {Li}},\ }\bibfield  {title} {\enquote {\bibinfo {title} {Rydberg superatoms: An artificial quantum system for quantum information processing and quantum optics},}\ }\href {\doibase 10.1063/5.0211071} {\bibfield  {journal} {\bibinfo  {journal} {Applied Physics Reviews}\ }\textbf {\bibinfo {volume} {11}} (\bibinfo {year} {2024}),\ 10.1063/5.0211071}\BibitemShut {NoStop}%
\bibitem [{\citenamefont {Hollerith}\ \emph {et~al.}(2022)\citenamefont {Hollerith}, \citenamefont {Srakaew}, \citenamefont {Wei}, \citenamefont {Rubio-Abadal}, \citenamefont {Adler}, \citenamefont {Weckesser}, \citenamefont {Kruckenhauser}, \citenamefont {Walther}, \citenamefont {van Bijnen}, \citenamefont {Rui}, \citenamefont {Gross}, \citenamefont {Bloch},\ and\ \citenamefont {Zeiher}}]{PhysRevLett.128.113602}%
  \BibitemOpen
  \bibfield  {author} {\bibinfo {author} {\bibfnamefont {Simon}\ \bibnamefont {Hollerith}}, \bibinfo {author} {\bibfnamefont {Kritsana}\ \bibnamefont {Srakaew}}, \bibinfo {author} {\bibfnamefont {David}\ \bibnamefont {Wei}}, \bibinfo {author} {\bibfnamefont {Antonio}\ \bibnamefont {Rubio-Abadal}}, \bibinfo {author} {\bibfnamefont {Daniel}\ \bibnamefont {Adler}}, \bibinfo {author} {\bibfnamefont {Pascal}\ \bibnamefont {Weckesser}}, \bibinfo {author} {\bibfnamefont {Andreas}\ \bibnamefont {Kruckenhauser}}, \bibinfo {author} {\bibfnamefont {Valentin}\ \bibnamefont {Walther}}, \bibinfo {author} {\bibfnamefont {Rick}\ \bibnamefont {van Bijnen}}, \bibinfo {author} {\bibfnamefont {Jun}\ \bibnamefont {Rui}}, \bibinfo {author} {\bibfnamefont {Christian}\ \bibnamefont {Gross}}, \bibinfo {author} {\bibfnamefont {Immanuel}\ \bibnamefont {Bloch}}, \ and\ \bibinfo {author} {\bibfnamefont {Johannes}\ \bibnamefont {Zeiher}},\ }\bibfield  {title} {\enquote {\bibinfo {title} {Realizing distance-selective interactions in a
  {Rydberg}-dressed atom array},}\ }\href {\doibase 10.1103/PhysRevLett.128.113602} {\bibfield  {journal} {\bibinfo  {journal} {Phys. Rev. Lett.}\ }\textbf {\bibinfo {volume} {128}},\ \bibinfo {pages} {113602} (\bibinfo {year} {2022})}\BibitemShut {NoStop}%
\bibitem [{\citenamefont {Cesa}\ and\ \citenamefont {Pichler}(2023)}]{PhysRevLett.131.170601}%
  \BibitemOpen
  \bibfield  {author} {\bibinfo {author} {\bibfnamefont {Francesco}\ \bibnamefont {Cesa}}\ and\ \bibinfo {author} {\bibfnamefont {Hannes}\ \bibnamefont {Pichler}},\ }\bibfield  {title} {\enquote {\bibinfo {title} {Universal quantum computation in globally driven {Rydberg} atom arrays},}\ }\href {\doibase 10.1103/PhysRevLett.131.170601} {\bibfield  {journal} {\bibinfo  {journal} {Phys. Rev. Lett.}\ }\textbf {\bibinfo {volume} {131}},\ \bibinfo {pages} {170601} (\bibinfo {year} {2023})}\BibitemShut {NoStop}%
\bibitem [{\citenamefont {Stojanovi\ifmmode~\acute{c}\else \'{c}\fi{}}(2021)}]{PhysRevA.103.022410}%
  \BibitemOpen
  \bibfield  {author} {\bibinfo {author} {\bibfnamefont {Vladimir~M.}\ \bibnamefont {Stojanovi\ifmmode~\acute{c}\else \'{c}\fi{}}},\ }\bibfield  {title} {\enquote {\bibinfo {title} {Scalable {$W$}-type entanglement resource in neutral-atom arrays with {Rydberg}-dressed resonant dipole-dipole interaction},}\ }\href {\doibase 10.1103/PhysRevA.103.022410} {\bibfield  {journal} {\bibinfo  {journal} {Phys. Rev. A}\ }\textbf {\bibinfo {volume} {103}},\ \bibinfo {pages} {022410} (\bibinfo {year} {2021})}\BibitemShut {NoStop}%
\bibitem [{\citenamefont {Schempp}\ \emph {et~al.}(2015)\citenamefont {Schempp}, \citenamefont {G\"unter}, \citenamefont {W\"uster}, \citenamefont {Weidem\"uller},\ and\ \citenamefont {Whitlock}}]{PhysRevLett.115.093002}%
  \BibitemOpen
  \bibfield  {author} {\bibinfo {author} {\bibfnamefont {H.}~\bibnamefont {Schempp}}, \bibinfo {author} {\bibfnamefont {G.}~\bibnamefont {G\"unter}}, \bibinfo {author} {\bibfnamefont {S.}~\bibnamefont {W\"uster}}, \bibinfo {author} {\bibfnamefont {M.}~\bibnamefont {Weidem\"uller}}, \ and\ \bibinfo {author} {\bibfnamefont {S.}~\bibnamefont {Whitlock}},\ }\bibfield  {title} {\enquote {\bibinfo {title} {Correlated exciton transport in {Rydberg}-dressed-atom spin chains},}\ }\href {\doibase 10.1103/PhysRevLett.115.093002} {\bibfield  {journal} {\bibinfo  {journal} {Phys. Rev. Lett.}\ }\textbf {\bibinfo {volume} {115}},\ \bibinfo {pages} {093002} (\bibinfo {year} {2015})}\BibitemShut {NoStop}%
\bibitem [{\citenamefont {Liu}\ \emph {et~al.}(2022)\citenamefont {Liu}, \citenamefont {Yang}, \citenamefont {Bienias}, \citenamefont {Iadecola},\ and\ \citenamefont {Gorshkov}}]{PhysRevLett.128.013603}%
  \BibitemOpen
  \bibfield  {author} {\bibinfo {author} {\bibfnamefont {Fangli}\ \bibnamefont {Liu}}, \bibinfo {author} {\bibfnamefont {Zhi-Cheng}\ \bibnamefont {Yang}}, \bibinfo {author} {\bibfnamefont {Przemyslaw}\ \bibnamefont {Bienias}}, \bibinfo {author} {\bibfnamefont {Thomas}\ \bibnamefont {Iadecola}}, \ and\ \bibinfo {author} {\bibfnamefont {Alexey~V.}\ \bibnamefont {Gorshkov}},\ }\bibfield  {title} {\enquote {\bibinfo {title} {Localization and criticality in antiblockaded two-dimensional {Rydberg} atom arrays},}\ }\href {\doibase 10.1103/PhysRevLett.128.013603} {\bibfield  {journal} {\bibinfo  {journal} {Phys. Rev. Lett.}\ }\textbf {\bibinfo {volume} {128}},\ \bibinfo {pages} {013603} (\bibinfo {year} {2022})}\BibitemShut {NoStop}%
\bibitem [{\citenamefont {Scholl}\ \emph {et~al.}(2022)\citenamefont {Scholl}, \citenamefont {Williams}, \citenamefont {Bornet}, \citenamefont {Wallner}, \citenamefont {Barredo}, \citenamefont {Henriet}, \citenamefont {Signoles}, \citenamefont {Hainaut}, \citenamefont {Franz}, \citenamefont {Geier}, \citenamefont {Tebben}, \citenamefont {Salzinger}, \citenamefont {Z\"urn}, \citenamefont {Lahaye}, \citenamefont {Weidem\"uller},\ and\ \citenamefont {Browaeys}}]{PRXQuantum.3.020303}%
  \BibitemOpen
  \bibfield  {author} {\bibinfo {author} {\bibfnamefont {P.}~\bibnamefont {Scholl}}, \bibinfo {author} {\bibfnamefont {H.~J.}\ \bibnamefont {Williams}}, \bibinfo {author} {\bibfnamefont {G.}~\bibnamefont {Bornet}}, \bibinfo {author} {\bibfnamefont {F.}~\bibnamefont {Wallner}}, \bibinfo {author} {\bibfnamefont {D.}~\bibnamefont {Barredo}}, \bibinfo {author} {\bibfnamefont {L.}~\bibnamefont {Henriet}}, \bibinfo {author} {\bibfnamefont {A.}~\bibnamefont {Signoles}}, \bibinfo {author} {\bibfnamefont {C.}~\bibnamefont {Hainaut}}, \bibinfo {author} {\bibfnamefont {T.}~\bibnamefont {Franz}}, \bibinfo {author} {\bibfnamefont {S.}~\bibnamefont {Geier}}, \bibinfo {author} {\bibfnamefont {A.}~\bibnamefont {Tebben}}, \bibinfo {author} {\bibfnamefont {A.}~\bibnamefont {Salzinger}}, \bibinfo {author} {\bibfnamefont {G.}~\bibnamefont {Z\"urn}}, \bibinfo {author} {\bibfnamefont {T.}~\bibnamefont {Lahaye}}, \bibinfo {author} {\bibfnamefont {M.}~\bibnamefont {Weidem\"uller}}, \ and\ \bibinfo {author} {\bibfnamefont
  {A.}~\bibnamefont {Browaeys}},\ }\bibfield  {title} {\enquote {\bibinfo {title} {Microwave engineering of programmable $xxz$ hamiltonians in arrays of {Rydberg} atoms},}\ }\href {\doibase 10.1103/PRXQuantum.3.020303} {\bibfield  {journal} {\bibinfo  {journal} {PRX Quantum}\ }\textbf {\bibinfo {volume} {3}},\ \bibinfo {pages} {020303} (\bibinfo {year} {2022})}\BibitemShut {NoStop}%
\bibitem [{\citenamefont {Wu}\ \emph {et~al.}(2022)\citenamefont {Wu}, \citenamefont {Yang}, \citenamefont {Yang}, \citenamefont {M\o{}lmer}, \citenamefont {Pohl}, \citenamefont {Tey},\ and\ \citenamefont {You}}]{PhysRevResearch.4.L032046}%
  \BibitemOpen
  \bibfield  {author} {\bibinfo {author} {\bibfnamefont {Xiaoling}\ \bibnamefont {Wu}}, \bibinfo {author} {\bibfnamefont {Fan}\ \bibnamefont {Yang}}, \bibinfo {author} {\bibfnamefont {Shuo}\ \bibnamefont {Yang}}, \bibinfo {author} {\bibfnamefont {Klaus}\ \bibnamefont {M\o{}lmer}}, \bibinfo {author} {\bibfnamefont {Thomas}\ \bibnamefont {Pohl}}, \bibinfo {author} {\bibfnamefont {Meng~Khoon}\ \bibnamefont {Tey}}, \ and\ \bibinfo {author} {\bibfnamefont {Li}~\bibnamefont {You}},\ }\bibfield  {title} {\enquote {\bibinfo {title} {Manipulating synthetic gauge fluxes via multicolor dressing of {Rydberg}-atom arrays},}\ }\href {\doibase 10.1103/PhysRevResearch.4.L032046} {\bibfield  {journal} {\bibinfo  {journal} {Phys. Rev. Res.}\ }\textbf {\bibinfo {volume} {4}},\ \bibinfo {pages} {L032046} (\bibinfo {year} {2022})}\BibitemShut {NoStop}%
\bibitem [{\citenamefont {Bravo}\ \emph {et~al.}(2022)\citenamefont {Bravo}, \citenamefont {Najafi}, \citenamefont {Gao},\ and\ \citenamefont {Yelin}}]{PRXQuantum.3.030325}%
  \BibitemOpen
  \bibfield  {author} {\bibinfo {author} {\bibfnamefont {Rodrigo~Araiza}\ \bibnamefont {Bravo}}, \bibinfo {author} {\bibfnamefont {Khadijeh}\ \bibnamefont {Najafi}}, \bibinfo {author} {\bibfnamefont {Xun}\ \bibnamefont {Gao}}, \ and\ \bibinfo {author} {\bibfnamefont {Susanne~F.}\ \bibnamefont {Yelin}},\ }\bibfield  {title} {\enquote {\bibinfo {title} {Quantum reservoir computing using arrays of {Rydberg} atoms},}\ }\href {\doibase 10.1103/PRXQuantum.3.030325} {\bibfield  {journal} {\bibinfo  {journal} {PRX Quantum}\ }\textbf {\bibinfo {volume} {3}},\ \bibinfo {pages} {030325} (\bibinfo {year} {2022})}\BibitemShut {NoStop}%
\bibitem [{\citenamefont {Byun}\ \emph {et~al.}(2024)\citenamefont {Byun}, \citenamefont {Jeong},\ and\ \citenamefont {Ahn}}]{PhysRevA.110.042612}%
  \BibitemOpen
  \bibfield  {author} {\bibinfo {author} {\bibfnamefont {Andrew}\ \bibnamefont {Byun}}, \bibinfo {author} {\bibfnamefont {Seokho}\ \bibnamefont {Jeong}}, \ and\ \bibinfo {author} {\bibfnamefont {Jaewook}\ \bibnamefont {Ahn}},\ }\bibfield  {title} {\enquote {\bibinfo {title} {Programming higher-order interactions of {Rydberg} atoms},}\ }\href {\doibase 10.1103/PhysRevA.110.042612} {\bibfield  {journal} {\bibinfo  {journal} {Phys. Rev. A}\ }\textbf {\bibinfo {volume} {110}},\ \bibinfo {pages} {042612} (\bibinfo {year} {2024})}\BibitemShut {NoStop}%
\bibitem [{\citenamefont {Adams}\ \emph {et~al.}(2020)\citenamefont {Adams}, \citenamefont {Pritchard},\ and\ \citenamefont {Shaffer}}]{adams2020rydberg}%
  \BibitemOpen
  \bibfield  {author} {\bibinfo {author} {\bibfnamefont {Charles~S}\ \bibnamefont {Adams}}, \bibinfo {author} {\bibfnamefont {Jonathan~D}\ \bibnamefont {Pritchard}}, \ and\ \bibinfo {author} {\bibfnamefont {James~P}\ \bibnamefont {Shaffer}},\ }\bibfield  {title} {\enquote {\bibinfo {title} {{Rydberg} atom quantum technologies},}\ }\href {\doibase 10.1088/1361-6455/ab52ef} {\bibfield  {journal} {\bibinfo  {journal} {Journal of Physics B: Atomic, Molecular and Optical Physics}\ }\textbf {\bibinfo {volume} {53}},\ \bibinfo {pages} {012002} (\bibinfo {year} {2020})}\BibitemShut {NoStop}%
\bibitem [{\citenamefont {Browaeys}\ and\ \citenamefont {Lahaye}(2020)}]{browaeys2020many}%
  \BibitemOpen
  \bibfield  {author} {\bibinfo {author} {\bibfnamefont {Antoine}\ \bibnamefont {Browaeys}}\ and\ \bibinfo {author} {\bibfnamefont {Thierry}\ \bibnamefont {Lahaye}},\ }\bibfield  {title} {\enquote {\bibinfo {title} {Many-body physics with individually controlled {Rydberg} atoms},}\ }\href {\doibase https://doi.org/10.1038/s41567-019-0733-z} {\bibfield  {journal} {\bibinfo  {journal} {Nature Physics}\ }\textbf {\bibinfo {volume} {16}},\ \bibinfo {pages} {132--142} (\bibinfo {year} {2020})}\BibitemShut {NoStop}%
\bibitem [{\citenamefont {Levine}\ \emph {et~al.}(2019)\citenamefont {Levine}, \citenamefont {Keesling}, \citenamefont {Semeghini}, \citenamefont {Omran}, \citenamefont {Wang}, \citenamefont {Ebadi}, \citenamefont {Bernien}, \citenamefont {Greiner}, \citenamefont {Vuleti\ifmmode~\acute{c}\else \'{c}\fi{}}, \citenamefont {Pichler},\ and\ \citenamefont {Lukin}}]{PhysRevLett.123.170503}%
  \BibitemOpen
  \bibfield  {author} {\bibinfo {author} {\bibfnamefont {Harry}\ \bibnamefont {Levine}}, \bibinfo {author} {\bibfnamefont {Alexander}\ \bibnamefont {Keesling}}, \bibinfo {author} {\bibfnamefont {Giulia}\ \bibnamefont {Semeghini}}, \bibinfo {author} {\bibfnamefont {Ahmed}\ \bibnamefont {Omran}}, \bibinfo {author} {\bibfnamefont {Tout~T.}\ \bibnamefont {Wang}}, \bibinfo {author} {\bibfnamefont {Sepehr}\ \bibnamefont {Ebadi}}, \bibinfo {author} {\bibfnamefont {Hannes}\ \bibnamefont {Bernien}}, \bibinfo {author} {\bibfnamefont {Markus}\ \bibnamefont {Greiner}}, \bibinfo {author} {\bibfnamefont {Vladan}\ \bibnamefont {Vuleti\ifmmode~\acute{c}\else \'{c}\fi{}}}, \bibinfo {author} {\bibfnamefont {Hannes}\ \bibnamefont {Pichler}}, \ and\ \bibinfo {author} {\bibfnamefont {Mikhail~D.}\ \bibnamefont {Lukin}},\ }\bibfield  {title} {\enquote {\bibinfo {title} {Parallel implementation of high-fidelity multiqubit gates with neutral atoms},}\ }\href {\doibase 10.1103/PhysRevLett.123.170503} {\bibfield  {journal} {\bibinfo
  {journal} {Phys. Rev. Lett.}\ }\textbf {\bibinfo {volume} {123}},\ \bibinfo {pages} {170503} (\bibinfo {year} {2019})}\BibitemShut {NoStop}%
\bibitem [{\citenamefont {Evered}\ \emph {et~al.}(2023)\citenamefont {Evered}, \citenamefont {Bluvstein}, \citenamefont {Kalinowski}, \citenamefont {Ebadi}, \citenamefont {Manovitz}, \citenamefont {Zhou}, \citenamefont {Li}, \citenamefont {Geim}, \citenamefont {Wang}, \citenamefont {Maskara} \emph {et~al.}}]{evered2023high}%
  \BibitemOpen
  \bibfield  {author} {\bibinfo {author} {\bibfnamefont {Simon~J}\ \bibnamefont {Evered}}, \bibinfo {author} {\bibfnamefont {Dolev}\ \bibnamefont {Bluvstein}}, \bibinfo {author} {\bibfnamefont {Marcin}\ \bibnamefont {Kalinowski}}, \bibinfo {author} {\bibfnamefont {Sepehr}\ \bibnamefont {Ebadi}}, \bibinfo {author} {\bibfnamefont {Tom}\ \bibnamefont {Manovitz}}, \bibinfo {author} {\bibfnamefont {Hengyun}\ \bibnamefont {Zhou}}, \bibinfo {author} {\bibfnamefont {Sophie~H}\ \bibnamefont {Li}}, \bibinfo {author} {\bibfnamefont {Alexandra~A}\ \bibnamefont {Geim}}, \bibinfo {author} {\bibfnamefont {Tout~T}\ \bibnamefont {Wang}}, \bibinfo {author} {\bibfnamefont {Nishad}\ \bibnamefont {Maskara}},  \emph {et~al.},\ }\bibfield  {title} {\enquote {\bibinfo {title} {High-fidelity parallel entangling gates on a neutral-atom quantum computer},}\ }\href {\doibase https://doi.org/10.1038/s41586-023-06481-y} {\bibfield  {journal} {\bibinfo  {journal} {Nature}\ }\textbf {\bibinfo {volume} {622}},\ \bibinfo {pages} {268--272}
  (\bibinfo {year} {2023})}\BibitemShut {NoStop}%
\bibitem [{\citenamefont {Guo}\ \emph {et~al.}(2025)\citenamefont {Guo}, \citenamefont {Su}, \citenamefont {Li},\ and\ \citenamefont {Shao}}]{PhysRevA.111.022420}%
  \BibitemOpen
  \bibfield  {author} {\bibinfo {author} {\bibfnamefont {F.~Q.}\ \bibnamefont {Guo}}, \bibinfo {author} {\bibfnamefont {Shi-Lei}\ \bibnamefont {Su}}, \bibinfo {author} {\bibfnamefont {Weibin}\ \bibnamefont {Li}}, \ and\ \bibinfo {author} {\bibfnamefont {X.~Q.}\ \bibnamefont {Shao}},\ }\bibfield  {title} {\enquote {\bibinfo {title} {Parity-controlled gate in a two-dimensional neutral-atom array},}\ }\href {\doibase 10.1103/PhysRevA.111.022420} {\bibfield  {journal} {\bibinfo  {journal} {Phys. Rev. A}\ }\textbf {\bibinfo {volume} {111}},\ \bibinfo {pages} {022420} (\bibinfo {year} {2025})}\BibitemShut {NoStop}%
\bibitem [{\citenamefont {Graham}\ \emph {et~al.}(2022)\citenamefont {Graham}, \citenamefont {Song}, \citenamefont {Scott}, \citenamefont {Poole}, \citenamefont {Phuttitarn}, \citenamefont {Jooya}, \citenamefont {Eichler}, \citenamefont {Jiang}, \citenamefont {Marra}, \citenamefont {Grinkemeyer} \emph {et~al.}}]{graham2022multi}%
  \BibitemOpen
  \bibfield  {author} {\bibinfo {author} {\bibfnamefont {TM}~\bibnamefont {Graham}}, \bibinfo {author} {\bibfnamefont {Y}~\bibnamefont {Song}}, \bibinfo {author} {\bibfnamefont {J}~\bibnamefont {Scott}}, \bibinfo {author} {\bibfnamefont {C}~\bibnamefont {Poole}}, \bibinfo {author} {\bibfnamefont {L}~\bibnamefont {Phuttitarn}}, \bibinfo {author} {\bibfnamefont {K}~\bibnamefont {Jooya}}, \bibinfo {author} {\bibfnamefont {P}~\bibnamefont {Eichler}}, \bibinfo {author} {\bibfnamefont {X}~\bibnamefont {Jiang}}, \bibinfo {author} {\bibfnamefont {A}~\bibnamefont {Marra}}, \bibinfo {author} {\bibfnamefont {B}~\bibnamefont {Grinkemeyer}},  \emph {et~al.},\ }\bibfield  {title} {\enquote {\bibinfo {title} {Multi-qubit entanglement and algorithms on a neutral-atom quantum computer},}\ }\href {\doibase https://doi.org/10.1038/s41586-022-04603-6} {\bibfield  {journal} {\bibinfo  {journal} {Nature}\ }\textbf {\bibinfo {volume} {604}},\ \bibinfo {pages} {457--462} (\bibinfo {year} {2022})}\BibitemShut {NoStop}%
\bibitem [{\citenamefont {Omran}\ \emph {et~al.}(2019)\citenamefont {Omran}, \citenamefont {Levine}, \citenamefont {Keesling}, \citenamefont {Semeghini}, \citenamefont {Wang}, \citenamefont {Ebadi}, \citenamefont {Bernien}, \citenamefont {Zibrov}, \citenamefont {Pichler}, \citenamefont {Choi} \emph {et~al.}}]{omran2019generation}%
  \BibitemOpen
  \bibfield  {author} {\bibinfo {author} {\bibfnamefont {Ahmed}\ \bibnamefont {Omran}}, \bibinfo {author} {\bibfnamefont {Harry}\ \bibnamefont {Levine}}, \bibinfo {author} {\bibfnamefont {Alexander}\ \bibnamefont {Keesling}}, \bibinfo {author} {\bibfnamefont {Giulia}\ \bibnamefont {Semeghini}}, \bibinfo {author} {\bibfnamefont {Tout~T}\ \bibnamefont {Wang}}, \bibinfo {author} {\bibfnamefont {Sepehr}\ \bibnamefont {Ebadi}}, \bibinfo {author} {\bibfnamefont {Hannes}\ \bibnamefont {Bernien}}, \bibinfo {author} {\bibfnamefont {Alexander~S}\ \bibnamefont {Zibrov}}, \bibinfo {author} {\bibfnamefont {Hannes}\ \bibnamefont {Pichler}}, \bibinfo {author} {\bibfnamefont {Soonwon}\ \bibnamefont {Choi}},  \emph {et~al.},\ }\bibfield  {title} {\enquote {\bibinfo {title} {Generation and manipulation of schr{\"o}dinger cat states in {Rydberg} atom arrays},}\ }\href {\doibase 10.1126/science.aax9743} {\bibfield  {journal} {\bibinfo  {journal} {Science}\ }\textbf {\bibinfo {volume} {365}},\ \bibinfo {pages} {570--574}
  (\bibinfo {year} {2019})}\BibitemShut {NoStop}%
\bibitem [{\citenamefont {Graham}\ \emph {et~al.}(2019)\citenamefont {Graham}, \citenamefont {Kwon}, \citenamefont {Grinkemeyer}, \citenamefont {Marra}, \citenamefont {Jiang}, \citenamefont {Lichtman}, \citenamefont {Sun}, \citenamefont {Ebert},\ and\ \citenamefont {Saffman}}]{PhysRevLett.123.230501}%
  \BibitemOpen
  \bibfield  {author} {\bibinfo {author} {\bibfnamefont {T.~M.}\ \bibnamefont {Graham}}, \bibinfo {author} {\bibfnamefont {M.}~\bibnamefont {Kwon}}, \bibinfo {author} {\bibfnamefont {B.}~\bibnamefont {Grinkemeyer}}, \bibinfo {author} {\bibfnamefont {Z.}~\bibnamefont {Marra}}, \bibinfo {author} {\bibfnamefont {X.}~\bibnamefont {Jiang}}, \bibinfo {author} {\bibfnamefont {M.~T.}\ \bibnamefont {Lichtman}}, \bibinfo {author} {\bibfnamefont {Y.}~\bibnamefont {Sun}}, \bibinfo {author} {\bibfnamefont {M.}~\bibnamefont {Ebert}}, \ and\ \bibinfo {author} {\bibfnamefont {M.}~\bibnamefont {Saffman}},\ }\bibfield  {title} {\enquote {\bibinfo {title} {Rydberg-mediated entanglement in a two-dimensional neutral atom qubit array},}\ }\href {\doibase 10.1103/PhysRevLett.123.230501} {\bibfield  {journal} {\bibinfo  {journal} {Phys. Rev. Lett.}\ }\textbf {\bibinfo {volume} {123}},\ \bibinfo {pages} {230501} (\bibinfo {year} {2019})}\BibitemShut {NoStop}%
\bibitem [{\citenamefont {Cesa}\ and\ \citenamefont {Martin}(2017)}]{PhysRevA.95.052330}%
  \BibitemOpen
  \bibfield  {author} {\bibinfo {author} {\bibfnamefont {A.}~\bibnamefont {Cesa}}\ and\ \bibinfo {author} {\bibfnamefont {J.}~\bibnamefont {Martin}},\ }\bibfield  {title} {\enquote {\bibinfo {title} {Two-qubit entangling gates between distant atomic qubits in a lattice},}\ }\href {\doibase 10.1103/PhysRevA.95.052330} {\bibfield  {journal} {\bibinfo  {journal} {Phys. Rev. A}\ }\textbf {\bibinfo {volume} {95}},\ \bibinfo {pages} {052330} (\bibinfo {year} {2017})}\BibitemShut {NoStop}%
\bibitem [{\citenamefont {Bluvstein}\ \emph {et~al.}(2022)\citenamefont {Bluvstein}, \citenamefont {Levine}, \citenamefont {Semeghini}, \citenamefont {Wang}, \citenamefont {Ebadi}, \citenamefont {Kalinowski}, \citenamefont {Keesling}, \citenamefont {Maskara}, \citenamefont {Pichler}, \citenamefont {Greiner} \emph {et~al.}}]{bluvstein2022quantum}%
  \BibitemOpen
  \bibfield  {author} {\bibinfo {author} {\bibfnamefont {Dolev}\ \bibnamefont {Bluvstein}}, \bibinfo {author} {\bibfnamefont {Harry}\ \bibnamefont {Levine}}, \bibinfo {author} {\bibfnamefont {Giulia}\ \bibnamefont {Semeghini}}, \bibinfo {author} {\bibfnamefont {Tout~T}\ \bibnamefont {Wang}}, \bibinfo {author} {\bibfnamefont {Sepehr}\ \bibnamefont {Ebadi}}, \bibinfo {author} {\bibfnamefont {Marcin}\ \bibnamefont {Kalinowski}}, \bibinfo {author} {\bibfnamefont {Alexander}\ \bibnamefont {Keesling}}, \bibinfo {author} {\bibfnamefont {Nishad}\ \bibnamefont {Maskara}}, \bibinfo {author} {\bibfnamefont {Hannes}\ \bibnamefont {Pichler}}, \bibinfo {author} {\bibfnamefont {Markus}\ \bibnamefont {Greiner}},  \emph {et~al.},\ }\bibfield  {title} {\enquote {\bibinfo {title} {A quantum processor based on coherent transport of entangled atom arrays},}\ }\href {\doibase https://doi.org/10.1038/s41586-022-04592-6} {\bibfield  {journal} {\bibinfo  {journal} {Nature}\ }\textbf {\bibinfo {volume} {604}},\ \bibinfo {pages}
  {451--456} (\bibinfo {year} {2022})}\BibitemShut {NoStop}%
\bibitem [{\citenamefont {Bornet}\ \emph {et~al.}(2023)\citenamefont {Bornet}, \citenamefont {Emperauger}, \citenamefont {Chen}, \citenamefont {Ye}, \citenamefont {Block}, \citenamefont {Bintz}, \citenamefont {Boyd}, \citenamefont {Barredo}, \citenamefont {Comparin}, \citenamefont {Mezzacapo} \emph {et~al.}}]{bornet2023scalable}%
  \BibitemOpen
  \bibfield  {author} {\bibinfo {author} {\bibfnamefont {Guillaume}\ \bibnamefont {Bornet}}, \bibinfo {author} {\bibfnamefont {Gabriel}\ \bibnamefont {Emperauger}}, \bibinfo {author} {\bibfnamefont {Cheng}\ \bibnamefont {Chen}}, \bibinfo {author} {\bibfnamefont {Bingtian}\ \bibnamefont {Ye}}, \bibinfo {author} {\bibfnamefont {Maxwell}\ \bibnamefont {Block}}, \bibinfo {author} {\bibfnamefont {Marcus}\ \bibnamefont {Bintz}}, \bibinfo {author} {\bibfnamefont {Jamie~A}\ \bibnamefont {Boyd}}, \bibinfo {author} {\bibfnamefont {Daniel}\ \bibnamefont {Barredo}}, \bibinfo {author} {\bibfnamefont {Tommaso}\ \bibnamefont {Comparin}}, \bibinfo {author} {\bibfnamefont {Fabio}\ \bibnamefont {Mezzacapo}},  \emph {et~al.},\ }\bibfield  {title} {\enquote {\bibinfo {title} {Scalable spin squeezing in a dipolar {Rydberg} atom array},}\ }\href {\doibase https://doi.org/10.1038/s41586-023-06414-9} {\bibfield  {journal} {\bibinfo  {journal} {Nature}\ }\textbf {\bibinfo {volume} {621}},\ \bibinfo {pages} {728--733} (\bibinfo {year}
  {2023})}\BibitemShut {NoStop}%
\bibitem [{\citenamefont {Carrera}\ \emph {et~al.}(2025)\citenamefont {Carrera}, \citenamefont {Erbin},\ and\ \citenamefont {Misguich}}]{wdqt-tpwz}%
  \BibitemOpen
  \bibfield  {author} {\bibinfo {author} {\bibfnamefont {Edison~S.}\ \bibnamefont {Carrera}}, \bibinfo {author} {\bibfnamefont {Harold}\ \bibnamefont {Erbin}}, \ and\ \bibinfo {author} {\bibfnamefont {Gr\'egoire}\ \bibnamefont {Misguich}},\ }\bibfield  {title} {\enquote {\bibinfo {title} {Preparing spin-squeezed states in {Rydberg} atom arrays via quantum optimal control},}\ }\href {\doibase 10.1103/wdqt-tpwz} {\bibfield  {journal} {\bibinfo  {journal} {Phys. Rev. A}\ }\textbf {\bibinfo {volume} {112}},\ \bibinfo {pages} {052615} (\bibinfo {year} {2025})}\BibitemShut {NoStop}%
\bibitem [{\citenamefont {Senoo}\ \emph {et~al.}(2026)\citenamefont {Senoo}, \citenamefont {Baumg{\"a}rtner}, \citenamefont {Lis}, \citenamefont {Vaidya}, \citenamefont {Zeng}, \citenamefont {Giudici}, \citenamefont {Pichler},\ and\ \citenamefont {Kaufman}}]{senoo2026high}%
  \BibitemOpen
  \bibfield  {author} {\bibinfo {author} {\bibfnamefont {Aruku}\ \bibnamefont {Senoo}}, \bibinfo {author} {\bibfnamefont {Alexander}\ \bibnamefont {Baumg{\"a}rtner}}, \bibinfo {author} {\bibfnamefont {Joanna~W}\ \bibnamefont {Lis}}, \bibinfo {author} {\bibfnamefont {Gaurav~M}\ \bibnamefont {Vaidya}}, \bibinfo {author} {\bibfnamefont {Zhongda}\ \bibnamefont {Zeng}}, \bibinfo {author} {\bibfnamefont {Giuliano}\ \bibnamefont {Giudici}}, \bibinfo {author} {\bibfnamefont {Hannes}\ \bibnamefont {Pichler}}, \ and\ \bibinfo {author} {\bibfnamefont {Adam~M}\ \bibnamefont {Kaufman}},\ }\bibfield  {title} {\enquote {\bibinfo {title} {High-fidelity entanglement and coherent multi-qubit mapping in an atom array},}\ }\href {\doibase https://doi.org/10.1038/s41567-026-03258-8} {\bibfield  {journal} {\bibinfo  {journal} {Nature Physics}\ ,\ \bibinfo {pages} {1--7}} (\bibinfo {year} {2026})}\BibitemShut {NoStop}%
\bibitem [{\citenamefont {Ebadi}\ \emph {et~al.}(2022)\citenamefont {Ebadi}, \citenamefont {Keesling}, \citenamefont {Cain}, \citenamefont {Wang}, \citenamefont {Levine}, \citenamefont {Bluvstein}, \citenamefont {Semeghini}, \citenamefont {Omran}, \citenamefont {Liu}, \citenamefont {Samajdar} \emph {et~al.}}]{ebadi2022quantum}%
  \BibitemOpen
  \bibfield  {author} {\bibinfo {author} {\bibfnamefont {Sepehr}\ \bibnamefont {Ebadi}}, \bibinfo {author} {\bibfnamefont {Alexander}\ \bibnamefont {Keesling}}, \bibinfo {author} {\bibfnamefont {Madelyn}\ \bibnamefont {Cain}}, \bibinfo {author} {\bibfnamefont {Tout~T}\ \bibnamefont {Wang}}, \bibinfo {author} {\bibfnamefont {Harry}\ \bibnamefont {Levine}}, \bibinfo {author} {\bibfnamefont {Dolev}\ \bibnamefont {Bluvstein}}, \bibinfo {author} {\bibfnamefont {Giulia}\ \bibnamefont {Semeghini}}, \bibinfo {author} {\bibfnamefont {Ahmed}\ \bibnamefont {Omran}}, \bibinfo {author} {\bibfnamefont {J-G}\ \bibnamefont {Liu}}, \bibinfo {author} {\bibfnamefont {Rhine}\ \bibnamefont {Samajdar}},  \emph {et~al.},\ }\bibfield  {title} {\enquote {\bibinfo {title} {Quantum optimization of maximum independent set using {Rydberg} atom arrays},}\ }\href {\doibase 10.1126/science.abo6587} {\bibfield  {journal} {\bibinfo  {journal} {Science}\ }\textbf {\bibinfo {volume} {376}},\ \bibinfo {pages} {1209--1215} (\bibinfo {year}
  {2022})}\BibitemShut {NoStop}%
\bibitem [{\citenamefont {Nguyen}\ \emph {et~al.}(2023)\citenamefont {Nguyen}, \citenamefont {Liu}, \citenamefont {Wurtz}, \citenamefont {Lukin}, \citenamefont {Wang},\ and\ \citenamefont {Pichler}}]{PRXQuantum.4.010316}%
  \BibitemOpen
  \bibfield  {author} {\bibinfo {author} {\bibfnamefont {Minh-Thi}\ \bibnamefont {Nguyen}}, \bibinfo {author} {\bibfnamefont {Jin-Guo}\ \bibnamefont {Liu}}, \bibinfo {author} {\bibfnamefont {Jonathan}\ \bibnamefont {Wurtz}}, \bibinfo {author} {\bibfnamefont {Mikhail~D.}\ \bibnamefont {Lukin}}, \bibinfo {author} {\bibfnamefont {Sheng-Tao}\ \bibnamefont {Wang}}, \ and\ \bibinfo {author} {\bibfnamefont {Hannes}\ \bibnamefont {Pichler}},\ }\bibfield  {title} {\enquote {\bibinfo {title} {Quantum optimization with arbitrary connectivity using {Rydberg} atom arrays},}\ }\href {\doibase 10.1103/PRXQuantum.4.010316} {\bibfield  {journal} {\bibinfo  {journal} {PRX Quantum}\ }\textbf {\bibinfo {volume} {4}},\ \bibinfo {pages} {010316} (\bibinfo {year} {2023})}\BibitemShut {NoStop}%
\bibitem [{\citenamefont {Mao}\ \emph {et~al.}(2026)\citenamefont {Mao}, \citenamefont {Liu}, \citenamefont {Chen}, \citenamefont {Zhang},\ and\ \citenamefont {Sun}}]{516n-6wtx}%
  \BibitemOpen
  \bibfield  {author} {\bibinfo {author} {\bibfnamefont {Rui}\ \bibnamefont {Mao}}, \bibinfo {author} {\bibfnamefont {Jin-Guo}\ \bibnamefont {Liu}}, \bibinfo {author} {\bibfnamefont {Shengminjie}\ \bibnamefont {Chen}}, \bibinfo {author} {\bibfnamefont {Jialin}\ \bibnamefont {Zhang}}, \ and\ \bibinfo {author} {\bibfnamefont {Xiaoming}\ \bibnamefont {Sun}},\ }\bibfield  {title} {\enquote {\bibinfo {title} {Embedding scheme for the maximum-independent-set problem on three-dimensional {Rydberg}-atom arrays},}\ }\href {\doibase 10.1103/516n-6wtx} {\bibfield  {journal} {\bibinfo  {journal} {Phys. Rev. A}\ }\textbf {\bibinfo {volume} {113}},\ \bibinfo {pages} {062420} (\bibinfo {year} {2026})}\BibitemShut {NoStop}%
\bibitem [{\citenamefont {Yeo}\ \emph {et~al.}(2025)\citenamefont {Yeo}, \citenamefont {Kim},\ and\ \citenamefont {Jeong}}]{yeo2025approximating}%
  \BibitemOpen
  \bibfield  {author} {\bibinfo {author} {\bibfnamefont {Hyeonjun}\ \bibnamefont {Yeo}}, \bibinfo {author} {\bibfnamefont {Ha~Eum}\ \bibnamefont {Kim}}, \ and\ \bibinfo {author} {\bibfnamefont {Kabgyun}\ \bibnamefont {Jeong}},\ }\bibfield  {title} {\enquote {\bibinfo {title} {Approximating maximum independent set on {Rydberg} atom arrays using local detunings},}\ }\href {\doibase https://doi.org/10.1002/qute.202400291} {\bibfield  {journal} {\bibinfo  {journal} {Advanced Quantum Technologies}\ }\textbf {\bibinfo {volume} {8}},\ \bibinfo {pages} {2400291} (\bibinfo {year} {2025})}\BibitemShut {NoStop}%
\bibitem [{\citenamefont {Cazals}\ \emph {et~al.}(2025)\citenamefont {Cazals}, \citenamefont {Sorondo}, \citenamefont {Onofre}, \citenamefont {Dalyac}, \citenamefont {Coelho},\ and\ \citenamefont {Vitale}}]{3sjz-gfmr}%
  \BibitemOpen
  \bibfield  {author} {\bibinfo {author} {\bibfnamefont {Pierre}\ \bibnamefont {Cazals}}, \bibinfo {author} {\bibfnamefont {Amalia}\ \bibnamefont {Sorondo}}, \bibinfo {author} {\bibfnamefont {Victor}\ \bibnamefont {Onofre}}, \bibinfo {author} {\bibfnamefont {Constantin}\ \bibnamefont {Dalyac}}, \bibinfo {author} {\bibfnamefont {Wesley}\ \bibnamefont {Coelho}}, \ and\ \bibinfo {author} {\bibfnamefont {Vittorio}\ \bibnamefont {Vitale}},\ }\bibfield  {title} {\enquote {\bibinfo {title} {Quantum optimization on {Rydberg}-atom arrays with arbitrary connectivity: Gadget limitations and a heuristic approach},}\ }\href {\doibase 10.1103/3sjz-gfmr} {\bibfield  {journal} {\bibinfo  {journal} {Phys. Rev. A}\ }\textbf {\bibinfo {volume} {112}},\ \bibinfo {pages} {062416} (\bibinfo {year} {2025})}\BibitemShut {NoStop}%
\bibitem [{\citenamefont {Bombieri}\ \emph {et~al.}(2025)\citenamefont {Bombieri}, \citenamefont {Zeng}, \citenamefont {Tricarico}, \citenamefont {Lin}, \citenamefont {Notarnicola}, \citenamefont {Cain}, \citenamefont {Lukin},\ and\ \citenamefont {Pichler}}]{PRXQuantum.6.020306}%
  \BibitemOpen
  \bibfield  {author} {\bibinfo {author} {\bibfnamefont {Lisa}\ \bibnamefont {Bombieri}}, \bibinfo {author} {\bibfnamefont {Zhongda}\ \bibnamefont {Zeng}}, \bibinfo {author} {\bibfnamefont {Roberto}\ \bibnamefont {Tricarico}}, \bibinfo {author} {\bibfnamefont {Rui}\ \bibnamefont {Lin}}, \bibinfo {author} {\bibfnamefont {Simone}\ \bibnamefont {Notarnicola}}, \bibinfo {author} {\bibfnamefont {Madelyn}\ \bibnamefont {Cain}}, \bibinfo {author} {\bibfnamefont {Mikhail~D.}\ \bibnamefont {Lukin}}, \ and\ \bibinfo {author} {\bibfnamefont {Hannes}\ \bibnamefont {Pichler}},\ }\bibfield  {title} {\enquote {\bibinfo {title} {Quantum adiabatic optimization with {Rydberg} arrays: Localization phenomena and encoding strategies},}\ }\href {\doibase 10.1103/PRXQuantum.6.020306} {\bibfield  {journal} {\bibinfo  {journal} {PRX Quantum}\ }\textbf {\bibinfo {volume} {6}},\ \bibinfo {pages} {020306} (\bibinfo {year} {2025})}\BibitemShut {NoStop}%
\bibitem [{\citenamefont {O’Rourke}\ and\ \citenamefont {Chan}(2023)}]{o2023entanglement}%
  \BibitemOpen
  \bibfield  {author} {\bibinfo {author} {\bibfnamefont {Matthew~J}\ \bibnamefont {O’Rourke}}\ and\ \bibinfo {author} {\bibfnamefont {Garnet Kin-Lic}\ \bibnamefont {Chan}},\ }\bibfield  {title} {\enquote {\bibinfo {title} {Entanglement in the quantum phases of an unfrustrated {Rydberg} atom array},}\ }\href {\doibase https://doi.org/10.1038/s41467-023-41166-0} {\bibfield  {journal} {\bibinfo  {journal} {Nature Communications}\ }\textbf {\bibinfo {volume} {14}},\ \bibinfo {pages} {5397} (\bibinfo {year} {2023})}\BibitemShut {NoStop}%
\bibitem [{\citenamefont {Semeghini}\ \emph {et~al.}(2021)\citenamefont {Semeghini}, \citenamefont {Levine}, \citenamefont {Keesling}, \citenamefont {Ebadi}, \citenamefont {Wang}, \citenamefont {Bluvstein}, \citenamefont {Verresen}, \citenamefont {Pichler}, \citenamefont {Kalinowski}, \citenamefont {Samajdar} \emph {et~al.}}]{semeghini2021probing}%
  \BibitemOpen
  \bibfield  {author} {\bibinfo {author} {\bibfnamefont {Giulia}\ \bibnamefont {Semeghini}}, \bibinfo {author} {\bibfnamefont {Harry}\ \bibnamefont {Levine}}, \bibinfo {author} {\bibfnamefont {Alexander}\ \bibnamefont {Keesling}}, \bibinfo {author} {\bibfnamefont {Sepehr}\ \bibnamefont {Ebadi}}, \bibinfo {author} {\bibfnamefont {Tout~T}\ \bibnamefont {Wang}}, \bibinfo {author} {\bibfnamefont {Dolev}\ \bibnamefont {Bluvstein}}, \bibinfo {author} {\bibfnamefont {Ruben}\ \bibnamefont {Verresen}}, \bibinfo {author} {\bibfnamefont {Hannes}\ \bibnamefont {Pichler}}, \bibinfo {author} {\bibfnamefont {Marcin}\ \bibnamefont {Kalinowski}}, \bibinfo {author} {\bibfnamefont {Rhine}\ \bibnamefont {Samajdar}},  \emph {et~al.},\ }\bibfield  {title} {\enquote {\bibinfo {title} {Probing topological spin liquids on a programmable quantum simulator},}\ }\href {\doibase 10.1126/science.abi8794} {\bibfield  {journal} {\bibinfo  {journal} {Science}\ }\textbf {\bibinfo {volume} {374}},\ \bibinfo {pages} {1242--1247} (\bibinfo
  {year} {2021})}\BibitemShut {NoStop}%
\bibitem [{\citenamefont {Giudici}\ \emph {et~al.}(2022)\citenamefont {Giudici}, \citenamefont {Lukin},\ and\ \citenamefont {Pichler}}]{PhysRevLett.129.090401}%
  \BibitemOpen
  \bibfield  {author} {\bibinfo {author} {\bibfnamefont {Giuliano}\ \bibnamefont {Giudici}}, \bibinfo {author} {\bibfnamefont {Mikhail~D.}\ \bibnamefont {Lukin}}, \ and\ \bibinfo {author} {\bibfnamefont {Hannes}\ \bibnamefont {Pichler}},\ }\bibfield  {title} {\enquote {\bibinfo {title} {Dynamical preparation of quantum spin liquids in {Rydberg} atom arrays},}\ }\href {\doibase 10.1103/PhysRevLett.129.090401} {\bibfield  {journal} {\bibinfo  {journal} {Phys. Rev. Lett.}\ }\textbf {\bibinfo {volume} {129}},\ \bibinfo {pages} {090401} (\bibinfo {year} {2022})}\BibitemShut {NoStop}%
\bibitem [{\citenamefont {Zhang}\ \emph {et~al.}(2025{\natexlab{a}})\citenamefont {Zhang}, \citenamefont {Cant{\'u}}, \citenamefont {Liu}, \citenamefont {Bylinskii}, \citenamefont {Braverman}, \citenamefont {Huber}, \citenamefont {Amato-Grill}, \citenamefont {Lukin}, \citenamefont {Gemelke}, \citenamefont {Keesling} \emph {et~al.}}]{zhang2025probing}%
  \BibitemOpen
  \bibfield  {author} {\bibinfo {author} {\bibfnamefont {Jin}\ \bibnamefont {Zhang}}, \bibinfo {author} {\bibfnamefont {Sergio~H}\ \bibnamefont {Cant{\'u}}}, \bibinfo {author} {\bibfnamefont {Fangli}\ \bibnamefont {Liu}}, \bibinfo {author} {\bibfnamefont {Alexei}\ \bibnamefont {Bylinskii}}, \bibinfo {author} {\bibfnamefont {Boris}\ \bibnamefont {Braverman}}, \bibinfo {author} {\bibfnamefont {Florian}\ \bibnamefont {Huber}}, \bibinfo {author} {\bibfnamefont {Jesse}\ \bibnamefont {Amato-Grill}}, \bibinfo {author} {\bibfnamefont {Alexander}\ \bibnamefont {Lukin}}, \bibinfo {author} {\bibfnamefont {Nathan}\ \bibnamefont {Gemelke}}, \bibinfo {author} {\bibfnamefont {Alexander}\ \bibnamefont {Keesling}},  \emph {et~al.},\ }\bibfield  {title} {\enquote {\bibinfo {title} {Probing quantum floating phases in {Rydberg} atom arrays},}\ }\href {\doibase https://doi.org/10.1038/s41467-025-55947-2} {\bibfield  {journal} {\bibinfo  {journal} {Nature communications}\ }\textbf {\bibinfo {volume} {16}},\ \bibinfo {pages} {712}
  (\bibinfo {year} {2025}{\natexlab{a}})}\BibitemShut {NoStop}%
\bibitem [{\citenamefont {De~L{\'e}s{\'e}leuc}\ \emph {et~al.}(2019)\citenamefont {De~L{\'e}s{\'e}leuc}, \citenamefont {Lienhard}, \citenamefont {Scholl}, \citenamefont {Barredo}, \citenamefont {Weber}, \citenamefont {Lang}, \citenamefont {B{\"u}chler}, \citenamefont {Lahaye},\ and\ \citenamefont {Browaeys}}]{de2019observation}%
  \BibitemOpen
  \bibfield  {author} {\bibinfo {author} {\bibfnamefont {Sylvain}\ \bibnamefont {De~L{\'e}s{\'e}leuc}}, \bibinfo {author} {\bibfnamefont {Vincent}\ \bibnamefont {Lienhard}}, \bibinfo {author} {\bibfnamefont {Pascal}\ \bibnamefont {Scholl}}, \bibinfo {author} {\bibfnamefont {Daniel}\ \bibnamefont {Barredo}}, \bibinfo {author} {\bibfnamefont {Sebastian}\ \bibnamefont {Weber}}, \bibinfo {author} {\bibfnamefont {Nicolai}\ \bibnamefont {Lang}}, \bibinfo {author} {\bibfnamefont {Hans~Peter}\ \bibnamefont {B{\"u}chler}}, \bibinfo {author} {\bibfnamefont {Thierry}\ \bibnamefont {Lahaye}}, \ and\ \bibinfo {author} {\bibfnamefont {Antoine}\ \bibnamefont {Browaeys}},\ }\bibfield  {title} {\enquote {\bibinfo {title} {Observation of a symmetry-protected topological phase of interacting bosons with {Rydberg} atoms},}\ }\href {\doibase 10.1126/science.aav9105} {\bibfield  {journal} {\bibinfo  {journal} {Science}\ }\textbf {\bibinfo {volume} {365}},\ \bibinfo {pages} {775--780} (\bibinfo {year} {2019})}\BibitemShut {NoStop}%
\bibitem [{\citenamefont {Lin}\ \emph {et~al.}(2020)\citenamefont {Lin}, \citenamefont {Calvera},\ and\ \citenamefont {Hsieh}}]{PhysRevB.101.220304}%
  \BibitemOpen
  \bibfield  {author} {\bibinfo {author} {\bibfnamefont {Cheng-Ju}\ \bibnamefont {Lin}}, \bibinfo {author} {\bibfnamefont {Vladimir}\ \bibnamefont {Calvera}}, \ and\ \bibinfo {author} {\bibfnamefont {Timothy~H.}\ \bibnamefont {Hsieh}},\ }\bibfield  {title} {\enquote {\bibinfo {title} {Quantum many-body scar states in two-dimensional {Rydberg} atom arrays},}\ }\href {\doibase 10.1103/PhysRevB.101.220304} {\bibfield  {journal} {\bibinfo  {journal} {Phys. Rev. B}\ }\textbf {\bibinfo {volume} {101}},\ \bibinfo {pages} {220304} (\bibinfo {year} {2020})}\BibitemShut {NoStop}%
\bibitem [{\citenamefont {Da\ifmmode~\breve{g}\else \u{g}\fi{}}\ \emph {et~al.}(2025)\citenamefont {Da\ifmmode~\breve{g}\else \u{g}\fi{}}, \citenamefont {Ma}, \citenamefont {Eugenio}, \citenamefont {Fang},\ and\ \citenamefont {Yelin}}]{jr7l-2cfb}%
  \BibitemOpen
  \bibfield  {author} {\bibinfo {author} {\bibfnamefont {Ceren~B.}\ \bibnamefont {Da\ifmmode~\breve{g}\else \u{g}\fi{}}}, \bibinfo {author} {\bibfnamefont {Hanzhen}\ \bibnamefont {Ma}}, \bibinfo {author} {\bibfnamefont {P.~Myles}\ \bibnamefont {Eugenio}}, \bibinfo {author} {\bibfnamefont {Fang}\ \bibnamefont {Fang}}, \ and\ \bibinfo {author} {\bibfnamefont {Susanne~F.}\ \bibnamefont {Yelin}},\ }\bibfield  {title} {\enquote {\bibinfo {title} {Emergent disorder and sub-ballistic dynamics in quantum simulations of the ising model using {Rydberg} atom arrays},}\ }\href {\doibase 10.1103/jr7l-2cfb} {\bibfield  {journal} {\bibinfo  {journal} {Phys. Rev. Lett.}\ }\textbf {\bibinfo {volume} {135}},\ \bibinfo {pages} {250403} (\bibinfo {year} {2025})}\BibitemShut {NoStop}%
\bibitem [{\citenamefont {Koyluoglu}\ \emph {et~al.}(2025)\citenamefont {Koyluoglu}, \citenamefont {Maskara}, \citenamefont {Feldmeier},\ and\ \citenamefont {Lukin}}]{5qhh-322q}%
  \BibitemOpen
  \bibfield  {author} {\bibinfo {author} {\bibfnamefont {Nazli~Ugur}\ \bibnamefont {Koyluoglu}}, \bibinfo {author} {\bibfnamefont {Nishad}\ \bibnamefont {Maskara}}, \bibinfo {author} {\bibfnamefont {Johannes}\ \bibnamefont {Feldmeier}}, \ and\ \bibinfo {author} {\bibfnamefont {Mikhail~D.}\ \bibnamefont {Lukin}},\ }\bibfield  {title} {\enquote {\bibinfo {title} {Floquet engineering of interactions and entanglement in periodically driven {Rydberg} chains},}\ }\href {\doibase 10.1103/5qhh-322q} {\bibfield  {journal} {\bibinfo  {journal} {Phys. Rev. Lett.}\ }\textbf {\bibinfo {volume} {135}},\ \bibinfo {pages} {113603} (\bibinfo {year} {2025})}\BibitemShut {NoStop}%
\bibitem [{\citenamefont {Tamura}\ \emph {et~al.}(2020)\citenamefont {Tamura}, \citenamefont {Yamakoshi},\ and\ \citenamefont {Nakagawa}}]{PhysRevA.101.043421}%
  \BibitemOpen
  \bibfield  {author} {\bibinfo {author} {\bibfnamefont {Hikaru}\ \bibnamefont {Tamura}}, \bibinfo {author} {\bibfnamefont {Tomotake}\ \bibnamefont {Yamakoshi}}, \ and\ \bibinfo {author} {\bibfnamefont {Ken'ichi}\ \bibnamefont {Nakagawa}},\ }\bibfield  {title} {\enquote {\bibinfo {title} {Analysis of coherent dynamics of a {Rydberg}-atom quantum simulator},}\ }\href {\doibase 10.1103/PhysRevA.101.043421} {\bibfield  {journal} {\bibinfo  {journal} {Phys. Rev. A}\ }\textbf {\bibinfo {volume} {101}},\ \bibinfo {pages} {043421} (\bibinfo {year} {2020})}\BibitemShut {NoStop}%
\bibitem [{\citenamefont {Labuhn}\ \emph {et~al.}(2016)\citenamefont {Labuhn}, \citenamefont {Barredo}, \citenamefont {Ravets}, \citenamefont {De~L{\'e}s{\'e}leuc}, \citenamefont {Macr{\`\i}}, \citenamefont {Lahaye},\ and\ \citenamefont {Browaeys}}]{labuhn2016tunable}%
  \BibitemOpen
  \bibfield  {author} {\bibinfo {author} {\bibfnamefont {Henning}\ \bibnamefont {Labuhn}}, \bibinfo {author} {\bibfnamefont {Daniel}\ \bibnamefont {Barredo}}, \bibinfo {author} {\bibfnamefont {Sylvain}\ \bibnamefont {Ravets}}, \bibinfo {author} {\bibfnamefont {Sylvain}\ \bibnamefont {De~L{\'e}s{\'e}leuc}}, \bibinfo {author} {\bibfnamefont {Tommaso}\ \bibnamefont {Macr{\`\i}}}, \bibinfo {author} {\bibfnamefont {Thierry}\ \bibnamefont {Lahaye}}, \ and\ \bibinfo {author} {\bibfnamefont {Antoine}\ \bibnamefont {Browaeys}},\ }\bibfield  {title} {\enquote {\bibinfo {title} {Tunable two-dimensional arrays of single {Rydberg} atoms for realizing quantum ising models},}\ }\href {\doibase https://doi.org/10.1038/nature18274} {\bibfield  {journal} {\bibinfo  {journal} {Nature}\ }\textbf {\bibinfo {volume} {534}},\ \bibinfo {pages} {667--670} (\bibinfo {year} {2016})}\BibitemShut {NoStop}%
\bibitem [{\citenamefont {Shen}\ \emph {et~al.}(2023)\citenamefont {Shen}, \citenamefont {Chen}, \citenamefont {Aliyu}, \citenamefont {Qin}, \citenamefont {Zhong}, \citenamefont {Loh},\ and\ \citenamefont {Lee}}]{PhysRevLett.131.080403}%
  \BibitemOpen
  \bibfield  {author} {\bibinfo {author} {\bibfnamefont {Ruizhe}\ \bibnamefont {Shen}}, \bibinfo {author} {\bibfnamefont {Tianqi}\ \bibnamefont {Chen}}, \bibinfo {author} {\bibfnamefont {Mohammad~Mujahid}\ \bibnamefont {Aliyu}}, \bibinfo {author} {\bibfnamefont {Fang}\ \bibnamefont {Qin}}, \bibinfo {author} {\bibfnamefont {Yin}\ \bibnamefont {Zhong}}, \bibinfo {author} {\bibfnamefont {Huanqian}\ \bibnamefont {Loh}}, \ and\ \bibinfo {author} {\bibfnamefont {Ching~Hua}\ \bibnamefont {Lee}},\ }\bibfield  {title} {\enquote {\bibinfo {title} {Proposal for observing yang-lee criticality in {Rydberg} atomic arrays},}\ }\href {\doibase 10.1103/PhysRevLett.131.080403} {\bibfield  {journal} {\bibinfo  {journal} {Phys. Rev. Lett.}\ }\textbf {\bibinfo {volume} {131}},\ \bibinfo {pages} {080403} (\bibinfo {year} {2023})}\BibitemShut {NoStop}%
\bibitem [{\citenamefont {Bluvstein}\ \emph {et~al.}(2021)\citenamefont {Bluvstein}, \citenamefont {Omran}, \citenamefont {Levine}, \citenamefont {Keesling}, \citenamefont {Semeghini}, \citenamefont {Ebadi}, \citenamefont {Wang}, \citenamefont {Michailidis}, \citenamefont {Maskara}, \citenamefont {Ho} \emph {et~al.}}]{bluvstein2021controlling}%
  \BibitemOpen
  \bibfield  {author} {\bibinfo {author} {\bibfnamefont {Dolev}\ \bibnamefont {Bluvstein}}, \bibinfo {author} {\bibfnamefont {Ahmed}\ \bibnamefont {Omran}}, \bibinfo {author} {\bibfnamefont {Harry}\ \bibnamefont {Levine}}, \bibinfo {author} {\bibfnamefont {Alexander}\ \bibnamefont {Keesling}}, \bibinfo {author} {\bibfnamefont {Giulia}\ \bibnamefont {Semeghini}}, \bibinfo {author} {\bibfnamefont {Sepehr}\ \bibnamefont {Ebadi}}, \bibinfo {author} {\bibfnamefont {Tout~T}\ \bibnamefont {Wang}}, \bibinfo {author} {\bibfnamefont {Alexios~A}\ \bibnamefont {Michailidis}}, \bibinfo {author} {\bibfnamefont {Nishad}\ \bibnamefont {Maskara}}, \bibinfo {author} {\bibfnamefont {Wen~Wei}\ \bibnamefont {Ho}},  \emph {et~al.},\ }\bibfield  {title} {\enquote {\bibinfo {title} {Controlling quantum many-body dynamics in driven {Rydberg} atom arrays},}\ }\href {\doibase 10.1126/science.abg2530} {\bibfield  {journal} {\bibinfo  {journal} {Science}\ }\textbf {\bibinfo {volume} {371}},\ \bibinfo {pages} {1355--1359} (\bibinfo {year}
  {2021})}\BibitemShut {NoStop}%
\bibitem [{\citenamefont {Bernien}\ \emph {et~al.}(2017)\citenamefont {Bernien}, \citenamefont {Schwartz}, \citenamefont {Keesling}, \citenamefont {Levine}, \citenamefont {Omran}, \citenamefont {Pichler}, \citenamefont {Choi}, \citenamefont {Zibrov}, \citenamefont {Endres}, \citenamefont {Greiner} \emph {et~al.}}]{bernien2017probing}%
  \BibitemOpen
  \bibfield  {author} {\bibinfo {author} {\bibfnamefont {Hannes}\ \bibnamefont {Bernien}}, \bibinfo {author} {\bibfnamefont {Sylvain}\ \bibnamefont {Schwartz}}, \bibinfo {author} {\bibfnamefont {Alexander}\ \bibnamefont {Keesling}}, \bibinfo {author} {\bibfnamefont {Harry}\ \bibnamefont {Levine}}, \bibinfo {author} {\bibfnamefont {Ahmed}\ \bibnamefont {Omran}}, \bibinfo {author} {\bibfnamefont {Hannes}\ \bibnamefont {Pichler}}, \bibinfo {author} {\bibfnamefont {Soonwon}\ \bibnamefont {Choi}}, \bibinfo {author} {\bibfnamefont {Alexander~S}\ \bibnamefont {Zibrov}}, \bibinfo {author} {\bibfnamefont {Manuel}\ \bibnamefont {Endres}}, \bibinfo {author} {\bibfnamefont {Markus}\ \bibnamefont {Greiner}},  \emph {et~al.},\ }\bibfield  {title} {\enquote {\bibinfo {title} {Probing many-body dynamics on a 51-atom quantum simulator},}\ }\href {\doibase https://doi.org/10.1038/nature24622} {\bibfield  {journal} {\bibinfo  {journal} {Nature}\ }\textbf {\bibinfo {volume} {551}},\ \bibinfo {pages} {579--584} (\bibinfo {year}
  {2017})}\BibitemShut {NoStop}%
\bibitem [{\citenamefont {Zhang}\ and\ \citenamefont {Cai}(2024)}]{PhysRevLett.132.206503}%
  \BibitemOpen
  \bibfield  {author} {\bibinfo {author} {\bibfnamefont {Tengzhou}\ \bibnamefont {Zhang}}\ and\ \bibinfo {author} {\bibfnamefont {Zi}~\bibnamefont {Cai}},\ }\bibfield  {title} {\enquote {\bibinfo {title} {Quantum slush state in {Rydberg} atom arrays},}\ }\href {\doibase 10.1103/PhysRevLett.132.206503} {\bibfield  {journal} {\bibinfo  {journal} {Phys. Rev. Lett.}\ }\textbf {\bibinfo {volume} {132}},\ \bibinfo {pages} {206503} (\bibinfo {year} {2024})}\BibitemShut {NoStop}%
\bibitem [{\citenamefont {Sable}\ \emph {et~al.}(2025)\citenamefont {Sable}, \citenamefont {Myers},\ and\ \citenamefont {Scarola}}]{746s-fv7x}%
  \BibitemOpen
  \bibfield  {author} {\bibinfo {author} {\bibfnamefont {Hrushikesh}\ \bibnamefont {Sable}}, \bibinfo {author} {\bibfnamefont {Nathan~M.}\ \bibnamefont {Myers}}, \ and\ \bibinfo {author} {\bibfnamefont {Vito~W.}\ \bibnamefont {Scarola}},\ }\bibfield  {title} {\enquote {\bibinfo {title} {Toward quantum analog simulation of many-body supersymmetry with {Rydberg} atom arrays},}\ }\href {\doibase 10.1103/746s-fv7x} {\bibfield  {journal} {\bibinfo  {journal} {Phys. Rev. Lett.}\ }\textbf {\bibinfo {volume} {135}},\ \bibinfo {pages} {033401} (\bibinfo {year} {2025})}\BibitemShut {NoStop}%
\bibitem [{\citenamefont {Scholl}\ \emph {et~al.}(2021)\citenamefont {Scholl}, \citenamefont {Schuler}, \citenamefont {Williams}, \citenamefont {Eberharter}, \citenamefont {Barredo}, \citenamefont {Schymik}, \citenamefont {Lienhard}, \citenamefont {Henry}, \citenamefont {Lang}, \citenamefont {Lahaye} \emph {et~al.}}]{scholl2021quantum}%
  \BibitemOpen
  \bibfield  {author} {\bibinfo {author} {\bibfnamefont {Pascal}\ \bibnamefont {Scholl}}, \bibinfo {author} {\bibfnamefont {Michael}\ \bibnamefont {Schuler}}, \bibinfo {author} {\bibfnamefont {Hannah~J}\ \bibnamefont {Williams}}, \bibinfo {author} {\bibfnamefont {Alexander~A}\ \bibnamefont {Eberharter}}, \bibinfo {author} {\bibfnamefont {Daniel}\ \bibnamefont {Barredo}}, \bibinfo {author} {\bibfnamefont {Kai-Niklas}\ \bibnamefont {Schymik}}, \bibinfo {author} {\bibfnamefont {Vincent}\ \bibnamefont {Lienhard}}, \bibinfo {author} {\bibfnamefont {Louis-Paul}\ \bibnamefont {Henry}}, \bibinfo {author} {\bibfnamefont {Thomas~C}\ \bibnamefont {Lang}}, \bibinfo {author} {\bibfnamefont {Thierry}\ \bibnamefont {Lahaye}},  \emph {et~al.},\ }\bibfield  {title} {\enquote {\bibinfo {title} {Quantum simulation of 2d antiferromagnets with hundreds of {Rydberg} atoms},}\ }\href {\doibase https://doi.org/10.1038/s41586-021-03585-1} {\bibfield  {journal} {\bibinfo  {journal} {Nature}\ }\textbf {\bibinfo {volume} {595}},\
  \bibinfo {pages} {233--238} (\bibinfo {year} {2021})}\BibitemShut {NoStop}%
\bibitem [{\citenamefont {Zhang}\ \emph {et~al.}(2025{\natexlab{b}})\citenamefont {Zhang}, \citenamefont {Wang}, \citenamefont {Zhang}, \citenamefont {Wang}, \citenamefont {Du}, \citenamefont {Li}, \citenamefont {Wu}, \citenamefont {Li}, \citenamefont {Hu}, \citenamefont {Zhai},\ and\ \citenamefont {Chen}}]{2gwz-65w1}%
  \BibitemOpen
  \bibfield  {author} {\bibinfo {author} {\bibfnamefont {Tao}\ \bibnamefont {Zhang}}, \bibinfo {author} {\bibfnamefont {Hanteng}\ \bibnamefont {Wang}}, \bibinfo {author} {\bibfnamefont {Wenjun}\ \bibnamefont {Zhang}}, \bibinfo {author} {\bibfnamefont {Yuqing}\ \bibnamefont {Wang}}, \bibinfo {author} {\bibfnamefont {Angrui}\ \bibnamefont {Du}}, \bibinfo {author} {\bibfnamefont {Ziqi}\ \bibnamefont {Li}}, \bibinfo {author} {\bibfnamefont {Yujia}\ \bibnamefont {Wu}}, \bibinfo {author} {\bibfnamefont {Chengshu}\ \bibnamefont {Li}}, \bibinfo {author} {\bibfnamefont {Jiazhong}\ \bibnamefont {Hu}}, \bibinfo {author} {\bibfnamefont {Hui}\ \bibnamefont {Zhai}}, \ and\ \bibinfo {author} {\bibfnamefont {Wenlan}\ \bibnamefont {Chen}},\ }\bibfield  {title} {\enquote {\bibinfo {title} {Observation of near-critical kibble-zurek scaling in {Rydberg} atom arrays},}\ }\href {\doibase 10.1103/2gwz-65w1} {\bibfield  {journal} {\bibinfo  {journal} {Phys. Rev. Lett.}\ }\textbf {\bibinfo {volume} {135}},\ \bibinfo {pages} {093403}
  (\bibinfo {year} {2025}{\natexlab{b}})}\BibitemShut {NoStop}%
\bibitem [{\citenamefont {Wang}\ \emph {et~al.}(2025{\natexlab{a}})\citenamefont {Wang}, \citenamefont {Li}, \citenamefont {Li}, \citenamefont {Gu},\ and\ \citenamefont {Liu}}]{zf2q-gxr1}%
  \BibitemOpen
  \bibfield  {author} {\bibinfo {author} {\bibfnamefont {Hanteng}\ \bibnamefont {Wang}}, \bibinfo {author} {\bibfnamefont {Chengshu}\ \bibnamefont {Li}}, \bibinfo {author} {\bibfnamefont {Xingyu}\ \bibnamefont {Li}}, \bibinfo {author} {\bibfnamefont {Yingfei}\ \bibnamefont {Gu}}, \ and\ \bibinfo {author} {\bibfnamefont {Shang}\ \bibnamefont {Liu}},\ }\bibfield  {title} {\enquote {\bibinfo {title} {Lattice defects in {Rydberg} atom arrays},}\ }\href {\doibase 10.1103/zf2q-gxr1} {\bibfield  {journal} {\bibinfo  {journal} {Phys. Rev. B}\ }\textbf {\bibinfo {volume} {112}},\ \bibinfo {pages} {205103} (\bibinfo {year} {2025}{\natexlab{a}})}\BibitemShut {NoStop}%
\bibitem [{\citenamefont {Euchner}\ and\ \citenamefont {Lesanovsky}(2025)}]{r54t-myhc}%
  \BibitemOpen
  \bibfield  {author} {\bibinfo {author} {\bibfnamefont {Simon}\ \bibnamefont {Euchner}}\ and\ \bibinfo {author} {\bibfnamefont {Igor}\ \bibnamefont {Lesanovsky}},\ }\bibfield  {title} {\enquote {\bibinfo {title} {Rydberg atom arrays as quantum simulators for molecular dynamics},}\ }\href {\doibase 10.1103/r54t-myhc} {\bibfield  {journal} {\bibinfo  {journal} {Phys. Rev. Res.}\ }\textbf {\bibinfo {volume} {7}},\ \bibinfo {pages} {L042009} (\bibinfo {year} {2025})}\BibitemShut {NoStop}%
\bibitem [{\citenamefont {Liu}\ \emph {et~al.}(2025)\citenamefont {Liu}, \citenamefont {Wu}, \citenamefont {Zhang},\ and\ \citenamefont {Yao}}]{6722-tf9c}%
  \BibitemOpen
  \bibfield  {author} {\bibinfo {author} {\bibfnamefont {Shuo}\ \bibnamefont {Liu}}, \bibinfo {author} {\bibfnamefont {Zhengzhi}\ \bibnamefont {Wu}}, \bibinfo {author} {\bibfnamefont {Shi-Xin}\ \bibnamefont {Zhang}}, \ and\ \bibinfo {author} {\bibfnamefont {Hong}\ \bibnamefont {Yao}},\ }\bibfield  {title} {\enquote {\bibinfo {title} {Supersymmetry dynamics on {Rydberg} atom arrays},}\ }\href {\doibase 10.1103/6722-tf9c} {\bibfield  {journal} {\bibinfo  {journal} {Phys. Rev. B}\ }\textbf {\bibinfo {volume} {112}},\ \bibinfo {pages} {L020301} (\bibinfo {year} {2025})}\BibitemShut {NoStop}%
\bibitem [{\citenamefont {Zeng}\ \emph {et~al.}(2025)\citenamefont {Zeng}, \citenamefont {Zhu}, \citenamefont {Chen},\ and\ \citenamefont {Shen}}]{ctwz-rhdv}%
  \BibitemOpen
  \bibfield  {author} {\bibinfo {author} {\bibfnamefont {Liang}\ \bibnamefont {Zeng}}, \bibinfo {author} {\bibfnamefont {Fei}\ \bibnamefont {Zhu}}, \bibinfo {author} {\bibfnamefont {Li}~\bibnamefont {Chen}}, \ and\ \bibinfo {author} {\bibfnamefont {Heng}\ \bibnamefont {Shen}},\ }\bibfield  {title} {\enquote {\bibinfo {title} {Rapid passage to ordered states in {Rydberg}-atom arrays},}\ }\href {\doibase 10.1103/ctwz-rhdv} {\bibfield  {journal} {\bibinfo  {journal} {Phys. Rev. A}\ }\textbf {\bibinfo {volume} {112}},\ \bibinfo {pages} {043124} (\bibinfo {year} {2025})}\BibitemShut {NoStop}%
\bibitem [{\citenamefont {Saffman}\ \emph {et~al.}(2010)\citenamefont {Saffman}, \citenamefont {Walker},\ and\ \citenamefont {M\o{}lmer}}]{RevModPhys.82.2313}%
  \BibitemOpen
  \bibfield  {author} {\bibinfo {author} {\bibfnamefont {M.}~\bibnamefont {Saffman}}, \bibinfo {author} {\bibfnamefont {T.~G.}\ \bibnamefont {Walker}}, \ and\ \bibinfo {author} {\bibfnamefont {K.}~\bibnamefont {M\o{}lmer}},\ }\bibfield  {title} {\enquote {\bibinfo {title} {Quantum information with {Rydberg} atoms},}\ }\href {\doibase 10.1103/RevModPhys.82.2313} {\bibfield  {journal} {\bibinfo  {journal} {Rev. Mod. Phys.}\ }\textbf {\bibinfo {volume} {82}},\ \bibinfo {pages} {2313--2363} (\bibinfo {year} {2010})}\BibitemShut {NoStop}%
\bibitem [{\citenamefont {Urban}\ \emph {et~al.}(2009)\citenamefont {Urban}, \citenamefont {Johnson}, \citenamefont {Henage}, \citenamefont {Isenhower}, \citenamefont {Yavuz}, \citenamefont {Walker},\ and\ \citenamefont {Saffman}}]{urban2009observation}%
  \BibitemOpen
  \bibfield  {author} {\bibinfo {author} {\bibfnamefont {E}~\bibnamefont {Urban}}, \bibinfo {author} {\bibfnamefont {Todd~A}\ \bibnamefont {Johnson}}, \bibinfo {author} {\bibfnamefont {T}~\bibnamefont {Henage}}, \bibinfo {author} {\bibfnamefont {L}~\bibnamefont {Isenhower}}, \bibinfo {author} {\bibfnamefont {DD}~\bibnamefont {Yavuz}}, \bibinfo {author} {\bibfnamefont {TG}~\bibnamefont {Walker}}, \ and\ \bibinfo {author} {\bibfnamefont {M}~\bibnamefont {Saffman}},\ }\bibfield  {title} {\enquote {\bibinfo {title} {Observation of {Rydberg} blockade between two atoms},}\ }\href {\doibase https://doi.org/10.1038/nphys1178} {\bibfield  {journal} {\bibinfo  {journal} {Nature Physics}\ }\textbf {\bibinfo {volume} {5}},\ \bibinfo {pages} {110--114} (\bibinfo {year} {2009})}\BibitemShut {NoStop}%
\bibitem [{\citenamefont {Feshbach}(1958)}]{feshbach1958unified}%
  \BibitemOpen
  \bibfield  {author} {\bibinfo {author} {\bibfnamefont {Herman}\ \bibnamefont {Feshbach}},\ }\bibfield  {title} {\enquote {\bibinfo {title} {Unified theory of nuclear reactions},}\ }\href {\doibase https://doi.org/10.1016/0003-4916(58)90007-1} {\bibfield  {journal} {\bibinfo  {journal} {Annals of Physics}\ }\textbf {\bibinfo {volume} {5}},\ \bibinfo {pages} {357--390} (\bibinfo {year} {1958})}\BibitemShut {NoStop}%
\bibitem [{\citenamefont {Chru\ifmmode \acute{s}\else \'{s}\fi{}ci\ifmmode~\acute{n}\else \'{n}\fi{}ski}\ and\ \citenamefont {Kossakowski}(2013)}]{PhysRevLett.111.050402}%
  \BibitemOpen
  \bibfield  {author} {\bibinfo {author} {\bibfnamefont {Dariusz}\ \bibnamefont {Chru\ifmmode \acute{s}\else \'{s}\fi{}ci\ifmmode~\acute{n}\else \'{n}\fi{}ski}}\ and\ \bibinfo {author} {\bibfnamefont {Andrzej}\ \bibnamefont {Kossakowski}},\ }\bibfield  {title} {\enquote {\bibinfo {title} {Feshbach projection formalism for open quantum systems},}\ }\href {\doibase 10.1103/PhysRevLett.111.050402} {\bibfield  {journal} {\bibinfo  {journal} {Phys. Rev. Lett.}\ }\textbf {\bibinfo {volume} {111}},\ \bibinfo {pages} {050402} (\bibinfo {year} {2013})}\BibitemShut {NoStop}%
\bibitem [{\citenamefont {Wu}\ \emph {et~al.}(2009)\citenamefont {Wu}, \citenamefont {Kurizki},\ and\ \citenamefont {Brumer}}]{PhysRevLett.102.080405}%
  \BibitemOpen
  \bibfield  {author} {\bibinfo {author} {\bibfnamefont {Lian-Ao}\ \bibnamefont {Wu}}, \bibinfo {author} {\bibfnamefont {Gershon}\ \bibnamefont {Kurizki}}, \ and\ \bibinfo {author} {\bibfnamefont {Paul}\ \bibnamefont {Brumer}},\ }\bibfield  {title} {\enquote {\bibinfo {title} {Master equation and control of an open quantum system with leakage},}\ }\href {\doibase 10.1103/PhysRevLett.102.080405} {\bibfield  {journal} {\bibinfo  {journal} {Phys. Rev. Lett.}\ }\textbf {\bibinfo {volume} {102}},\ \bibinfo {pages} {080405} (\bibinfo {year} {2009})}\BibitemShut {NoStop}%
\bibitem [{\citenamefont {Jing}\ \emph {et~al.}(2014)\citenamefont {Jing}, \citenamefont {Wu}, \citenamefont {Yu}, \citenamefont {You}, \citenamefont {Wang},\ and\ \citenamefont {Garcia}}]{PhysRevA.89.032110}%
  \BibitemOpen
  \bibfield  {author} {\bibinfo {author} {\bibfnamefont {Jun}\ \bibnamefont {Jing}}, \bibinfo {author} {\bibfnamefont {Lian-Ao}\ \bibnamefont {Wu}}, \bibinfo {author} {\bibfnamefont {Ting}\ \bibnamefont {Yu}}, \bibinfo {author} {\bibfnamefont {J.~Q.}\ \bibnamefont {You}}, \bibinfo {author} {\bibfnamefont {Zhao-Ming}\ \bibnamefont {Wang}}, \ and\ \bibinfo {author} {\bibfnamefont {Lluc}\ \bibnamefont {Garcia}},\ }\bibfield  {title} {\enquote {\bibinfo {title} {One-component dynamical equation and noise-induced adiabaticity},}\ }\href {\doibase 10.1103/PhysRevA.89.032110} {\bibfield  {journal} {\bibinfo  {journal} {Phys. Rev. A}\ }\textbf {\bibinfo {volume} {89}},\ \bibinfo {pages} {032110} (\bibinfo {year} {2014})}\BibitemShut {NoStop}%
\bibitem [{\citenamefont {Lienhard}\ \emph {et~al.}(2020)\citenamefont {Lienhard}, \citenamefont {Scholl}, \citenamefont {Weber}, \citenamefont {Barredo}, \citenamefont {de~L\'es\'eleuc}, \citenamefont {Bai}, \citenamefont {Lang}, \citenamefont {Fleischhauer}, \citenamefont {B\"uchler}, \citenamefont {Lahaye},\ and\ \citenamefont {Browaeys}}]{PhysRevX.10.021031}%
  \BibitemOpen
  \bibfield  {author} {\bibinfo {author} {\bibfnamefont {Vincent}\ \bibnamefont {Lienhard}}, \bibinfo {author} {\bibfnamefont {Pascal}\ \bibnamefont {Scholl}}, \bibinfo {author} {\bibfnamefont {Sebastian}\ \bibnamefont {Weber}}, \bibinfo {author} {\bibfnamefont {Daniel}\ \bibnamefont {Barredo}}, \bibinfo {author} {\bibfnamefont {Sylvain}\ \bibnamefont {de~L\'es\'eleuc}}, \bibinfo {author} {\bibfnamefont {Rukmani}\ \bibnamefont {Bai}}, \bibinfo {author} {\bibfnamefont {Nicolai}\ \bibnamefont {Lang}}, \bibinfo {author} {\bibfnamefont {Michael}\ \bibnamefont {Fleischhauer}}, \bibinfo {author} {\bibfnamefont {Hans~Peter}\ \bibnamefont {B\"uchler}}, \bibinfo {author} {\bibfnamefont {Thierry}\ \bibnamefont {Lahaye}}, \ and\ \bibinfo {author} {\bibfnamefont {Antoine}\ \bibnamefont {Browaeys}},\ }\bibfield  {title} {\enquote {\bibinfo {title} {Realization of a density-dependent peierls phase in a synthetic, spin-orbit coupled {Rydberg} system},}\ }\href {\doibase 10.1103/PhysRevX.10.021031} {\bibfield  {journal}
  {\bibinfo  {journal} {Phys. Rev. X}\ }\textbf {\bibinfo {volume} {10}},\ \bibinfo {pages} {021031} (\bibinfo {year} {2020})}\BibitemShut {NoStop}%
\bibitem [{\citenamefont {Li}\ \emph {et~al.}(2022)\citenamefont {Li}, \citenamefont {You}, \citenamefont {Shao},\ and\ \citenamefont {Li}}]{PhysRevA.105.032417}%
  \BibitemOpen
  \bibfield  {author} {\bibinfo {author} {\bibfnamefont {X.~X.}\ \bibnamefont {Li}}, \bibinfo {author} {\bibfnamefont {J.~B.}\ \bibnamefont {You}}, \bibinfo {author} {\bibfnamefont {X.~Q.}\ \bibnamefont {Shao}}, \ and\ \bibinfo {author} {\bibfnamefont {Weibin}\ \bibnamefont {Li}},\ }\bibfield  {title} {\enquote {\bibinfo {title} {Coherent ground-state transport of neutral atoms},}\ }\href {\doibase 10.1103/PhysRevA.105.032417} {\bibfield  {journal} {\bibinfo  {journal} {Phys. Rev. A}\ }\textbf {\bibinfo {volume} {105}},\ \bibinfo {pages} {032417} (\bibinfo {year} {2022})}\BibitemShut {NoStop}%
\bibitem [{\citenamefont {Scully}\ \emph {et~al.}(1998)\citenamefont {Scully}, \citenamefont {Zubairy},\ and\ \citenamefont {Milonni}}]{scully1998quantum}%
  \BibitemOpen
  \bibfield  {author} {\bibinfo {author} {\bibfnamefont {Marlan~O}\ \bibnamefont {Scully}}, \bibinfo {author} {\bibfnamefont {M~Suhail}\ \bibnamefont {Zubairy}}, \ and\ \bibinfo {author} {\bibfnamefont {Peter~W}\ \bibnamefont {Milonni}},\ }\href@noop {} {\enquote {\bibinfo {title} {Quantum optics},}\ } (\bibinfo {year} {1998})\BibitemShut {NoStop}%
\bibitem [{\citenamefont {Wind}\ \emph {et~al.}(2025)\citenamefont {Wind}, \citenamefont {Nill}, \citenamefont {Gamper}, \citenamefont {Germer}, \citenamefont {Mauth}, \citenamefont {Alt}, \citenamefont {Lesanovsky},\ and\ \citenamefont {Hofferberth}}]{wind2025entanglement}%
  \BibitemOpen
  \bibfield  {author} {\bibinfo {author} {\bibfnamefont {Cedric}\ \bibnamefont {Wind}}, \bibinfo {author} {\bibfnamefont {Chris}\ \bibnamefont {Nill}}, \bibinfo {author} {\bibfnamefont {Julia}\ \bibnamefont {Gamper}}, \bibinfo {author} {\bibfnamefont {Samuel}\ \bibnamefont {Germer}}, \bibinfo {author} {\bibfnamefont {Valerie}\ \bibnamefont {Mauth}}, \bibinfo {author} {\bibfnamefont {Wolfgang}\ \bibnamefont {Alt}}, \bibinfo {author} {\bibfnamefont {Igor}\ \bibnamefont {Lesanovsky}}, \ and\ \bibinfo {author} {\bibfnamefont {Sebastian}\ \bibnamefont {Hofferberth}},\ }\bibfield  {title} {\enquote {\bibinfo {title} {Entanglement of mechanical oscillators mediated by a {Rydberg} tweezer chain},}\ }\href {\doibase https://doi.org/10.48550/arXiv.2510.08371} {\bibfield  {journal} {\bibinfo  {journal} {arXiv preprint arXiv:2510.08371}\ } (\bibinfo {year} {2025}),\ https://doi.org/10.48550/arXiv.2510.08371}\BibitemShut {NoStop}%
\bibitem [{\citenamefont {Vidal}\ and\ \citenamefont {Werner}(2002)}]{PhysRevA.65.032314}%
  \BibitemOpen
  \bibfield  {author} {\bibinfo {author} {\bibfnamefont {G.}~\bibnamefont {Vidal}}\ and\ \bibinfo {author} {\bibfnamefont {R.~F.}\ \bibnamefont {Werner}},\ }\bibfield  {title} {\enquote {\bibinfo {title} {Computable measure of entanglement},}\ }\href {\doibase 10.1103/PhysRevA.65.032314} {\bibfield  {journal} {\bibinfo  {journal} {Phys. Rev. A}\ }\textbf {\bibinfo {volume} {65}},\ \bibinfo {pages} {032314} (\bibinfo {year} {2002})}\BibitemShut {NoStop}%
\bibitem [{\citenamefont {Horodecki}\ \emph {et~al.}(2009)\citenamefont {Horodecki}, \citenamefont {Horodecki}, \citenamefont {Horodecki},\ and\ \citenamefont {Horodecki}}]{RevModPhys.81.865}%
  \BibitemOpen
  \bibfield  {author} {\bibinfo {author} {\bibfnamefont {Ryszard}\ \bibnamefont {Horodecki}}, \bibinfo {author} {\bibfnamefont {Pawe\l{}}\ \bibnamefont {Horodecki}}, \bibinfo {author} {\bibfnamefont {Micha\l{}}\ \bibnamefont {Horodecki}}, \ and\ \bibinfo {author} {\bibfnamefont {Karol}\ \bibnamefont {Horodecki}},\ }\bibfield  {title} {\enquote {\bibinfo {title} {Quantum entanglement},}\ }\href {\doibase 10.1103/RevModPhys.81.865} {\bibfield  {journal} {\bibinfo  {journal} {Rev. Mod. Phys.}\ }\textbf {\bibinfo {volume} {81}},\ \bibinfo {pages} {865--942} (\bibinfo {year} {2009})}\BibitemShut {NoStop}%
\bibitem [{\citenamefont {Barredo}\ \emph {et~al.}(2015)\citenamefont {Barredo}, \citenamefont {Labuhn}, \citenamefont {Ravets}, \citenamefont {Lahaye}, \citenamefont {Browaeys},\ and\ \citenamefont {Adams}}]{PhysRevLett.114.113002}%
  \BibitemOpen
  \bibfield  {author} {\bibinfo {author} {\bibfnamefont {Daniel}\ \bibnamefont {Barredo}}, \bibinfo {author} {\bibfnamefont {Henning}\ \bibnamefont {Labuhn}}, \bibinfo {author} {\bibfnamefont {Sylvain}\ \bibnamefont {Ravets}}, \bibinfo {author} {\bibfnamefont {Thierry}\ \bibnamefont {Lahaye}}, \bibinfo {author} {\bibfnamefont {Antoine}\ \bibnamefont {Browaeys}}, \ and\ \bibinfo {author} {\bibfnamefont {Charles~S.}\ \bibnamefont {Adams}},\ }\bibfield  {title} {\enquote {\bibinfo {title} {Coherent excitation transfer in a spin chain of three {Rydberg} atoms},}\ }\href {\doibase 10.1103/PhysRevLett.114.113002} {\bibfield  {journal} {\bibinfo  {journal} {Phys. Rev. Lett.}\ }\textbf {\bibinfo {volume} {114}},\ \bibinfo {pages} {113002} (\bibinfo {year} {2015})}\BibitemShut {NoStop}%
\bibitem [{\citenamefont {Browaeys}\ \emph {et~al.}(2016)\citenamefont {Browaeys}, \citenamefont {Barredo},\ and\ \citenamefont {Lahaye}}]{browaeys2016experimental}%
  \BibitemOpen
  \bibfield  {author} {\bibinfo {author} {\bibfnamefont {Antoine}\ \bibnamefont {Browaeys}}, \bibinfo {author} {\bibfnamefont {Daniel}\ \bibnamefont {Barredo}}, \ and\ \bibinfo {author} {\bibfnamefont {Thierry}\ \bibnamefont {Lahaye}},\ }\bibfield  {title} {\enquote {\bibinfo {title} {Experimental investigations of dipole--dipole interactions between a few {Rydberg} atoms},}\ }\href {\doibase 10.1088/0953-4075/49/15/152001} {\bibfield  {journal} {\bibinfo  {journal} {Journal of Physics B: Atomic, Molecular and Optical Physics}\ }\textbf {\bibinfo {volume} {49}},\ \bibinfo {pages} {152001} (\bibinfo {year} {2016})}\BibitemShut {NoStop}%
\bibitem [{\citenamefont {Marcuzzi}\ \emph {et~al.}(2017)\citenamefont {Marcuzzi}, \citenamefont {Min\'a\ifmmode~\check{r}\else \v{r}\fi{}}, \citenamefont {Barredo}, \citenamefont {de~L\'es\'eleuc}, \citenamefont {Labuhn}, \citenamefont {Lahaye}, \citenamefont {Browaeys}, \citenamefont {Levi},\ and\ \citenamefont {Lesanovsky}}]{PhysRevLett.118.063606}%
  \BibitemOpen
  \bibfield  {author} {\bibinfo {author} {\bibfnamefont {Matteo}\ \bibnamefont {Marcuzzi}}, \bibinfo {author} {\bibfnamefont {Ji\ifmmode \check{r}\else~\v{r}\fi{}\'{\i}}\ \bibnamefont {Min\'a\ifmmode~\check{r}\else \v{r}\fi{}}}, \bibinfo {author} {\bibfnamefont {Daniel}\ \bibnamefont {Barredo}}, \bibinfo {author} {\bibfnamefont {Sylvain}\ \bibnamefont {de~L\'es\'eleuc}}, \bibinfo {author} {\bibfnamefont {Henning}\ \bibnamefont {Labuhn}}, \bibinfo {author} {\bibfnamefont {Thierry}\ \bibnamefont {Lahaye}}, \bibinfo {author} {\bibfnamefont {Antoine}\ \bibnamefont {Browaeys}}, \bibinfo {author} {\bibfnamefont {Emanuele}\ \bibnamefont {Levi}}, \ and\ \bibinfo {author} {\bibfnamefont {Igor}\ \bibnamefont {Lesanovsky}},\ }\bibfield  {title} {\enquote {\bibinfo {title} {Facilitation dynamics and localization phenomena in {Rydberg} lattice gases with position disorder},}\ }\href {\doibase 10.1103/PhysRevLett.118.063606} {\bibfield  {journal} {\bibinfo  {journal} {Phys. Rev. Lett.}\ }\textbf {\bibinfo {volume} {118}},\
  \bibinfo {pages} {063606} (\bibinfo {year} {2017})}\BibitemShut {NoStop}%
\bibitem [{\citenamefont {Valencia-Tortora}\ \emph {et~al.}(2024)\citenamefont {Valencia-Tortora}, \citenamefont {Pancotti}, \citenamefont {Fleischhauer}, \citenamefont {Bernien},\ and\ \citenamefont {Marino}}]{PhysRevLett.132.223201}%
  \BibitemOpen
  \bibfield  {author} {\bibinfo {author} {\bibfnamefont {Riccardo~J.}\ \bibnamefont {Valencia-Tortora}}, \bibinfo {author} {\bibfnamefont {Nicola}\ \bibnamefont {Pancotti}}, \bibinfo {author} {\bibfnamefont {Michael}\ \bibnamefont {Fleischhauer}}, \bibinfo {author} {\bibfnamefont {Hannes}\ \bibnamefont {Bernien}}, \ and\ \bibinfo {author} {\bibfnamefont {Jamir}\ \bibnamefont {Marino}},\ }\bibfield  {title} {\enquote {\bibinfo {title} {{Rydberg} platform for nonergodic chiral quantum dynamics},}\ }\href {\doibase 10.1103/PhysRevLett.132.223201} {\bibfield  {journal} {\bibinfo  {journal} {Phys. Rev. Lett.}\ }\textbf {\bibinfo {volume} {132}},\ \bibinfo {pages} {223201} (\bibinfo {year} {2024})}\BibitemShut {NoStop}%
\bibitem [{\citenamefont {Wang}\ \emph {et~al.}(2025{\natexlab{b}})\citenamefont {Wang}, \citenamefont {Wang}, \citenamefont {Panja}, \citenamefont {Wang},\ and\ \citenamefont {Liang}}]{PhysRevResearch.7.L022035}%
  \BibitemOpen
  \bibfield  {author} {\bibinfo {author} {\bibfnamefont {Yupeng}\ \bibnamefont {Wang}}, \bibinfo {author} {\bibfnamefont {Junjie}\ \bibnamefont {Wang}}, \bibinfo {author} {\bibfnamefont {Aishik}\ \bibnamefont {Panja}}, \bibinfo {author} {\bibfnamefont {Xinghan}\ \bibnamefont {Wang}}, \ and\ \bibinfo {author} {\bibfnamefont {Qi-Yu}\ \bibnamefont {Liang}},\ }\bibfield  {title} {\enquote {\bibinfo {title} {Directional transport in {Rydberg} atom arrays via kinetic constraints and temporal modulation},}\ }\href {\doibase 10.1103/PhysRevResearch.7.L022035} {\bibfield  {journal} {\bibinfo  {journal} {Phys. Rev. Res.}\ }\textbf {\bibinfo {volume} {7}},\ \bibinfo {pages} {L022035} (\bibinfo {year} {2025}{\natexlab{b}})}\BibitemShut {NoStop}%
\bibitem [{\citenamefont {Zhao}\ \emph {et~al.}(2025)\citenamefont {Zhao}, \citenamefont {Datla}, \citenamefont {Tian}, \citenamefont {Aliyu},\ and\ \citenamefont {Loh}}]{PhysRevX.15.011035}%
  \BibitemOpen
  \bibfield  {author} {\bibinfo {author} {\bibfnamefont {Luheng}\ \bibnamefont {Zhao}}, \bibinfo {author} {\bibfnamefont {Prithvi~Raj}\ \bibnamefont {Datla}}, \bibinfo {author} {\bibfnamefont {Weikun}\ \bibnamefont {Tian}}, \bibinfo {author} {\bibfnamefont {Mohammad~Mujahid}\ \bibnamefont {Aliyu}}, \ and\ \bibinfo {author} {\bibfnamefont {Huanqian}\ \bibnamefont {Loh}},\ }\bibfield  {title} {\enquote {\bibinfo {title} {Observation of quantum thermalization restricted to hilbert space fragments and ${\mathbb{z}}_{2k}$ scars},}\ }\href {\doibase 10.1103/PhysRevX.15.011035} {\bibfield  {journal} {\bibinfo  {journal} {Phys. Rev. X}\ }\textbf {\bibinfo {volume} {15}},\ \bibinfo {pages} {011035} (\bibinfo {year} {2025})}\BibitemShut {NoStop}%
\bibitem [{\citenamefont {Dalibard}\ \emph {et~al.}(1992)\citenamefont {Dalibard}, \citenamefont {Castin},\ and\ \citenamefont {M\o{}lmer}}]{PhysRevLett.68.580}%
  \BibitemOpen
  \bibfield  {author} {\bibinfo {author} {\bibfnamefont {Jean}\ \bibnamefont {Dalibard}}, \bibinfo {author} {\bibfnamefont {Yvan}\ \bibnamefont {Castin}}, \ and\ \bibinfo {author} {\bibfnamefont {Klaus}\ \bibnamefont {M\o{}lmer}},\ }\bibfield  {title} {\enquote {\bibinfo {title} {Wave-function approach to dissipative processes in quantum optics},}\ }\href {\doibase 10.1103/PhysRevLett.68.580} {\bibfield  {journal} {\bibinfo  {journal} {Phys. Rev. Lett.}\ }\textbf {\bibinfo {volume} {68}},\ \bibinfo {pages} {580--583} (\bibinfo {year} {1992})}\BibitemShut {NoStop}%
\bibitem [{\citenamefont {M{\o}lmer}\ \emph {et~al.}(1993)\citenamefont {M{\o}lmer}, \citenamefont {Castin},\ and\ \citenamefont {Dalibard}}]{molmer1993monte}%
  \BibitemOpen
  \bibfield  {author} {\bibinfo {author} {\bibfnamefont {Klaus}\ \bibnamefont {M{\o}lmer}}, \bibinfo {author} {\bibfnamefont {Yvan}\ \bibnamefont {Castin}}, \ and\ \bibinfo {author} {\bibfnamefont {Jean}\ \bibnamefont {Dalibard}},\ }\bibfield  {title} {\enquote {\bibinfo {title} {Monte carlo wave-function method in quantum optics},}\ }\href {\doibase https://doi.org/10.1364/JOSAB.10.000524} {\bibfield  {journal} {\bibinfo  {journal} {Journal of the Optical Society of America B}\ }\textbf {\bibinfo {volume} {10}},\ \bibinfo {pages} {524--538} (\bibinfo {year} {1993})}\BibitemShut {NoStop}%
\bibitem [{\citenamefont {Bohorquez}\ \emph {et~al.}(2023)\citenamefont {Bohorquez}, \citenamefont {Chinnarasu}, \citenamefont {Isaacs}, \citenamefont {Booth}, \citenamefont {Beck}, \citenamefont {McDermott},\ and\ \citenamefont {Saffman}}]{PhysRevA.108.022805}%
  \BibitemOpen
  \bibfield  {author} {\bibinfo {author} {\bibfnamefont {J.~C.}\ \bibnamefont {Bohorquez}}, \bibinfo {author} {\bibfnamefont {R.}~\bibnamefont {Chinnarasu}}, \bibinfo {author} {\bibfnamefont {J.}~\bibnamefont {Isaacs}}, \bibinfo {author} {\bibfnamefont {D.}~\bibnamefont {Booth}}, \bibinfo {author} {\bibfnamefont {M.}~\bibnamefont {Beck}}, \bibinfo {author} {\bibfnamefont {R.}~\bibnamefont {McDermott}}, \ and\ \bibinfo {author} {\bibfnamefont {M.}~\bibnamefont {Saffman}},\ }\bibfield  {title} {\enquote {\bibinfo {title} {Reducing {Rydberg}-state dc polarizability by microwave dressing},}\ }\href {\doibase 10.1103/PhysRevA.108.022805} {\bibfield  {journal} {\bibinfo  {journal} {Phys. Rev. A}\ }\textbf {\bibinfo {volume} {108}},\ \bibinfo {pages} {022805} (\bibinfo {year} {2023})}\BibitemShut {NoStop}%
\end{thebibliography}%
\end{document}